%% file: paperDraft.tex
 \documentclass[final,1p,times]{siamart220329}
\input{ex_shared}

\usepackage{tabularx}

\ifpdf
\hypersetup{
  pdftitle={A simulation method for the wetting dynamics of liquid droplets
on deformable membranes},
  pdfauthor={M. Mokbel, D. Mokbel, S. Liese, C. A. Weber, and S. Aland}
}
\fi

\begin{document}
\maketitle

\begin{abstract}
Biological cells utilize membranes and liquid-like droplets, known as biomolecular condensates, to structure their interior. 
The interaction of droplets and membranes, despite being involved in several key biological processes, is so far little understood. 
Here, we present a first numerical method to simulate the continuum dynamics of droplets interacting with deformable membranes via wetting. 
The method combines the advantages of the phase-field method for multi-phase flow simulation and the arbitrary Lagrangian-Eulerian (ALE) method for an explicit description of the elastic surface. 
The model is thermodynamically consistent, coupling bulk hydrodynamics with capillary forces, as well as bending, tension, and stretching of a thin membrane. The method is validated by comparing simulations for single droplets to theoretical results of shape equations, and its capabilities are illustrated in 2D and 3D axisymmetric scenarios. 

\end{abstract}




\section{Introduction}\label{sec:Introduction} 

{Liquid droplets provide a mechanism for the spatial organisation in living cells \cite{shin2017,alberti2021biomolecular,banani2017biomolecular}.} These droplets, referred to as biomolecular condensates, arise from condensation of protein-rich material through liquid-liquid phase separation.
Examples of such condensates are nucleoli, Cajal bodies, P-bodies, and stress granules \cite{banani2017biomolecular,brangwynne2009germline,alberti2021biomolecular}. 
One particularly interesting aspect of biomolecular condensates is their interaction with biological membranes and vesicles {via wetting \cite{brangwynne2009germline,zhao2020}}. 
Recent studies already suggest that membrane-droplet interactions are involved in several key biological processes \cite{beutel2019phase,snead2022membrane,lee2020endoplasmic,ambroggio2021dengue,su2016phase,day2021liquid}.
Understanding the dynamics of wetting of biomolecular condensates on biological membranes requires developing a physical model for the  interplay between liquid droplets and deformable surfaces.  

When a droplet and a membrane interact, surface tension $\sigma_{\text{f}}$ causes a reduction in liquid/liquid surface area, which is typically achieved by increasing membrane curvature. This is counteracted by the bending rigidity $K_B$ and the balance of the two determines the elastocapillary length scale $\sqrt{K_B/\sigma_{\text{f}}}$ of the observed curvature radius \cite{lipowsky2019,mondal2023membrane,gouveia2022capillary,mangiarotti2023}. Subsequently, membrane bending results in an apparent contact angle that deviates from droplets wetting a planar surface. 
Theoretical descriptions of membrane-droplet interactions were studied focusing on equilibrium configurations and single droplets interacting with one vesicle \cite{lipowsky2019,lipowsky2022}. A full numerical method to simulate the wetting dynamics on deformable membranes is yet missing.

Even in the absence of wetting, the continuum description of a moving elastic membrane is a highly nonlinear, nonlocal moving boundary problem. 
Over the past 15 years, various mathematical modeling approaches have been proposed to describe deformable surfaces immersed in a fluid, including immersed boundary methods \cite{kruger2011efficient, shi2014three, ong2020immersed}, level-set methods \cite{laadhari2014computing, salac2011level}, mesh-free methods \cite{rosolen2013adaptive}, particle methods \cite{noguchi2004fluid, iyer2023dynamic} and the phase-field method \cite{Aland2014b}. 
Nowadays, efficient and unconditionally stable methods exist \cite{ barrett2016stable, yang2022totally, guillen2018unconditionally, yang2017efficient}.
In addition, in the last decade numerical methods to simulate wetting on rigid structures have matured, yielding higher order and energy stable schemes to describe complex engineering applications \cite{shen2015efficient, yang2018efficient, zhu2020phase, kusumaatmaja2007modeling, aland2015}. 
However, the combination of these two fields, namely the continuum simulation of wetting of deformable membranes is so far unexplored. Computational methods are currently limited to molecular dynamics simulations which fail to simulate the most relevant time and length scales \cite{lipowsky2023remodeling, ghosh2023different}, and a gradient descent minimization of surface energies \cite{satarifard2023mutual}, which fails to produce the correct time evolution or hydrodynamics. 
{
To our knowledge only one preliminary attempt has been made to construct a dynamic simulation method \cite{pepona2021modeling}. This approach, based on the Lattice-Boltzmann method, was however not tailored to fluidic biological membranes, turned out to be relatively unstable and inaccurate, and produced spurious phases and a temporal lag between membrane and fluid movement. }

In this paper we present a first stable numerical method to describe the wetting dynamics of liquid droplets on deformable membranes. 
We represent the membrane as dimensionally reduced (hyper-) surface, which enables the accurate resolution of the high membrane curvature that is often observed as a cusp in experimental images.
On the other hand, droplets are described by a phase-field model with a diffuse interface between the two liquids. This diffusive nature regularizes the stress singularity at the contact line, making phase-field models a very natural approach to describe wetting phenomena. 
The rigorous energy-based structure of the model allows for consistent modeling of topological transitions (e.g., Reference \cite{kim2012}), and enables energy stable discrete formulations and robust time discretizations \cite{shen2015efficient, yang2018efficient,aland2015,aland2014time}.

{ 
In \autoref{sec:model}, we introduce the model equations in a thermodynamically consistent way. In \autoref{sec:discretization}, we present the numerical method. Benchmark tests and further illustrative simulations are carried out in \autoref{sec:Results}. Conclusions are drawn in \autoref{sec:Conclusion}.
}

\section{Model}
\label{sec:model}

\begin{figure}
    \centering    \includegraphics[width=0.6\textwidth]{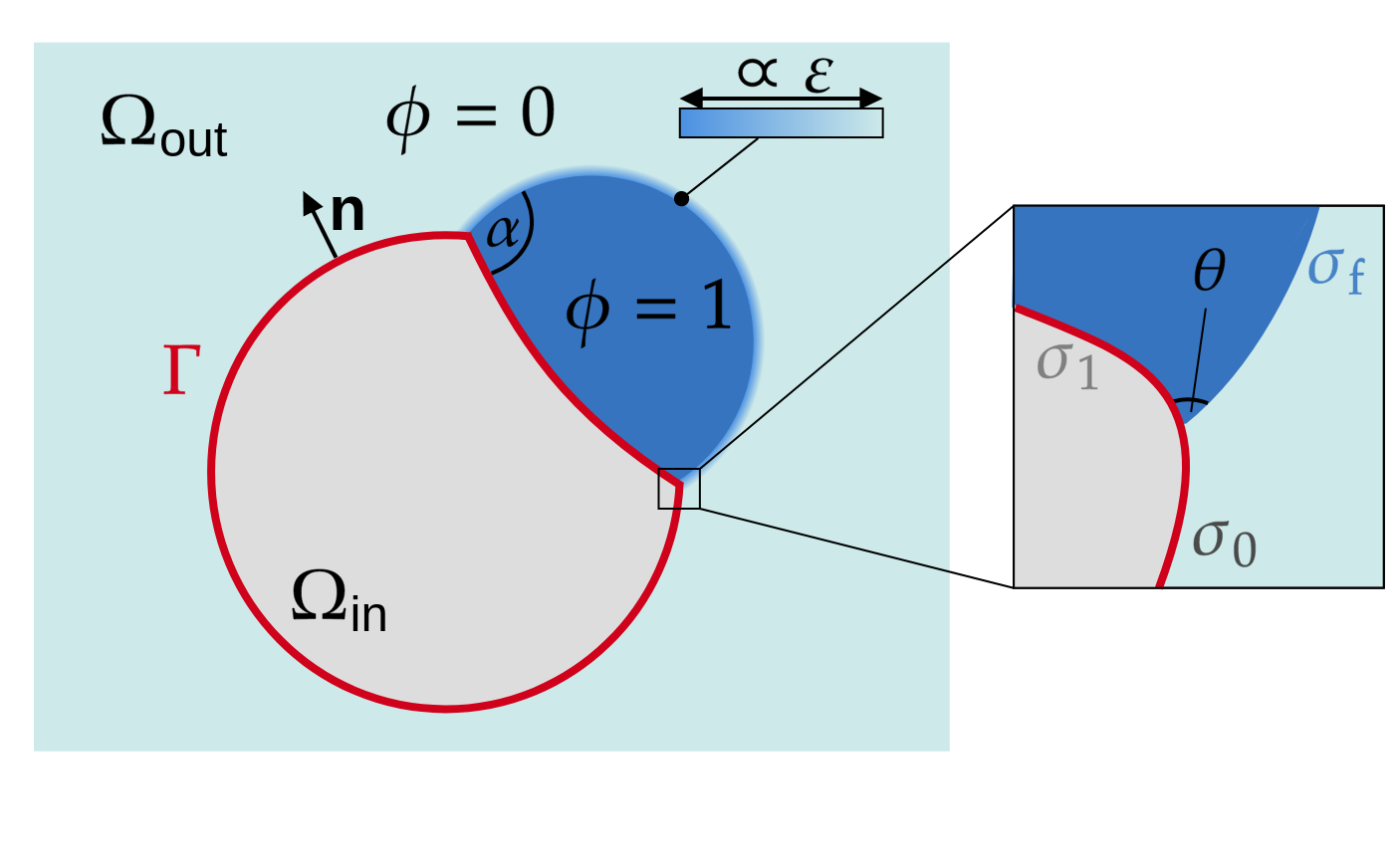}
    \caption{Illustration a droplet wetting a vesicle. A deformable closed membrane $\Gamma$ separates the fluid domain $\Omega_{\text{in}}$ and a two-phase fluid domain $\Omega_{\text{out}}$. The two fluids in $\Omega_{\text{out}}$ are indicated by the value of the phase-field function $\phi$. The fluid-fluid interface is diffuse with a thickness $\varepsilon$ and surface tension $\sigma_{\text{f}}$. The membrane $\Gamma$ has two distinct surface tensions $\sigma_0$ and $\sigma_1$ depending on whether it is adjacent to the dilute or dense phase, respectively (see inset). 
    Two different contact angles can be considered, where $\alpha$ is the macroscopic Neumann angle, that can be measured in experiments, while $\theta$ is the local Young's angle between the droplet interface and the locally flat membrane.}
    \label{fig:buddingSchematic}
\end{figure}

We investigate an elastic, closed lipid bilayer membrane immersed in and surrounded by fluids, see \autoref{fig:buddingSchematic}. The membrane is considered to have negligible thickness. The fluids undergo phase separation, resulting in two distinct phases: a condensed (droplet) phase and a dilute (ambient) phase. 
The interplay of surface tension and bending rigidity causes the membrane to bend.
%
%

Mathematically, the fluid domain $\Omega\in\mathrm{R}^d$ is separated by the membrane $\Gamma$ into two subdomains $\Omega_{\text{out}}$ and $\Omega_{\text{in}}$, referring to the outside and the inside of the membrane respectively.  \autoref{fig:buddingSchematic} depicts  the scenario of a droplet wetting a deformable vesicle
for $d=2$.  The thin membrane is represented as a hypersurface of dimension $d-1$. 
An order parameter $\phi$ is introduced to describe the two-phase fluid, i.e., we define $\phi = 0$ in the ambient phase and $\phi=1$ in the droplet phase.
For the ease of notation, we present the method for the case that phase separation occurs only on one side of the membrane. Without loss of generality, we choose
$\Omega_{\textrm{out}}$ as the phase separating domain. The opposite case of phase separation in the interior of the membrane can be equivalently handled by the proposed method, as also demonstrated in the results section.
%

To investigate the wetting and deformation of the membrane, two different contact angles are considered in this work. 
The apparent contact angle $\alpha$, which is generally easier to determine experimentally, is obtained by describing the membrane as having a kink at the contact point with the droplet interface. 
%
However, since the membrane has a bending rigidity, a kink at the contact point ($d=2$) between fluid-fluid interface and membrane does not occur. Instead, the membrane is curved smoothly (see \autoref{fig:buddingSchematic}, inset). This gives rise to the local contact angle $\theta$ between membrane and fluid-fluid interface at the contact point. 
The reaction of the membrane to bending deformations depends on the bending stiffness $K_B$ of the membrane material, which is assumed to be constant along the membrane.

\subsection{Free energy}
We consider the exemplary system for droplets which are outside the membrane. The derivation for the general case of droplets being inside or on both sides of $\Gamma$ is analogous.
The total free energy of the system $E$ is composed of contributions in the bulk ($\Omega_{\text{out}}$ and $\Omega_{\text{in}}$) and on the membrane ($\Gamma$), respectively:
\begin{align}  
E &= E_{\Omega} + E_{\Gamma} \, ,\label{eq:E_tot}\\ 
\intertext{with}
E_{\Omega}&=E_{\text{kin}}+E_{\sigma_{\text{f}}} \,,\\ 
E_{\Gamma}&=E_{\text{bend}}+E_{\sigma_{\text{m}}}+E_{\text{stretch}}\,.
\end{align}

The bulk energy $E_{\Omega}$ contains two contributions originating from the kinetic energy of the bulk fluid $E_{\text{kin}}$ and from the droplet interface $E_{\sigma_{\text{f}}}$. The membrane free energy $E_{\Gamma}$ accounts for the bending energy $E_{\text{bend}}$, the surface tension along the membrane, which leads to $E_{\sigma_{\text{m}}}$, and the stretching energy $E_{\text{stretch}}$.


The various energy contributions are tightly coupled. The fluid velocity is influenced by both the phase field and the membrane forces. Concurrently, the deformation of the membrane is affected by the velocity of the two fluids and the force response of the membrane. The membrane forces depend on the fluid in contact with the respective membrane region, i.e., they are dependent on the phase field.

We model the individual energy terms as described in the following. The kinetic energy is
\begin{align}
E_{\text{kin}}&=\int_{\Omega}\frac{\rho}{2}\vert\textbf{v}\vert^2 \,\text{d}V\, 
\end{align}
where $\rho$ denotes the fluid density,  
$\textbf{v}$ is the hydrodynamic velocity field that is continuous across $\Gamma$. The different interfaces in the system are characterized by three distinct surface tensions. The fluid-fluid tension $\sigma_{\text{f}}$, 
and the two membrane-fluid tensions, $\sigma_0$ and $\sigma_1$ (see \autoref{fig:buddingSchematic}).
Furthermore, the phase-field function $\phi$ has to vary smoothly across the fluid-fluid interface leading to a thin diffuse interface of width $\varepsilon$.
The phase-field function $\phi$  can correspond to the concentration
of molecules and the interface profile to a physical fluid-fluid interface \cite{jacqmin2000, vanDerWaals1893}.
We follow Ref.~\cite{cahn2004} and write the corresponding free energy: 
\begin{align}
E_{\sigma_{\text{f}}}&=6\sqrt{2}\sigma_{\text{f}}\int_{\Omega_{\text{out}}}
\left(
\frac{\varepsilon}{2}\vert\nabla\phi\vert^2+\frac{1}{\varepsilon}W(\phi ) \right) \,\text{d}V \, ,
 \end{align}   
where $\sigma_{\text{f}}$ is the surface tension of the interface between the droplet and the ambient liquid, 
and $W(\phi)$ is the bulk free energy density, for which we 
choose a double-well potential $W(\phi)=\frac{1}{4}\phi^2\left(1-\phi\right)^2$. 
This double-well potential is a simple
model for phase transitions in mixed systems, in particular describing the phase separation in a binary mixture. 
The prefactor $6\sqrt{2}$ originates from our choice for $W(\phi)$ \cite{aland_diss}. We use the abbreviation  $\tilde{\sigma}_{\text{f}}=6\sqrt{2}\sigma_{\text{f}}$ in the following. 

Moreover, the bending, tension and stretching energy for the deformable membrane are given as\cite{dziuk08,mokbel2021}: 
 \begin{align}  E_{\text{bend}}&=\int_{\Gamma}\frac{K_B}{2}\kappa^2\,\text{d}A \, , \\    E_{\sigma_{\text{m}}}&=\int_{\Gamma}\sigma_{\text{m}}(\phi )\,\text{d}A \, , \\
E_{\text{stretch}}&=\int_{\Gamma}\frac{K_A}{2}\left(J-1\right)^2 \,\text{d}A \, .\label{eq:Estretch}
\end{align}
Here, $K_B$ is the membrane bending stiffness and $\kappa$ is the total curvature of the membrane. The surface tension $\sigma_m\left(\phi\right)$ represents the tension along $\Gamma$ between the membrane and the fluid indicated by the value of the phase field. We use the following differentiable function of the phase field $\sigma_m\left(\phi\right)=(\sigma_1-\sigma_0)\phi^2(3-2\phi)+\sigma_0$, which implies $\sigma_m(0)=\sigma_0$ and $\sigma_m(1)=\sigma_1$. 
For flat surfaces this choice of $\sigma_m$ is known to result in correct contact angles for all level sets of the phase-field.
Finally, $K_A$ is the area dilation modulus and $J$ the determinant of the deformation gradient tensor. The latter gives the local area stretch of an infinitesimal surface  area segment and is equal to $1$ for an undilated  surface. 

\textbf{Remark:}
In 3D elasticity theory, the response of an isotropic elastic body to elastic deformations can be described by two material-specific parameters: Young's modulus $E$ and Poisson's ratio $\nu$. 
For a thin elastic material of thickness $d$, these parameters are typically reformulated into surface parameters, for example, the area dilation modulus $K_A$ and area shear modulus $K_S$ and the bending modulus $K_B$ \cite{poisson2019}. Considering a rectangular surface element, $K_A$ describes the response of the membrane to in-plane area changes with a constant aspect ratio of the surface element. $K_S$ provides information about the response to in-plane shear deformations with constant area of the surface element. However, for lipid bilayer membranes, no shear forces occur and hence, $K_S=0$. The elastic surface parameters can be calculated directly from Young's modulus, Poisson ratio, and membrane thickness:
\begin{align}
K_A = \frac{dE}{2(1-\nu)},\quad
K_B = \frac{d^3E}{24(1-\nu^2)} \, . 
\label{eq:moduli}
\end{align}

\subsection{Model derivation}

To derive the dynamics of the system in a thermodynamically consistent manner, we compute the time variation of the total energy \ref{eq:E_tot}. The details of these calculations can be found in the Supplementary Material Appendix I. 
Assuming incompressible fluid flow ($\nabla \cdot \textbf{v}=0$) and constant mass density $\rho$, we obtain
\begin{align}
    d_t E=&~~~\int_{\Gamma}\partial^{\bullet}_t\phi\left(\frac{\delta E_{\sigma_m}}{\delta\phi} + \tilde{\sigma}_{\text{f}}\varepsilon\, \textbf{n}\cdot\nabla\phi\right)\notag\\
    &\qquad\ \ +\textbf{v}\cdot\left(\frac{\delta E_{\sigma_m}}{\delta\Gamma}+\frac{\delta E_{\text{bend}}}{\delta\Gamma}+\frac{\delta E_{\text{stretch}}}{\delta\Gamma}-\tilde{\sigma}_{\rm f}\varepsilon \, \nabla\phi(\nabla\phi\cdot{\bf n})\right) \,\text{d}A \notag \\
    &+\int_{\Omega_{\text{out}}}\partial^{\bullet}_t\phi\frac{\delta E_{\sigma_{\text{f}}}}{\delta\phi} + {\bf v}\cdot\left(\nabla\cdot(\tilde{\sigma}_{\rm f}\varepsilon\nabla\phi\otimes\nabla\phi)\right)\,\text{d}V \notag \\
    &+\int_{\Omega_{\text{in}}\cup\Omega_{\text{out}}}\rho{\bf v}\cdot\partial_t^ \bullet {\bf v}\,\text{d}V, \label{eq:dtE main}
\end{align}
where $\partial_t^\bullet = \partial_t + {\bf v}\cdot\nabla$ denotes the material time derivative and $\textbf{n}$ denotes the normal vector of $\Gamma$ pointing to $\Omega_{\text{out}}$. The variational derivatives are
\begin{align}
    \frac{\delta E_{\sigma_{\text{f}}}}{\delta\phi} &= \tilde{\sigma}_{\text{f}}\left(\frac{1}{\varepsilon}W'(\phi) - \varepsilon\Delta\phi\right) \, ,\\
    \frac{\delta E_{\sigma_m}}{\delta\phi} &= \sigma_m'(\phi)\, , \\
    \frac{\delta E_{\sigma_m}}{\delta\Gamma} &= \kappa\sigma_{\text{m}}(\phi) \, \textbf{n}-\nabla_{\Gamma}\sigma_{\text{m}}(\phi)  \, ,  \\        
    \frac{\delta E_{\text{bend}}}{\delta\Gamma}&=K_B\left(\Delta_{\Gamma}\kappa-2K_g\kappa+\frac{1}{2}\kappa^3 \right)\textbf{n} \, ,\label{eq:bendingForce}\\
    \frac{\delta E_{\text{stretch}}}{\delta\Gamma}&\approx K_A\kappa(J-1) \, \textbf{n}-K_A\nabla_{\Gamma}(J-1) \, .
\end{align}
The last term is a linear approximation, which is quite accurate for small stretching deformations. Since the area dilation modulus $K_A$ is large for membranes, taking into account small stretching deformations is sufficient  ($J\approx 1$). 
Consistent with other derivations~\cite{cates_lecture, reichl2016modern, mokbel2021,vanBrummelen2017}, we recover in Eq.~\ref{eq:dtE main} the capillary stress of a diffuse interface  $\textbf{S}_{\sigma_{\text{f}}}:=-\tilde{\sigma}_{\text{f}}\varepsilon\nabla\phi\otimes\nabla\phi$.
We note that our derivation considered the membrane to be impermeable; a condition which will be relaxed later to account for slow water flux through the membrane \cite{fettiplace1980water}.

Using the usual viscous and pressure stress ${\bf S}=\eta(\nabla{\bf v}+\nabla{\bf v}^T)-p{\bf I}$, we arrive at the following evolution equations consistent with the second law of thermodynamics: 
\begin{align}
    \partial_t\phi + \textbf{v}\cdot\nabla\phi     
    &=M\Delta\frac{\delta E_{\sigma_{\text{f}}}}{\delta\phi}\, ,&\text{ in }\Omega_{\text{out}}\, , \label{eq:govEq1}\\
    \tilde{\sigma}_{\text{f}}\varepsilon\, \textbf{n}\cdot\nabla\phi&=-\frac{\delta E_{\sigma_m}}{\delta\phi} \, ,&\text{ on }\Gamma  \, ,\label{eq:govEq2}\\    
    {\bf n}\cdot \nabla\frac{\delta E_{\sigma_{\text{f}}}}{\delta\phi} &=0 \, , &\text{ on }\Gamma\, , \label{eq:govEq3}\\
    \rho(\partial_t\textbf{v}+\textbf{v}\cdot\nabla\textbf{v})&=\nabla\cdot\left(\textbf{S}+\chi_{\text{out}}\textbf{S}_{\sigma_{\text{f}}}\right)\, , &\text{ in }\Omega_{\text{in}}\cup\Omega_{\text{out}} \, ,\\
    \nabla\cdot\textbf{v}&=0\, ,&\text{ in }\Omega_{\text{in}}\cup\Omega_{\text{out}} \, ,\\
    \textbf{n}\cdot\left[\textbf{S}+\chi_{\text{out}}\textbf{S}_{\sigma_{\text{f}}}\right]^{\text{out}}_{\text{in}}&=-
    \frac{\delta E_{\sigma_m}}{\delta\Gamma}
    -\frac{\delta E_{\text{bend}}}{\delta\Gamma}-\frac{\delta E_{\text{stretch}}}{\delta\Gamma}\, ,&\text{ on }\Gamma \, .\label{eq:govEq6}
\end{align}
Here, the  $M$ is a mobility coefficient that is considered to be constant and $\chi_{\text{out}}$ is the characteristic function of $\Omega_{\text{out}}$, i.e., $\chi_{\text{out}}=1$ in $\Omega_{\text{out}}$ and $\chi_{\text{out}}=0$ in $\Omega\setminus\Omega_{\text{out}}$. Furthermore, $\left[\cdot\right]_{\text{in}}^{\text{out}}$ denotes the jump of the quantity in brackets across $\Gamma$.
Note, that a boundary condition for velocity $\textbf{v}$ across the membrane is not required, due to our assumption of a continuous velocity field, which implies a no-slip condition along both sides of the membrane.

Eqs.~\ref{eq:govEq1}-\ref{eq:govEq6} correspond to the incompressible Navier-Stokes equations including a capillary stress, and thereby coupled to the Cahn-Hilliard equation for the binary fluid domain, together with an interfacial balance of forces.
Paired with Eq.~\ref{eq:dtE main} we obtain non-increasing total free energy, 
\begin{align}
    d_t E=-\int_{\Omega_{\text{out}}} M\bigg|\nabla\frac{\delta E_{\sigma_{\text{f}}}}{\delta\phi}\bigg| ^2\,\text{d}V-\int_{\Omega_{\text{in}}\cup\Omega_{\text{out}}}\frac{\eta}{2}\lVert\nabla\textbf{v}+\nabla\textbf{v}^T\rVert^2_{\text{F}}\,\text{d}V\leq 0\,.
\end{align}

Hence, Eqs.~\ref{eq:govEq1}-\ref{eq:govEq6} are consistent governing equations of a membrane wetted on one side. The opposing case of wetting inside the membrane follows by interchanging the indices $()_{\textrm{in}}$ and  $()_{\textrm{out}}$.
Wetting on both sides can be handled by introducing one distinct phase field for each side of the membrane, together with a second membrane surface energy and a second capillary stress tensor.

\section{Discretization}\label{sec:discretization}
\subsection{ALE discretization}\label{sec:ALEDiscretization}
We use a fitted finite element method to discretize the system in space. The two subdomains $\Omega_{\text{out}}$ and $\Omega_{\text{in}}$ are discretized on two separate but connected, moving numerical grid partitions.  An arbitrary Lagrangian-Eulerian (ALE) method is used to advect the grid. Thereby, grid points on $\Gamma$ move with the material velocity, that is, they are material points. In contrast, the grid points in the fluids move with a continuous harmonic extension of this velocity in order to keep a proper shape of the mesh. This grid velocity $\mathbf{w}$ is calculated in the two subdomains by solving the Laplace problem
\begin{align}
    \Delta\mathbf{w} &= 0  \, , &&\text{in}\ \Omega_{\rm in}\cup\Omega_{\rm out}\, , \nonumber\\
    \mathbf{w} &= \mathbf{v} - P\,p_{\text{diff}}\,\phi_d\, \textbf{n}\, , &&\text{on}\ \Gamma \, ,\nonumber\\
    \mathbf{w} &= 0 \, , &&\text{on}\ \partial\Omega/\Gamma \, .\label{eq:ALELaplaceProblem}
\end{align}
A slight membrane permeability is incorporated into the model by the term $P\,p_{\text{diff}}\,\phi_d\textbf{n}$ from the boundary condition on $\Gamma$, with the material specific permeability $P$ of the membrane.
For zero permeability ($P=0$), the approach corresponds to a standard ALE method. However, membranes can be permeable to water on large time scales \cite{fettiplace1980water} which is regulated by $P>0$. 
Water flux across the membrane is driven by the pressure difference $p_{\text{diff}}=(p_{\text{out}}-p_{\text{in}}) - p_{\text{eq}}$ with the pressures $p_i$ in $\Omega_i$. 
The pressure difference $p_{\text{eq}}$ is the equilibrium pressure difference which may result from osmotic  pressure. Here, we choose $p_{\text{eq}}$ equal to mechanical pressure jump $(p_{\text{out}}-p_{\text{in}})$ in the spherical equilibrium state. Consequently, for small droplets in 2D, $p_{\text{eq}}$ is approximated by $p_{\text{eq}}=\sigma_0/r_{\text{m}} + K_B/r_{\text{m}}^3$, where $r_{\text{m}}$ is the radius of a spherical membrane with the same volume as the present membrane. In 3D, the bending force of a sphere vanishes, and hence $p_{\text{eq}}=2\sigma_0/r_{\text{m}} $. 
The indicator function $\phi_d(x)$ is designed to ensure, that the droplets themselves are not permeable. The permeability condition should only be imposed at a distance $d$ away from the droplet interface to prevent instabilities caused by high pressure differences in the vicinity of the phase-field interface. 
This distance is dependent on the physical properties of the system (shell size, droplet number, droplet size, etc.) and can be obtained empirically. Hence, we use
\begin{align}
    \phi_d(x)=\left\{\begin{array}{cl}
         1,&\text{if } x\in\Gamma\cap\left\{x | \phi(x)<0.5\right\}\cap\left\{x | \underset{{x_{0.5}}\in\Gamma_{0.5} }{\min}\|x-x_{0.5}\| > d\right\}  \\
         0,&\text{else} 
    \end{array}\right. \, ,
\end{align}
where $\Gamma_{0.5} = \Gamma\cap\left\{x|\phi(x)=0.5\right\}$. Further, a typical value for $d$ is $3\varepsilon$. 

Finally, the calculated grid velocity $\mathbf{w}$ is subtracted in all convective terms of the governing equations. Therefore, the material derivate $\partial_t^\bullet$ is replaced by
\begin{align}
 \partial_t^\bullet \to   \partial_{t,x^*} + \left(\mathbf{v}-\mathbf{w}\right)\cdot\nabla \, ,
\end{align}
where $\partial_{t,x^*}$ defines the time derivative of a quantity along a moving grid point.

\subsection{Time discretization}
The problem is discretized with equidistant time steps of size $\tau$. At each time step, the general workflow of the numerical solution procedure is as follows. First, the coupled system of momentum balance, mass balance, and phase-field evolution, \autoref{eq:govEq1}-\ref{eq:govEq6} is solved in one monolithic system. Afterwards the grid velocity $\mathbf{w}$ is computed as explained in \autoref{sec:ALEDiscretization}. Then in a last step, the grid is updated, that is, each grid point is moved by the corresponding value of $\mathbf{w}$. An IMEX (implicit/explicit) Euler method is used to formulate a time discretization of \autoref{eq:govEq1}-\ref{eq:govEq6}, which is linear in the solution variables. We denote quantities on discrete time points by a superscript, where $()^{n-1}$ refers to the previous time step and $()^n$ refers to the current time step. 
In order to compute bending, stretching, and surface tension forces, we compute the curvature vector $\textbf{\textkappa}=-\kappa\bm{n}$ for the current time step as a solution variable of the system, where we use the fact that $\Delta_\Gamma\text{id}_\Gamma = -\kappa\mathbf{n}$ with $\text{id}_\Gamma$ the identity map on $\Gamma$. This way, the stability of the system is increased by an implicit treatment of all membrane forces.

The capillary stress $\bm{S}_{\sigma_{\text{f}}}$ is taken semi-implicitly by using $\tilde{\sigma}_{\text{f}}\varepsilon\nabla\phi^n\otimes\nabla\phi^{n-1}$. Note that we define the involved tensor product such that \[\nabla\cdot\left(\nabla\phi^n\otimes\nabla\phi^{n-1}\right)_i = \sum_j\partial_j\left(\partial_j\phi^n\partial_i\phi^{n-1}\right).\]  This treatment must be accompanied by using the new velocity field $\mathbf{v}^n$ in the advective term of the phase field. Further, the nonlinear derivative of the double well potential $W'(\phi)$ is linearized by a first order Taylor expansion. We end up with the time discrete system:

In each time step $n$, find $\mathbf{v}^n$, $p_{\text{out}}^n$, $p_{\text{in}}^n$, $\phi^n$, $\mu^n, \textbf{\textkappa}^n,\mathbf{f}_\Gamma^n$ such that
\begin{align}
\begin{rcases}
\begin{aligned}
      \overline{\rho}^{n-1}\left(\frac{\mathbf{v}^n-\mathbf{v}^{n-1}}{\tau} + \left(\mathbf{v}^{n-1}-\mathbf{w}^{n-1}\right)\cdot\nabla\mathbf{v}^n\right)= \\
    -\nabla\left(\chi_{\text{out}}^{n-1} p_{\text{out}}^n + \chi_{\text{in}}^{n-1} p_{\text{in}}^n\right) + \nabla\cdot\left(\overline{\eta}^{n-1}\left(\nabla\mathbf{v}^n + \left(\nabla\mathbf{v}^n\right)^T\right)\right)\\
    - \nabla\cdot\left(\chi_{\text{out}}^{n-1}\tilde{\sigma}_{\text{f}}\varepsilon\nabla\phi^n\otimes\nabla\phi^{n-1}\right) - \delta_\Gamma^{n-1}\mathbf{f}_\Gamma^n \\
    \nabla\cdot\mathbf{v}^n = 0  
\end{aligned}
\end{rcases} &\quad\text{in}\ \Omega^{n-1}\\
\begin{rcases}
\begin{aligned}
\left(\frac{\phi^n-\phi^{n-1}}{\tau} + \mathbf{v}^{n}\cdot\nabla\phi^{n-1} -\mathbf{w}^{n-1}\cdot\nabla\phi^{n}\right) = M\Delta\mu^n\\
\mu^n = \tilde{\sigma}_{\text{f}}\varepsilon^{-1}\left(W'\left(\phi^{n-1}\right) + W''\left(\phi^{n-1}\right)\left(\phi^n - \phi^{n-1}\right)\right) + \tilde{\sigma}_{\text{f}}\varepsilon\Delta\phi^n
\end{aligned}
\end{rcases} &\quad\text{in}\ \Omega_{\text{out}}^{n-1} \label{eq:mu}\\
\begin{rcases}
\begin{aligned}
\text{id}_\Gamma^{n} = \tau\left(\mathbf{v}^n - P\,p_{\text{diff}}^{n-1}\,\phi_d^{n-1}\textbf{n}^{n-1} \right) + \text{id}_\Gamma^{n-1}\\
\textbf{\textkappa}^n = \Delta_\Gamma\text{id}_\Gamma^{n} \\
\mathbf{f}_\Gamma^n = \sigma_s(\phi^{n-1})\textbf{\textkappa}^n + \sigma_s'(\phi^{n-1})\nabla_\Gamma\phi^n - \left(\frac{\delta E_{\text{bend}}}{\delta \Gamma}\right)^{n-\frac{1}{2}} - \left(\frac{\delta E_{\text{stretch}}}{\delta \Gamma}\right)^{n-\frac{1}{2}}
\end{aligned}
\end{rcases}  &\quad\text{on}\ \Gamma
\end{align}
Here, $\mu=\frac{\delta E_{\sigma_{\text{f}}}}{\delta\phi}$ denotes the chemical potential,  $\mathbf{v}^{n-1}$ and $\phi^{n-1}$ denote the corresponding quantities from the last time step, but after the applied change of the grid point coordinates due to the mesh update. The bending force $\left(\frac{\delta E_{\text{bend}}}{\delta \Gamma}\right)^{n-\frac{1}{2}}$ is computed with a semi implicit scheme which will be described in \autoref{sec:membraneForceSystem}, where linear occurrences of the curvature vector $\textbf{\textkappa}$ are taken from the current time step. 
The stretching force $\left(\frac{\delta E_{\text{stretch}}}{\delta \Gamma}\right)^{n-\frac{1}{2}}$  mainly depends on the curvature vector and the deformation gradient determinant $J$. 
An implicit prediction of $J$ is crucial to eliminate the numerical stiffness of the coupling between shape evolution and hydrodynamics. Accordingly, we introduce the evolution equation
\begin{align}
    \partial^{\bullet}_t J = J\nabla_\Gamma\cdot\mathbf{v}\, .
\end{align}
Thus, an implicit prediction of $J^n$ is given by updating the current value $J^{n-1}$ according to 
\begin{align}
    J^n = J^{n-1} + \tau J^{n-1}\nabla_\Gamma\cdot\mathbf{v}^n \, ,
\end{align}
where $J^{n-1}$ is computed from coordinates of the surface grid. 
The density $\overline{\rho}$ depends on the subdomain and on the phase field, i.e. 
\begin{align}
    \overline{\rho} = 
    \begin{cases}
        \rho_{\text{in}} & \text{in}\ \Omega_{\text{in}} \\
        \rho_{\text{out}}(\phi) = \rho_{\text{out},1}\phi + \rho_{\text{out},0}(1 - \phi) & \text{in}\ \Omega_{\text{out}} \\
    \end{cases}
\end{align}
where $\rho_{\text{in}}$, $\rho_{\text{out,1}}$, and $\rho_{\text{out,2}}$ are prescribed constants. The viscosity $\overline{\nu}$ is defined similarly. 

\subsection{Space discretization}
\begin{figure}[!ht]
    \centering
    \includegraphics[width=0.5\textwidth]{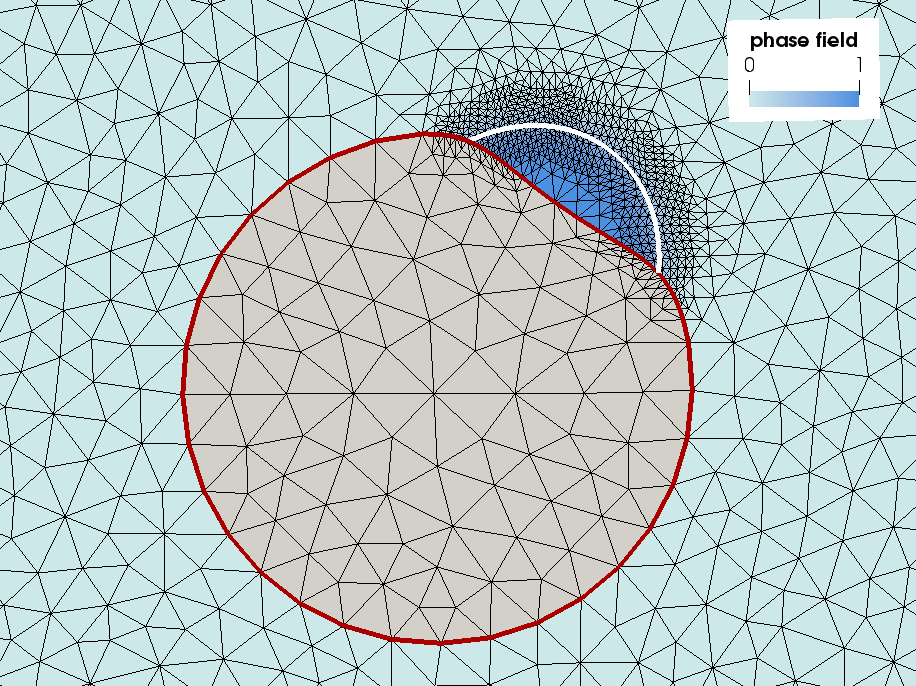}
    \caption{Illustration of the numerical mesh. The membrane triangulation $T_{\Gamma}$ (red) is fitted to the triangulations $T_{\text{in}}$ (gray) and $T_{\text{out}}$. The phase field $\phi$ representing the droplet is defined in $T_{\text{out}}$ and controls the adaptive refinement. The white line indicates the fluid-fluid interface as 0.5-level set of $\phi$.}
    \label{fig:meshExample}
\end{figure}

\noindent
In our finite-element approach, we consider the triangulations $T_{\text{out}}$ of $\Omega_{\text{out}}$ and $T_{\text{in}}$ of $\Omega_{\text{in}}$, whereby the membrane $\Gamma$, that connects the two domains, is triangulated by $T_\Gamma =T_{\text{out}}\cap T_{\text{in}}$. In particular, the connection is ensured by the fact that every grid point on the interface of $\Omega_{\text{out}}$ is also a grid point on the interface of $\Omega_{\text{in}}$. This fitted grid approach allows us to exactly enforce continuity of velocity across $\Gamma$. The jump conditions of the stress are implemented by definition of separate finite element spaces for $T_{\text{out}}$ and $T_{\text{in}}$, s.t. each degree of freedom on the membrane $T_\Gamma$ exists twice - one belonging to $T_{\text{out}}$ and one belonging to $T_{\text{in}}$. An example for the numerical mesh is shown in \autoref{fig:meshExample}.

We now present the combined discrete system of Equations \ref{eq:govEq1}-\ref{eq:govEq6} 
in the weak form. Adopting the approach from Reference \cite{mokbel2020}, we introduce the following finite element spaces:
\begin{align}
V_h &:= \left\{v\in C^0(\overline{\Omega})\cap H^1_0(\Omega) \,|\, v_{|k} \in P_2(k), k\in T_{\text{out}}\cup T_{\text{in}}\right\} \,,\nonumber\\
P_{h,i} &:= \left\{q\in C^0(\overline{\Omega}_i)\cap L^2_0(\Omega_i) \,|\, q_{|k} \in P_1(k), k\in T_i\right\} ,\quad i = \text{out, in}\, , \nonumber\\
C_{h} &:= \left\{c\in C^0(\overline{\Omega}_{\text{out}})\,|\, c_{|k} \in P_2(k), k\in T_{\text{out}}\right\} \, , \nonumber \\
S_{h} &:= \left\{f\in C^0(\Gamma)\cap L^2_0(\Gamma),|\, f_{|k} \in P_1(k), k\in T_{\Gamma}\right\} \, . \label{eq:finiteElementSpaces}
\end{align}
Here, $V_h$ is the finite element space for the components of the velocity $\mathbf{v}$. It ensures continuity of the respective variables across $T_\Gamma$. Note that in \autoref{eq:finiteElementSpaces}, $i$ denotes a placeholder for the distinction between the two fluid domains. The use of two separate spaces, $P_{h,\text{out}}$ and $P_{h,\text{in}}$, is motivated by the discontinuity of pressure across $\Gamma$ caused by the closed elastic membrane. The use of standard finite element spaces for the discretization of the discontinuous pressure leads to poor numerical properties, with an approximation order of only $\mathcal{O}(\sqrt{h})$ w.r.t. to the $L^2$ norm \cite{gross2007}. Accordingly, the usage of two separate spaces, $P_{h,\text{out}}$ and $P_{h,\text{in}}$, extends the standard Taylor-Hood finite element space by additional degrees of freedom of the pressure at the interface, such that the discontinuity can be exactly resolved. The finite element space $C_h$ refers to $\phi$ and $\mu$ in Equations \ref{eq:mu}. 
The remaining finite element space $S_h$ is used to compute the force exerted by the elasticity of the membrane, namely surface tension, in-plane stretching, and bending stiffness.

While all equations are addressed within a unified coupled system, we opt to present them individually in the following, segregated into fluid motion, phase separation, and membrane forces, to enhance readability.

\subsubsection{Navier-Stokes system}
\noindent
With the previous arguments, we can establish a uniform weak formulation of the momentum and mass conservation equation for the combined domain $\Omega$. The weak form reads:

\noindent
Find $(\mathbf{v}^n,p^n_{\text{out}},p^n_{\text{in}}) \in V^d_h\times P_{h,\text{out}}\times P_{h,\text{in}}$, such that $\forall (\mathbf{z},q_{\text{out}},q_{\text{in}})\in V^d_h\times P_{h,\text{out}}\times P_{h,\text{in}}$:
\begin{align}
0 =&~~~ \int_{\Omega^{n-1}} \overline{\rho}^{n-1}\left(\frac{\mathbf{v}^n-\mathbf{v}^{n-1}}{\tau} + \left(\mathbf{v}^{n-1}-\mathbf{w}^{n-1}\right)\cdot\nabla\mathbf{v}^n\right)\cdot\mathbf{z} \nonumber\\
 &~~~~~~~~+\left(\overline{\eta}^{n-1}\left(\nabla\mathbf{v}^n + \left(\nabla\mathbf{v}^n\right)^T\right)\right):\nabla\mathbf{z}\  \text{d}V\nonumber\\
& - \int_{\Omega_{\text{out}}^{n-1}} \tilde{\sigma}_{\text{f}}\varepsilon\nabla\phi^n\otimes\nabla\phi^{n-1}\, :\, \nabla\mathbf{z} + p^n_{\text{out}}\nabla\cdot\mathbf{z}\ \text{d}V \nonumber\\
& - \int_{\Omega_{\text{in}}^{n-1}}  p^n_{\text{in}}\nabla\cdot\mathbf{z}\ \text{d}V \nonumber\\
& - \int_{\Gamma^{n-1}} \mathbf{f}_{\Gamma}^n\cdot\mathbf{z}\ \text{d}A \label{eq:navierStokesWeakCoupled}\, ,\\
0=&~~~ \int_{\Omega^{n-1}_{\text{out}}} q_{\text{out}}\nabla\cdot\mathbf{v}^n \ \text{d}V +\int_{\Omega^{n-1}_{\text{in}}} q_{\text{in}}\nabla\cdot\mathbf{v}^n \ \text{d}V\, .
\end{align}

\subsubsection{Cahn-Hilliard system}
\noindent
The weak form of Eq. \ref{eq:mu} equipped with boundary conditions \ref{eq:govEq2}-\ref{eq:govEq3} reads

Find ($\phi^n, \mu^n)\in C_h\times C_h$ such that $\forall (\psi_1,\psi_2)\in C_h\times C_h$:
\begin{align}
    0 = & \int_{\Omega_{\text{out}}^{n-1}} \left(\frac{\phi^n-\phi^{n-1}}{\tau} + \mathbf{v}^{n}\cdot\nabla\phi^{n-1} -\mathbf{w}^{n-1}\cdot\nabla\phi^{n}\right)\psi_1 + M\nabla\mu^n\cdot\nabla\psi_1 \ \text{d}V \label{eq:cahnHilliardWeakCoupled}\\
    0= & \int_{\Omega_{\text{out}}^{n-1}}  \mu^n\psi_2 -  \tilde{\sigma}_{\text{f}}\varepsilon\nabla\phi^n\cdot\nabla\psi_2 - \frac{\tilde{\sigma}_{\text{f}}}{\varepsilon}\left(W'\left(\phi^{n-1}\right) + W''\left(\phi^{n-1}\right)\left(\phi^n - \phi^{n-1}\right)\right)\psi_2 \ \text{d}V\\
    & - \int_{\Gamma^{n-1}} \sigma_s'(\phi^{n-1})\psi_2 \ \text{d}A \, .\nonumber
\end{align}
Let us note that the Cahn-Hilliard system is still directly coupled to the Navier-Stokes system by the presence of $\mathbf{v}^n$ in \autoref{eq:cahnHilliardWeakCoupled} and the presence of $\phi^n$ in \autoref{eq:navierStokesWeakCoupled}. This coupling removes time step restrictions for small interface thickness \cite{aland2014time}. 

\subsubsection{Membrane force system}\label{sec:membraneForceSystem}
\noindent
The bending force \autoref{eq:bendingForce} includes the total surface curvature and its derivatives. In order to formulate a stable discretization, these terms need to be included implicitly to the system. Therefore, we use the weak formulation from Dziuk \cite{dziuk08}:
\begin{align}
    \int_\Gamma \frac{\delta E_{\text{bend}}}{\delta\Gamma}{\textbf{\textpsi}} = &K_B\int_\Gamma \frac{1}{2}|\textbf{\textkappa}|^2\nabla_\Gamma\cdot\textbf{\textpsi} +  \nabla_\Gamma{\textbf{\textkappa}}:\nabla_\Gamma\textbf{\textpsi} +  \nabla_\Gamma\cdot\textbf{\textkappa}\nabla_\Gamma\cdot\textbf{\textpsi}  \nonumber\\ &~~~~~~~~-  \left(\nabla_\Gamma\textbf{\textpsi} + \nabla_\Gamma\textbf{\textpsi}^T\right)\mathbf{P}:\nabla_\Gamma\textbf{\textkappa}~\mathrm{d}A\label{eq:bendingForceWeakForm}
\end{align}
for any vector valued test function $\textbf{\textpsi}$.  Derivations can be found in \cite{dziuk08,barrett2020}. However, we also derived \autoref{eq:bendingForceWeakForm} in the present notation in the Supplementary Material Appendix II. The resulting weak form of the combined membrane force system reads:

Find $(\textbf{\textkappa}^n,\mathbf{f}^n_\Gamma)\in S_h^d\times S_h^d$, such that $\forall (\mathbf{s}_1,\mathbf{s}_2)\in S_h^d\times S_h^d$:
\begin{align}
    0= & \int_{\Gamma^{n-1}} \textbf{\textkappa}^n\cdot\mathbf{s}_1 + \nabla_\Gamma \left(\text{id}_\Gamma^{n-1} + \tau\mathbf{v}^n\right) : \nabla_\Gamma\mathbf{s}_1 \ \text{d}A \label{eq:curvatureVectorWeakForm} \, , \\
    0 = & \int_{\Gamma^{n-1}} \mathbf{f}^n_{\Gamma}\cdot\mathbf{s}_2 -  \left(\sigma_s(\phi^{n-1})\textbf{\textkappa}^n + \sigma_s'(\phi^{n-1})\nabla_{\Gamma}\phi^n\right)\cdot\mathbf{s}_2 -K_B\left[-\frac{1}{2}\left( \textbf{\textkappa}^n\cdot\textbf{\textkappa}^{n-1}\nabla_{\Gamma}\cdot\mathbf{s}_2\right) \right.\nonumber\\ 
    &\qquad \left.- \nabla_{\Gamma}{\textbf{\textkappa}^n}:\nabla_{\Gamma}\mathbf{s}_2 -  \nabla_{\Gamma}\cdot\textbf{\textkappa}^n\nabla_{\Gamma}\cdot\mathbf{s}_2  +  \left(\nabla_{\Gamma}\mathbf{s}_2 + \nabla_{\Gamma}\mathbf{s}_2^T\right)\mathbf{P}^{n-1}:\nabla_{\Gamma}\textbf{\textkappa}^n\right] \nonumber\\
   &\qquad-K_A\left[(J^{n}-1)\textbf{\textkappa}^{n-1}\cdot\mathbf{s}_2+\nabla_{\Gamma}(J^{n}-1)\cdot\mathbf{s}_2\right]\ \text{d}A \, . \label{eq:membraneForceWeakForm}
\end{align}
Note that the resulting force $\mathbf{f}^n_\Gamma$ is as well directly coupled to the Navier-Stokes-Cahn-Hilliard system by the presence of $\mathbf{v}^n$ in \autoref{eq:curvatureVectorWeakForm} and $\phi^n$ in \autoref{eq:membraneForceWeakForm}. The force itself is plugged into the Navier-Stokes system in \autoref{eq:navierStokesWeakCoupled}. As the discrete system is linear, Eqs. \ref{eq:navierStokesWeakCoupled}-\ref{eq:membraneForceWeakForm} can be monolithically assembled and no subiterations are needed.

\subsection{Remeshing}
\noindent
While the grid update method outlined in \autoref{sec:ALEDiscretization} is well-suited for minor deformations of the membrane, it encounters challenges when substantial membrane deformations occur. Particularly in the proximity of the contact region between the droplet and the membrane, elements may be pushed towards each other, resulting in distorted triangles/tetrahedra with small minimum angles, undermining accuracy and stability of the finite element method. Consequently, we use a mechanism to reconstruct the mesh whenever the minimum angle of any triangle falls below a specified threshold  $\alpha_{min}$ (in the simulations, $\alpha_{min}$ has been chosen between 5° and 10°). The following retriangulation steps are performed:
\begin{enumerate}
    \item Extract the membrane grid points. Note that the adaptively refined grid has to be used here in order to retain accurate information on the membrane shape.
    \item Generate a new (unrefined) mesh with \texttt{gmsh} \cite{gmsh}, using the outer boundary box and the membrane grid points from the old grid.
    \item Interpolate the data of the solution $\phi$ and the deformation gradient determinant $J$ to the new grid (see the step below). $J$ is needed to preserve the information of the unstretched initial state of the membrane. Refine the grid afterwards by refinement of all elements that are on the fluid-droplet interface (i.e. where values of $\phi$ are in between $0.05$ and $0.95$). Repeat the step until the refinement level of the old grid is reached. Always exclude elements from the refinement, which have a size smaller than a prescribed minimum (typically in the order of the average element volume on the interface of the old grid). This has to be done since for the new unrefined grid, grid points on the membrane from the old refined grid have been used.
    \item Interpolate the data of the solution $\mathbf{v}$ by solution of the following problem, which ensures that the velocity remains divergence free on the new grid:
    \begin{align*}
        \mathbf{v}_{\text{new}} - \nabla L &= \mathbf{v}_{\text{old}} \, ,\\
        \nabla\cdot\mathbf{v}_{\text{new}} &= 0 \, ,  
    \end{align*}
    where $L$ acts as a Lagrange multiplier. The problem is solved with finite elements on the new grid. However, the right hand side operator in the weak form is of the form $\int_{T_{\text{new}}}\mathbf{v}_{\text{old}}\cdot\textbf{\textpsi}$ with the triangulation $T_{\text{new}}$ based on the new grid and test functions $\textbf{\textpsi}$ defined on the new grid. In contrast, the ansatz functions are defined on the old grid.  Computation of such an integral involves identifying all intersections that an element on the new grid has with the old grid. These intersections are then triangulated, and this triangulation is utilized to calculate the integral on the respective element on $T_{\text{new}}$. In the implementation, this process is carried out within the \texttt{dune} module \texttt{dune-grid-glue}. Further details can be found in \cite{dune-grid-glue}. 
    \item Update all data structures to the new grid and solve the next time step with the old solutions from step 4.
\end{enumerate}

\begin{center}
\begin{table}
    \caption{\label{tab:model_value}Characteristic parameters of membrane vesicles and biomolecular coacervates}
    \begin{tabular}{| p{0.55\textwidth}p{0.3\textwidth}p{0.15\textwidth}| }
    \hline
    \textbf{Parameter name} & \textbf{value} & \textbf{reference} 
    \\ \hline
     bending rigidity $K_{B}$ &  $10^{-19}-10^{-18}\,Nm$  & \cite{etthakafy2017,lee2001} \\
     surface tension coacervate $\sigma_f$ &  $10^{-4}-1 \,mN/m$  & \cite{wang2021surface} \\
    viscosity $\eta$ &    &  \\
    \hspace{0.5cm} cytosol &  $1-10 \,Pa\,s$  &  \cite{wirtz2009} \\
    \hspace{0.5cm} coacervate &  $0.1-10^3 \,Pa\,s$  & \cite{wang2021surface} \\
    interface width $\epsilon$ &  $5-10 \,nm$  & \cite{tsanai2021} \\
    cytosol density $\rho$ &  $\sim 10^3\,kg/m^3$  & \cite{moran2010} \\
    area dilation modulus $K_A$ &  $100-300\,mN/m$  & \cite{evans1990,picas2012} \\
    scaled permeability $Pp_{\rm diff}$ &  $10^{-7}-10^{-5}\,m/s$  & \cite{olbrich2000,sacerdote2005} \\
    \hline
    \end{tabular}
    \end{table}
\end{center}

\section{Numerical Test} \label{sec:Results}
\noindent
We perform numerical tests to validate the proposed method and to illustrate its potential. 
The majority of the physical parameters are selected within the realm of realistic biological systems (see Table \ref{tab:model_value}). Although some parameters are well-established, others, such as viscosity and surface tension of condensates, can exhibit significant variability, with observed ranges spanning four orders of magnitude. 
Given that the aim of this paper is not to concentrate on a particular biological system, we intentionally vary parameters across a broad spectrum. Despite the numerical method's capability to handle phase-dependent viscosity \cite{aland2012benchmark}, we choose constant viscosity throughout this work for simplicity.

\subsection{Validation} \label{sec:validation}
\noindent
To validate the numerical method, we consider membrane remodeling induced by a single droplet in 2D. Assuming an initially flat membrane sheet, which is significantly larger than the droplet, enables the analytical derivation of the stationary morphology based on shape equations (see Supplementary Material Appendix III). 
To compare with these analytical shapes, we setup the numerical method to simulate the evolution up to the stationary state. 
The membrane ends are connected to the boundary of the domain with a free-slip condition. Membrane stretching and permeability are neglected ($K_A=P=0$).

\begin{figure}[!ht]
    \centering
    \vspace{0.25cm}
    \begin{tabular}{ccc}
        {\includegraphics[width=0.3\textwidth]{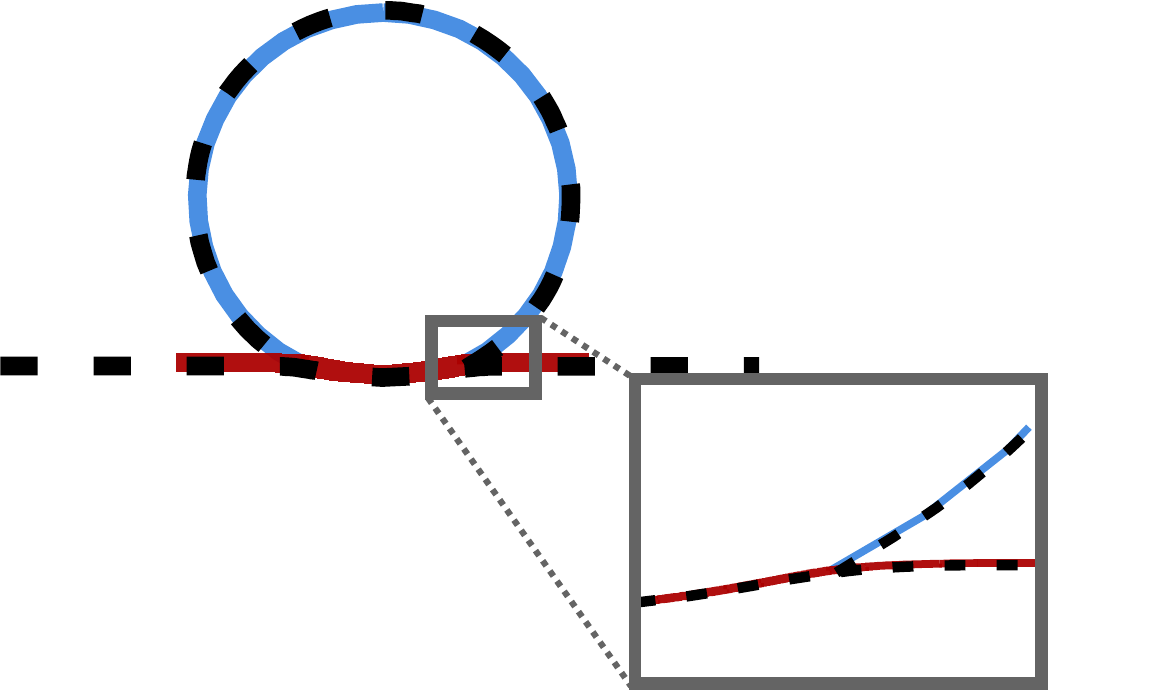}} & {\includegraphics[width=0.3\textwidth]{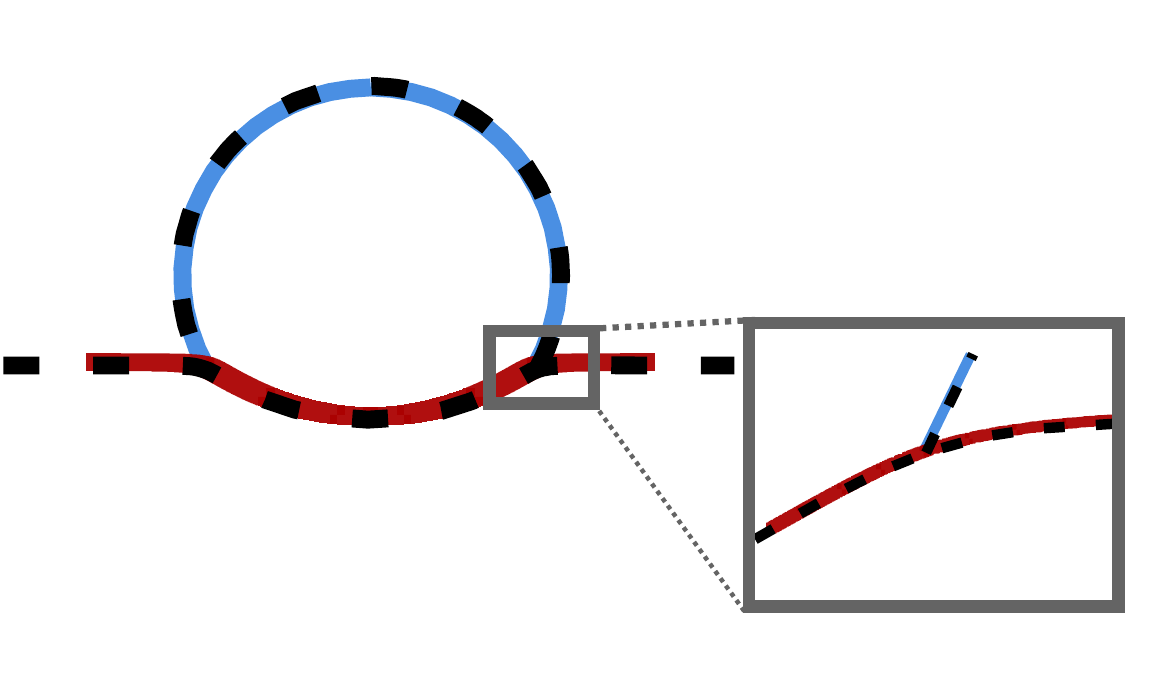}} & {\includegraphics[width=0.3\textwidth]{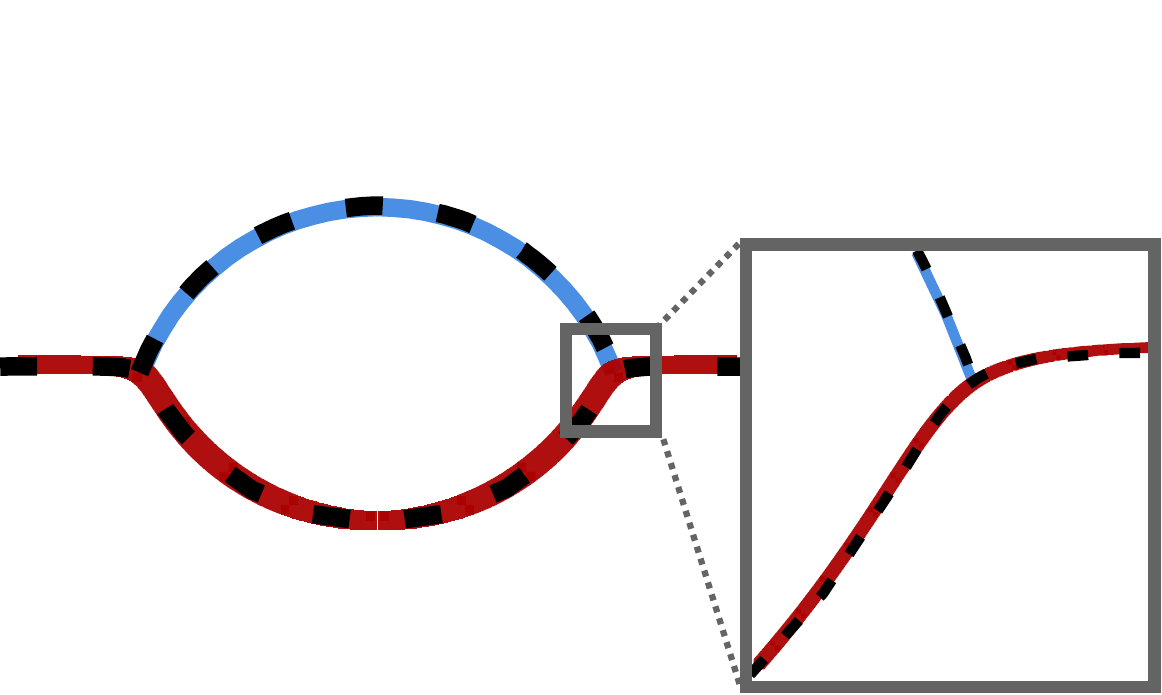}}\\
        \footnotesize{$\cos\theta = -0.95$, $\frac{\sigma_0}{\sigma_{\text{f}}}=0.75$,}   & \footnotesize{$\cos\theta = -0.75$, $\frac{\sigma_0}{\sigma_{\text{f}}}=0.75$,}  & \footnotesize{$\cos\theta = -0.25$, $\frac{\sigma_0}{\sigma_{\text{f}}}=0.75$,}  \\
        \footnotesize{$\alpha=168.15^\circ$} & \footnotesize{$\alpha=153.62^\circ$} & \footnotesize{$\alpha=135.95^\circ$} \\ 
        {\includegraphics[width=0.3\textwidth]{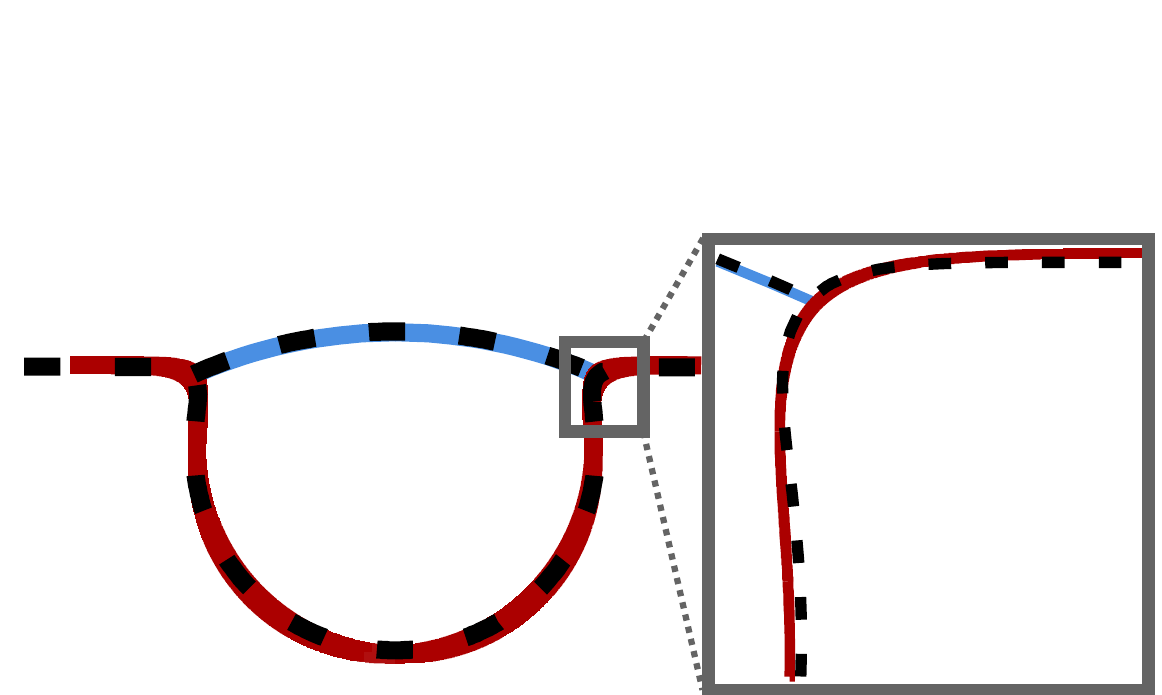}} & {\includegraphics[width=0.3\textwidth]{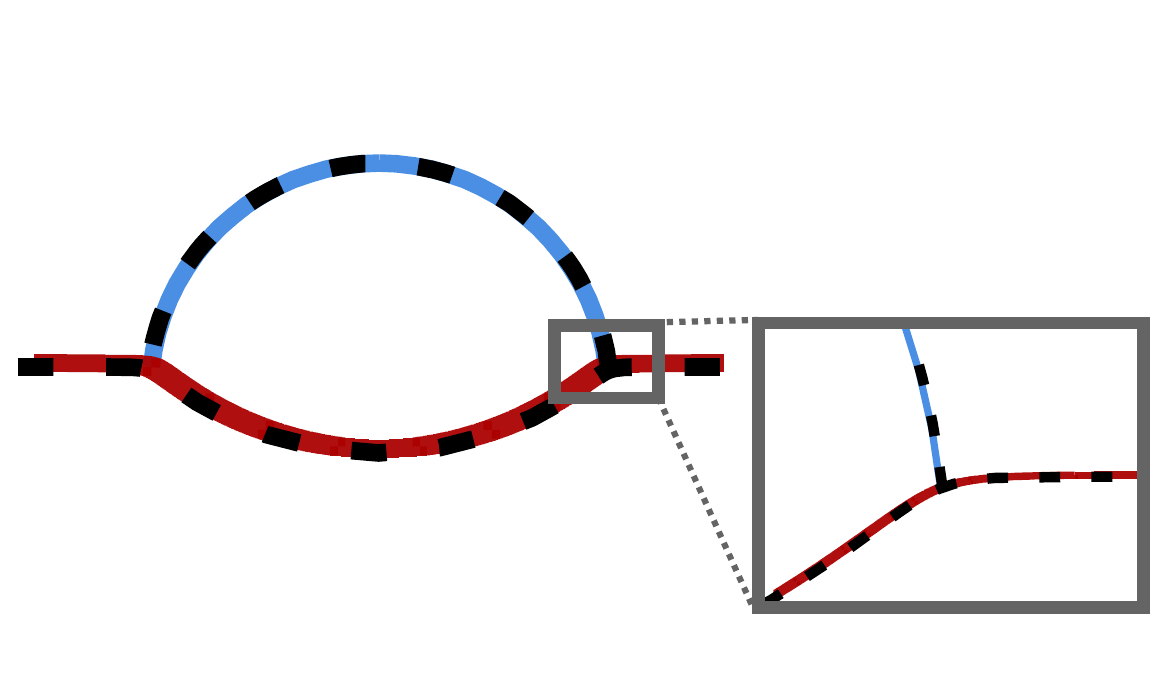}} & {\includegraphics[width=0.3\textwidth]{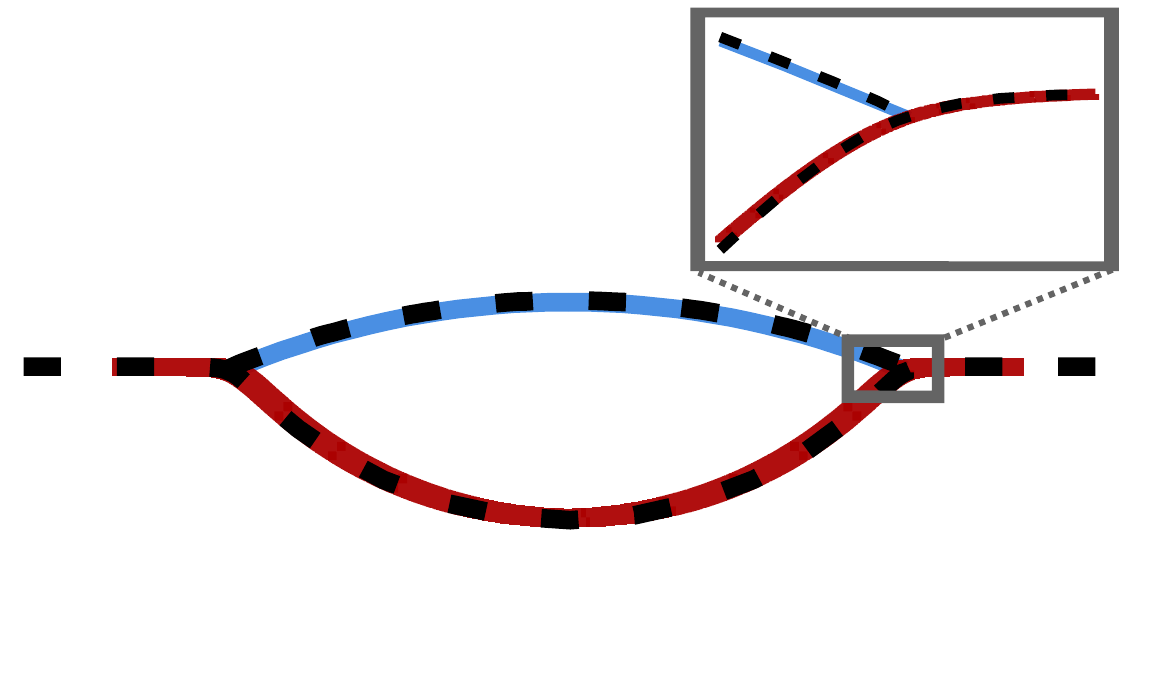}} \\
        \footnotesize{$\cos\theta = 0.35$, $\frac{\sigma_0}{\sigma_{\text{f}}}=0.75$, } & \footnotesize{$\cos\theta = -0.25$, $\frac{\sigma_0}{\sigma_{\text{f}}}=1.25$, } & \footnotesize{$\cos\theta = 0.75$, $\frac{\sigma_0}{\sigma_{\text{f}}}=1.25$, } \\
        \footnotesize{$\alpha=138.32^\circ$} & \footnotesize{$\alpha=124.22^\circ$} & \footnotesize{$\alpha=71.79^\circ$}  
    \end{tabular}
    \caption{Comparison of stationary shapes obtained by simulations (dashed lines) and theoretical model (solid lines) for a single droplet (blue) on an initially flat  membrane (red). Compelling agreement between the two solutions is even visible in the close-up around the three-phase contact point. \textbf{Parameters}: Initial droplet shape is spherical (radius $50$nm) in the theoretical model, and half-spherical cap (radius $70.71$nm) in the simulations. The used surface tensions can be determined from \autoref{eq:theta} with $\sigma_{\text{f}}=15\mu$N/m. 
    Further, $K_A=0, K_B=8\cdot 10^{-20}$Nm, $P=0$.}
    \label{fig:validation}
\end{figure}

\autoref{fig:validation} shows the comparison of stationary shapes between numerical and analytical results.
The given local contact angle $\theta$ 
is related to the surface tensions through the Young-Dupré-law,
\begin{align}
    \cos\theta = \frac{\sigma_0-\sigma_1}{\sigma_{\text{f}}} \, ,\label{eq:theta}
\end{align}
and differs from the apparent contact angle $\alpha$ (see \autoref{fig:buddingSchematic}), which in the limit of both large droplets (compared to the capillarity length) and large vesicles (compared to the droplet) is 
given by Neumann's law \cite{kusumaatmaja2011}:
\begin{align}
    \cos\alpha = \frac{\sigma_0^2-\sigma_1^2-\sigma_{\text{f}}^2}{2\sigma_{\text{f}}\sigma_1}.
    \label{eq: angle alpha Neumann}
\end{align}

The surface tensions have been chosen s.t. different resulting shapes can be compared, from a nearly spherical droplet with very low bending of the membrane \autoref{fig:validation}(a) over strong indentation of the membrane by the droplet \autoref{fig:validation}(d) to a lens-like shape of the droplet on the membrane \autoref{fig:validation}(f). For all different parameter configurations chosen, we observe a perfect agreement between numerical results and the results of the shape equations. 
This verifies that the numerical method accurately incorporates all surface forces and resolves the locally high membrane curvature in the triple contact point by the adaptive grid. 

The hydrodynamics of the numerical method and the underlying flow solver have been previously validated in various applications on cells and shells in fluid flow \cite{Mokbel,mokbel2023}. Furthermore, benchmark validations against other numerical codes \cite{aland2012benchmark}, experimental data \cite{Aland2013,aland2014time}, and convergence studies \cite{aland2012benchmark, mokbel2020} have been successfully conducted. 
In addition, the wetting boundary conditions were validated for dynamic situations and applied to scenarios of moving contact lines \cite{MokbelStickSlip}. 
Consequently, the method exhibits precise capabilities in capturing dynamic phenomena, which strongly suggest that it is well-suited for application in future studies leveraging realistic experimental data. 

\subsection{Droplets on a vesicle}
\noindent
This chapter aims to showcase the outcomes of numerical investigations involving droplets adhering to an initially spherical membrane. We start with single droplets on the membrane. 
Afterwards, multiple droplets on the membrane are considered. To begin, we examine the interaction dynamics of two closely positioned droplets and explore how the bending stiffness of the membrane influences the interplay between these two droplets. Finally, we consider liquid-liquid phase separation around the membrane with condensates of low, neutral and high wettability, respectively. These simulations illustrate the capabilites of the present model.

\subsubsection{Single droplet} 
In the following we consider a single droplet situated outside a spherical vesicle membrane. Three distinct combinations of the surface tensions were selected to explore shape evolutions into different categories of stationary states, as introduced in \cite{tiemei2022}: adhesion, lens shape, and partially wrapped/endocytosis.

\begin{figure}
    \centering
    {\includegraphics[width=0.24\textwidth]{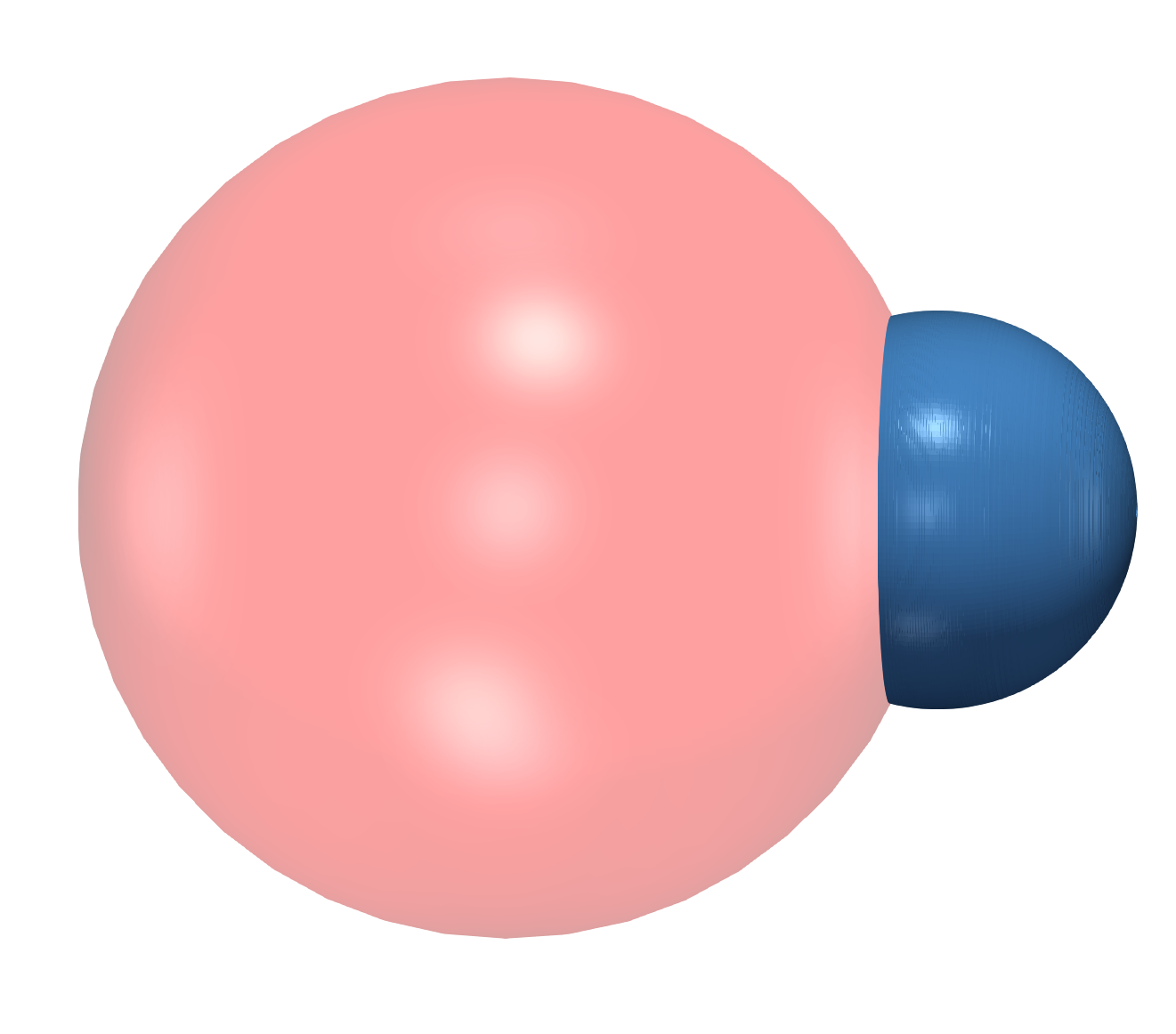}}

    {\footnotesize $t=0~$s}
    \vspace{0.25cm}
    
    \textbf{Adhesion}: $ \sigma_{\text{f}} = 30\,\mu\text{N/m},\qquad \sigma_0 = 15\,\mu\text{N/m},\qquad \sigma_1 = 30\,\mu\text{N/m}$ \vspace{0.5cm}
    \begin{tabular}{cccc}
        {\includegraphics[width=0.24\textwidth]{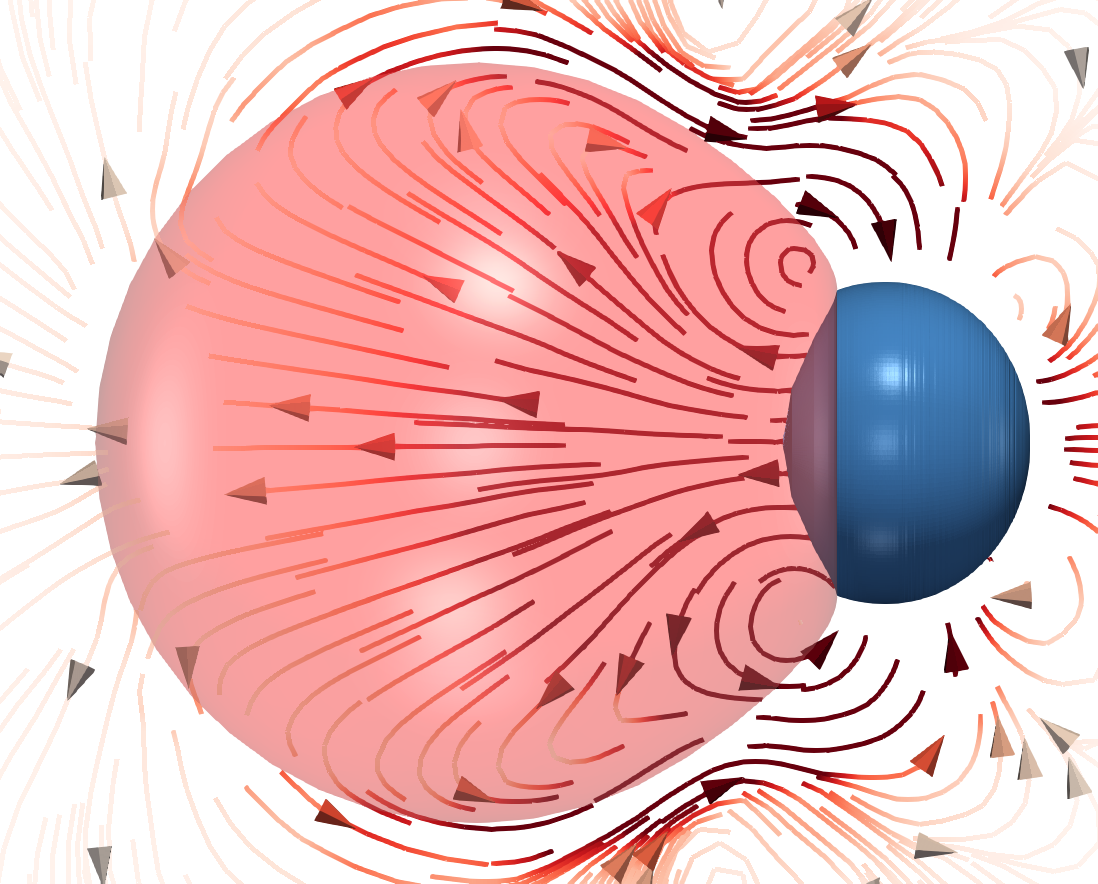}}  &
        {\includegraphics[width=0.24\textwidth]{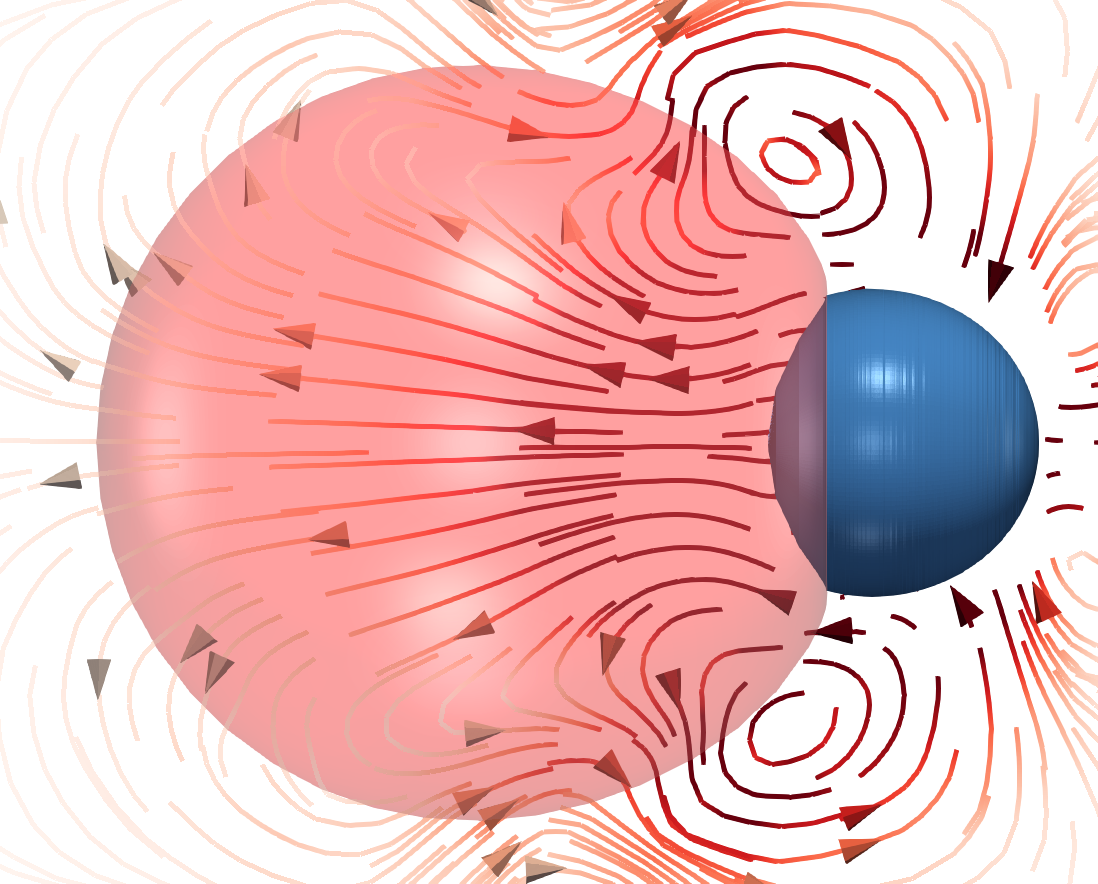}}&
        {\includegraphics[width=0.24\textwidth]{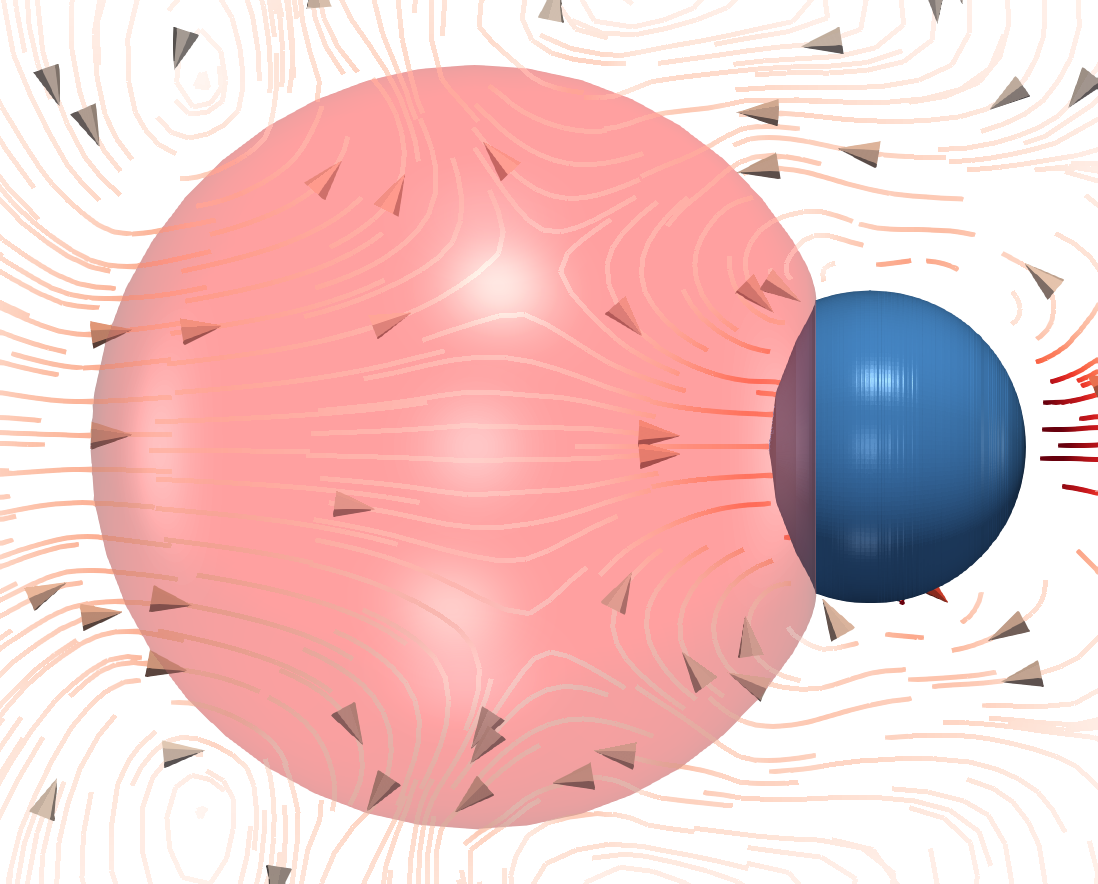}}
    &{\includegraphics[width=0.24\textwidth]{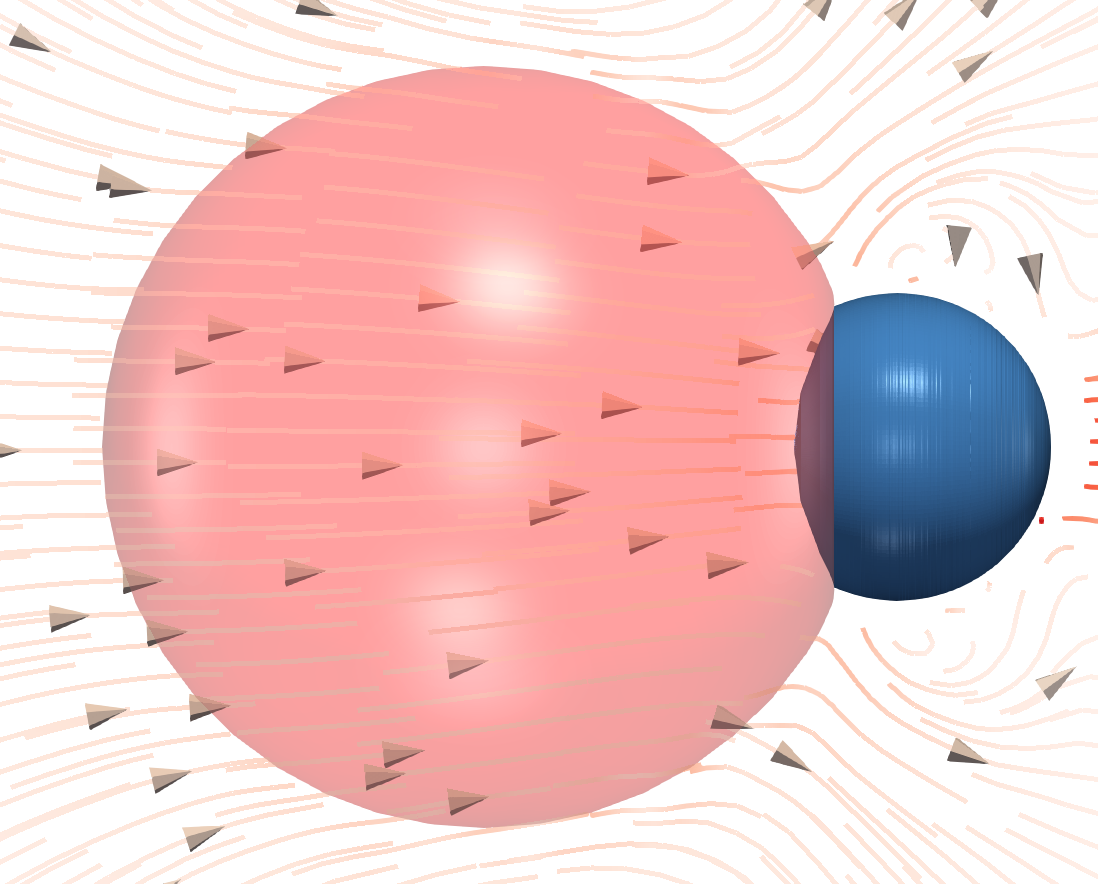}}  \\  
    \footnotesize $t=0.5~$s & 
        \footnotesize $t=1~$s & 
        \footnotesize $t=3~$s & 
        \footnotesize $t=18~$s
    \end{tabular} 
    
    \vspace{0.1cm}
   \textbf{Lens shape}: $ \sigma_{\text{f}} = 30\,\mu\text{N/m},\qquad \sigma_0 = 30\,\mu\text{N/m},\qquad \sigma_1 = 15\,\mu\text{N/m}$ \vspace{0.1cm}
    \begin{tabular}{cccc}
     {\includegraphics[width=0.24\textwidth]{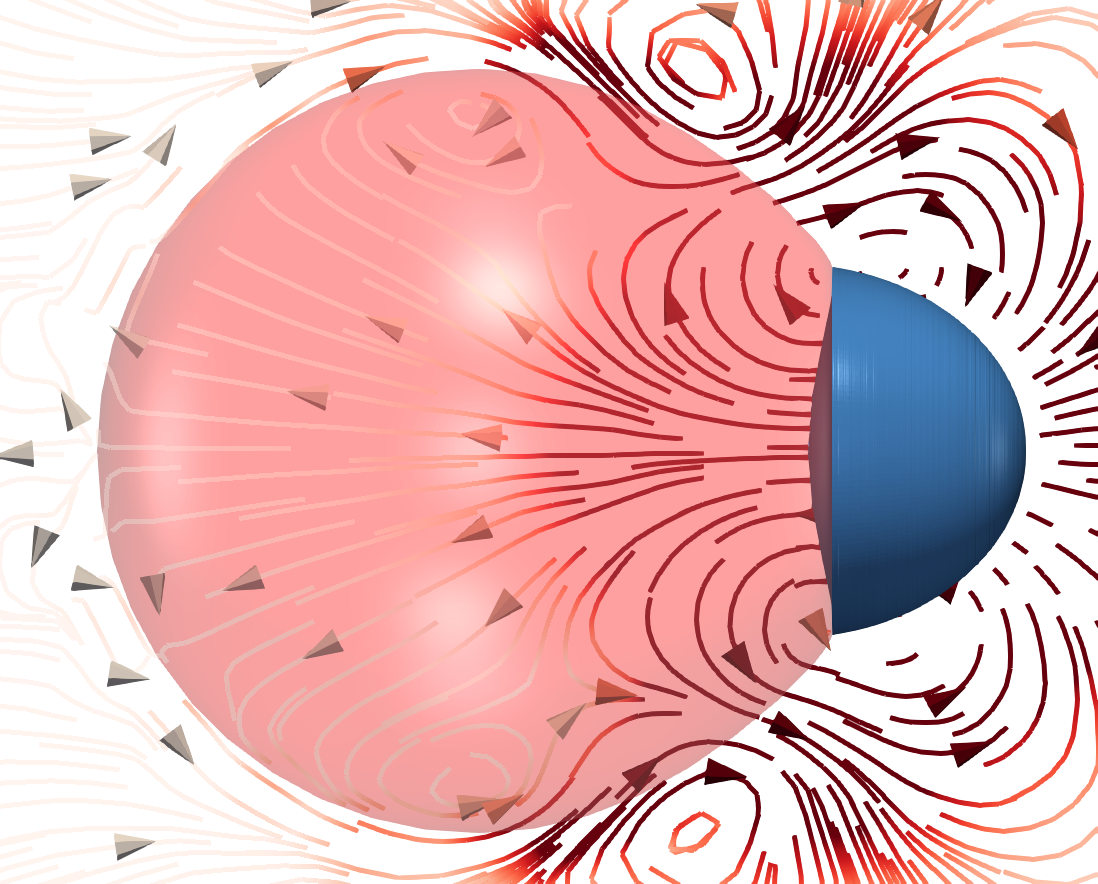}}  &
        {\includegraphics[width=0.24\textwidth]{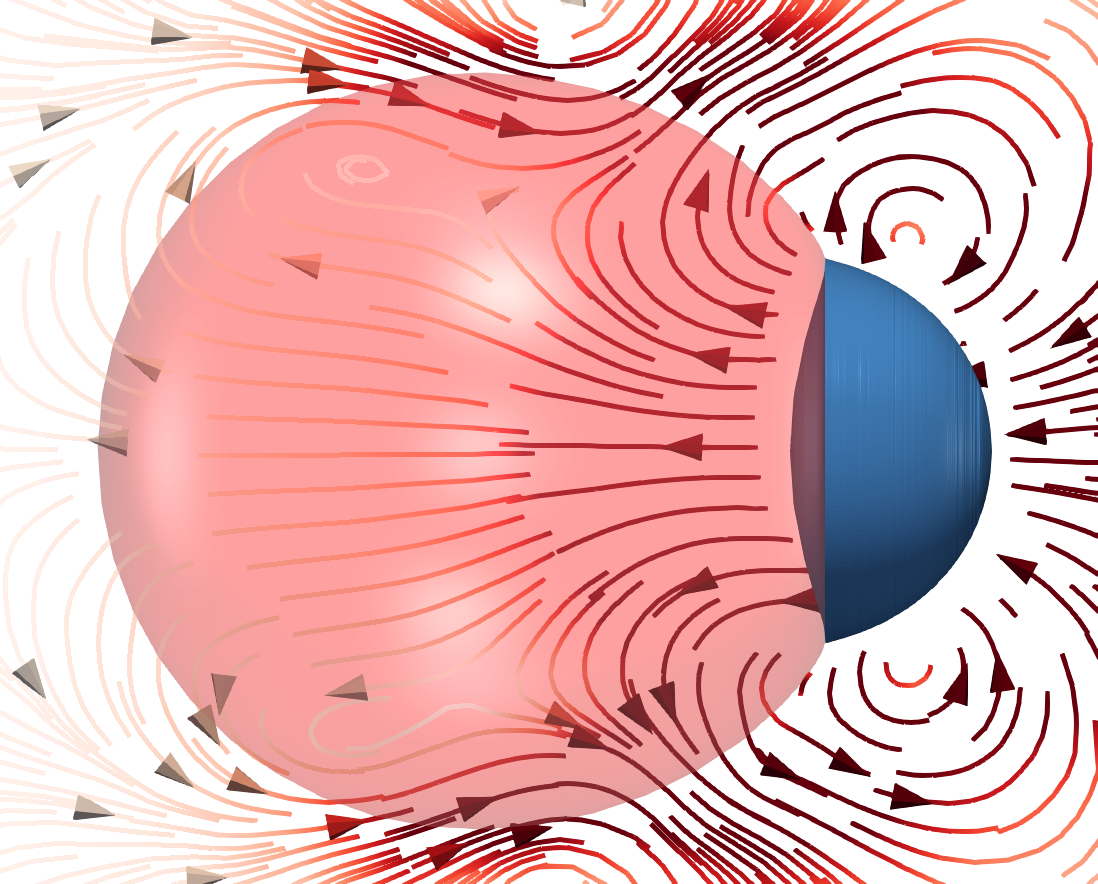}}&
        {\includegraphics[width=0.24\textwidth]{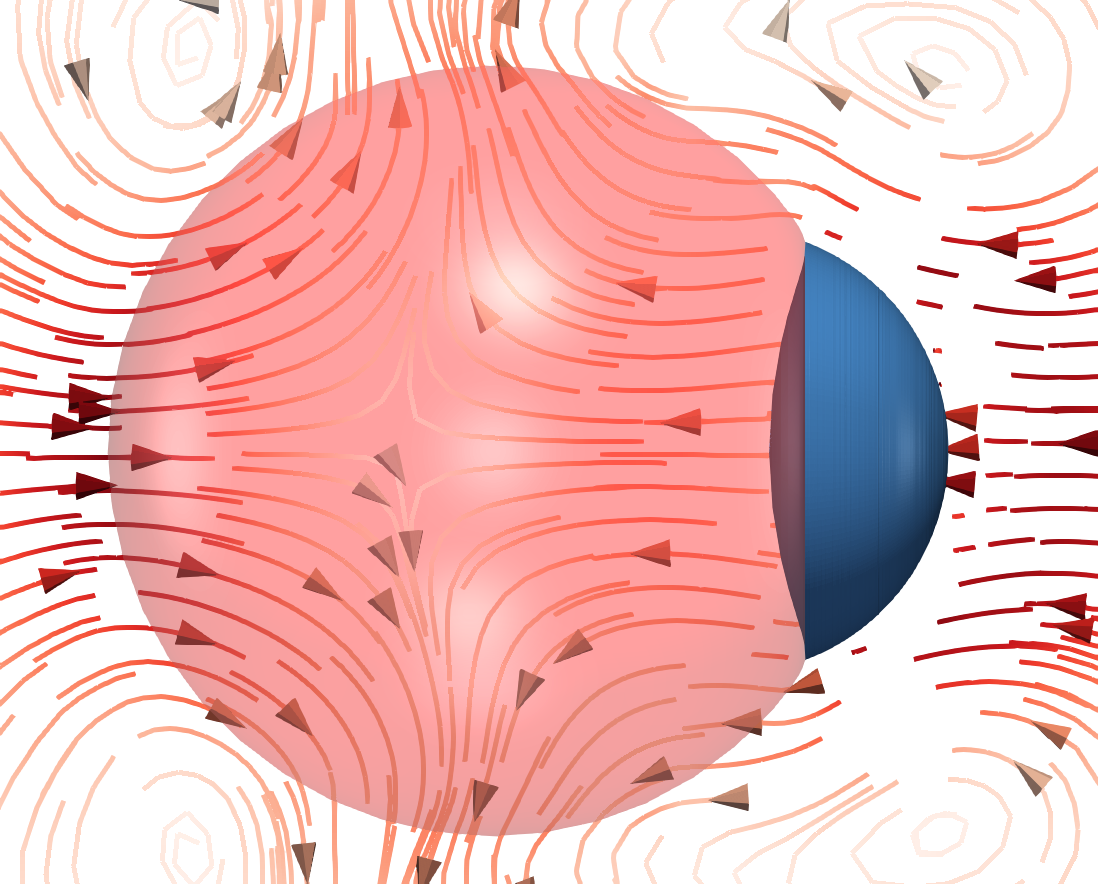}}
    &{\includegraphics[width=0.24\textwidth]{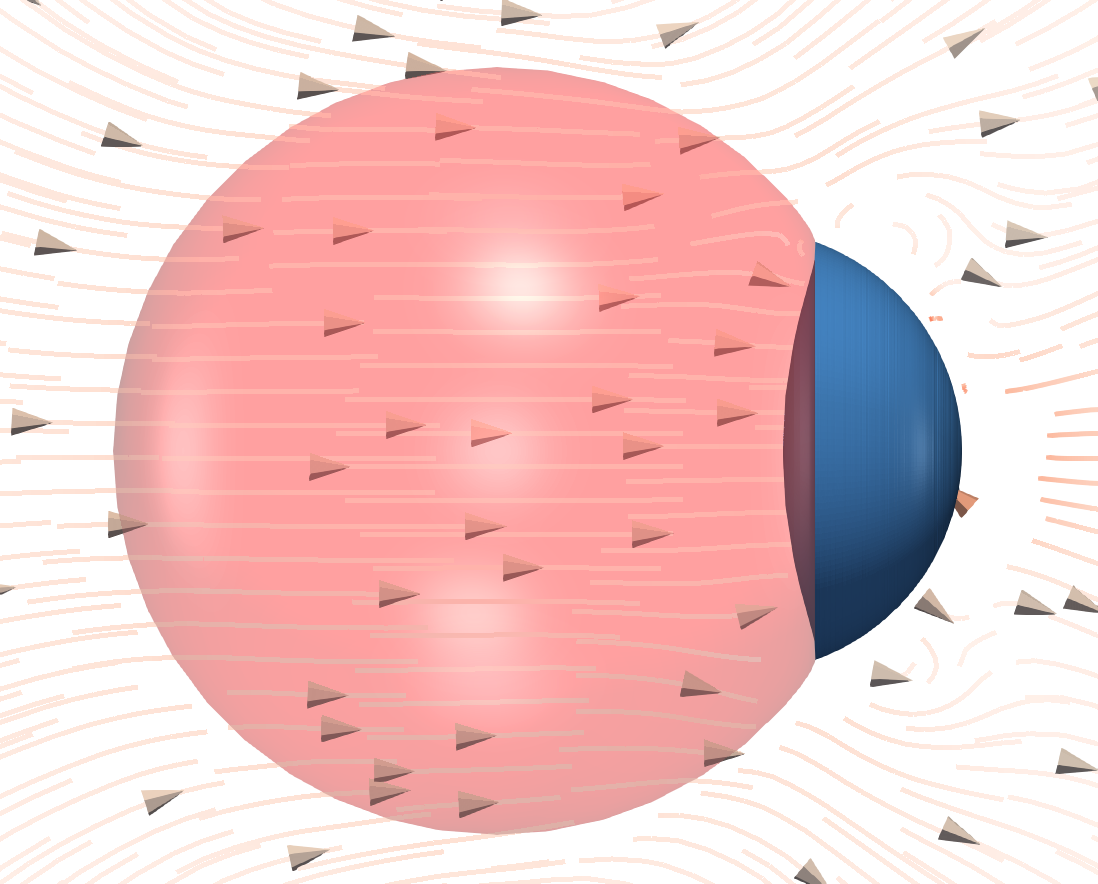}}  \\
        \footnotesize $t=0.5~$s & 
        \footnotesize $t=1~$s & 
        \footnotesize $t=3~$s & 
        \footnotesize $t=18~$s
    \end{tabular} 
    
    \vspace{0.1cm}
 \textbf{Wrapping}:  $ \sigma_{\text{f}} = 30\,\mu\text{N/m},\qquad \sigma_0 = 16\,\mu\text{N/m}, \qquad\sigma_1 = 1\,\mu\text{N/m}$ \vspace{0.1cm}
    \begin{tabular}{cccc}
     {\includegraphics[width=0.24\textwidth]{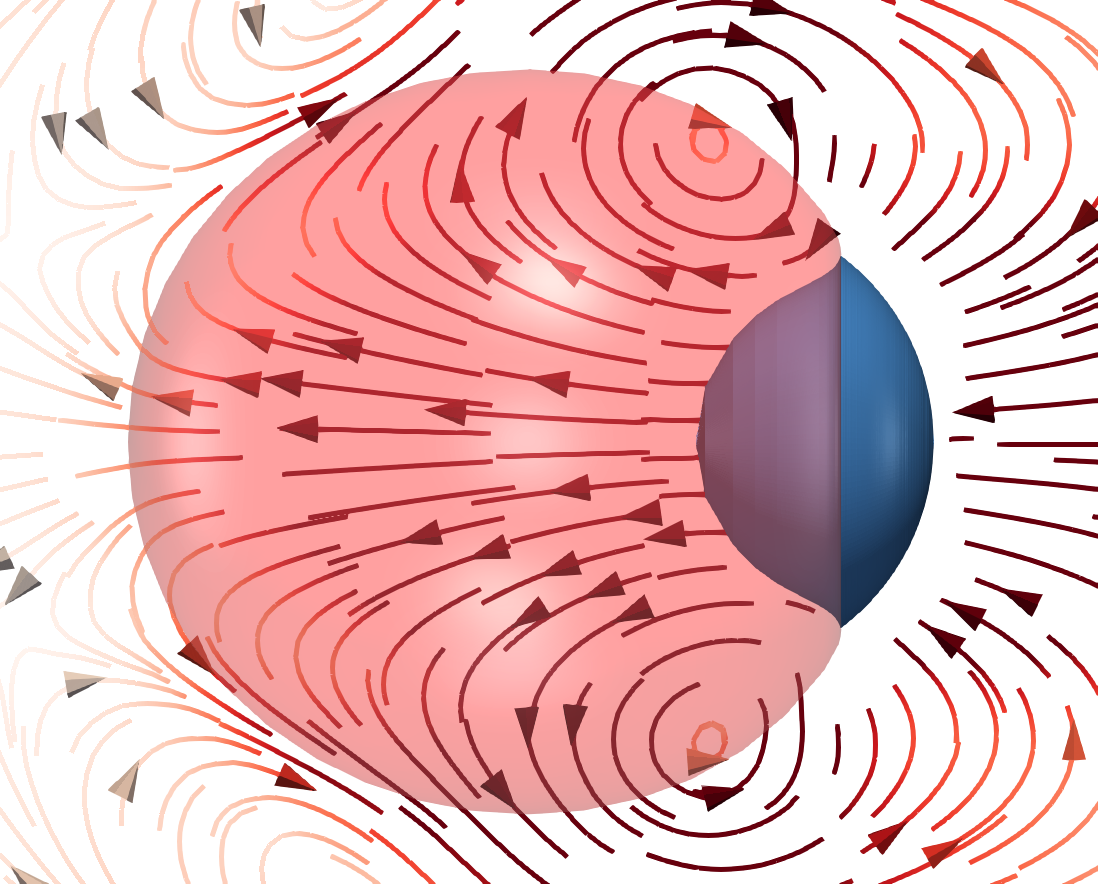}}  &
        {\includegraphics[width=0.24\textwidth]{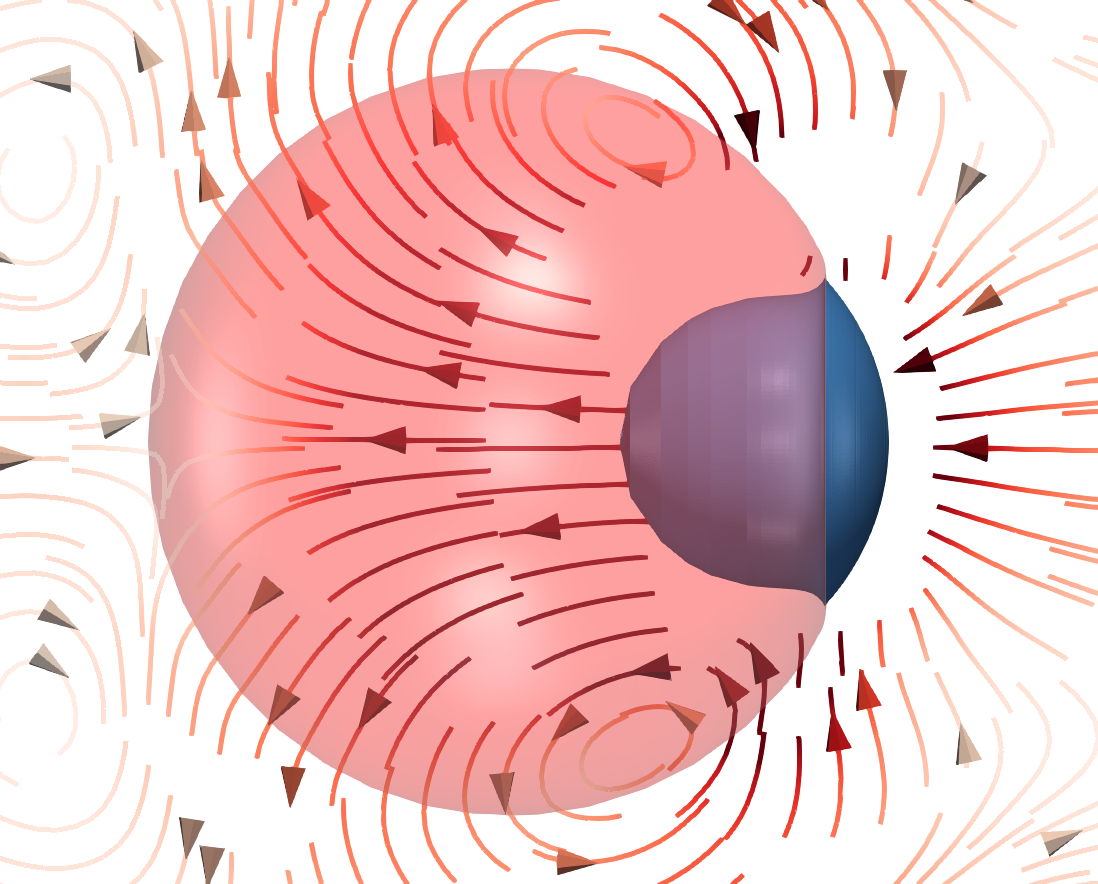}}&
        {\includegraphics[width=0.24\textwidth]{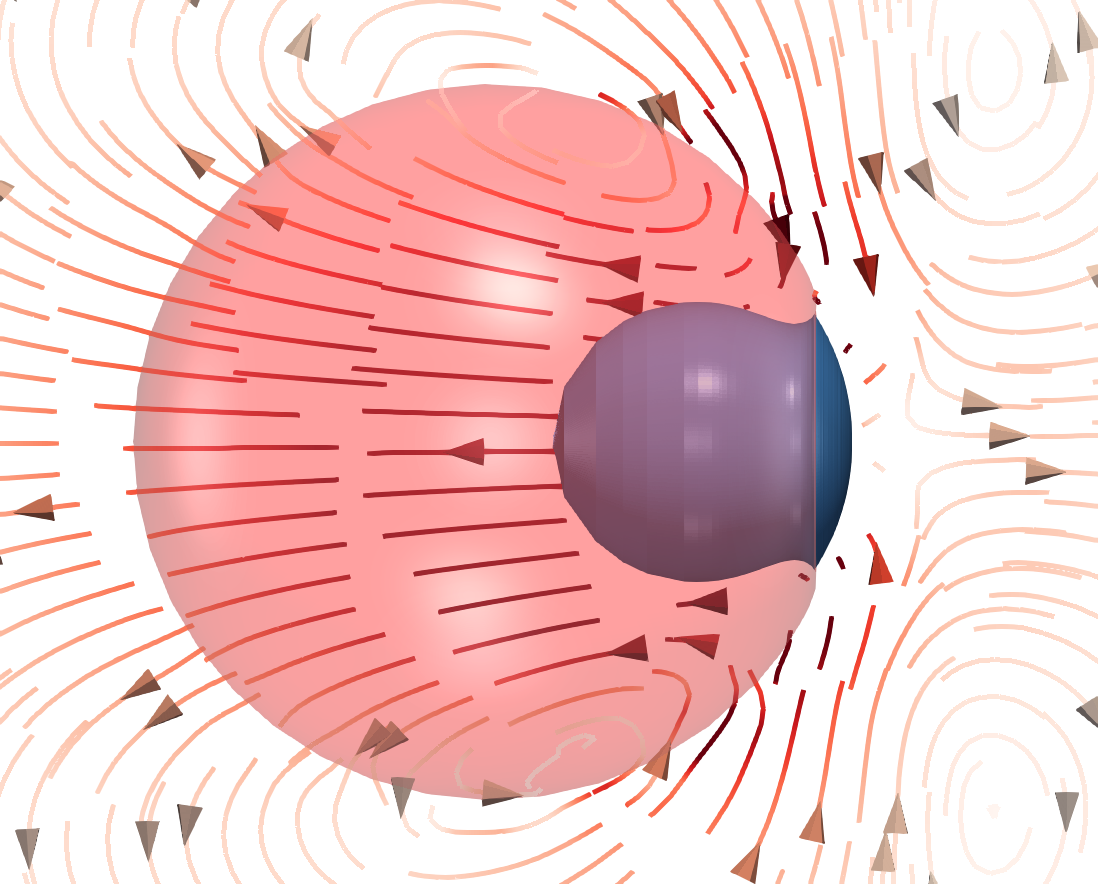}}
    &{\includegraphics[width=0.24\textwidth]{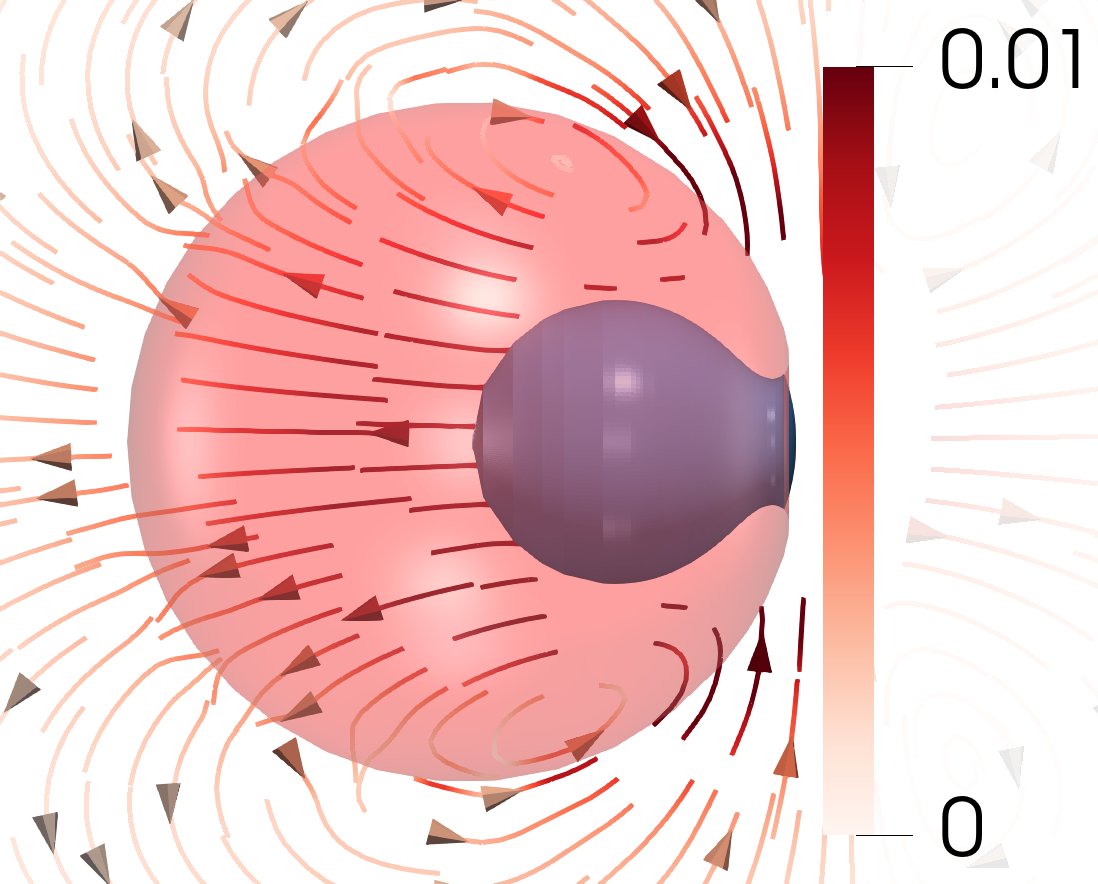}}  \\
        \footnotesize $t=2.5~$s & 
        \footnotesize $t=5~$s & 
        \footnotesize $t=10~$s & 
        \footnotesize $t=18~$s
    \end{tabular}        
    
    \caption{Time evolution for three different parameter configurations exhibit different categories of droplet-membrane interaction: adhesion (top row), lens shape (middle row), and wrapped/endocytosis (bottom row). 
    The single top snapshot shows the initial configuration of droplet (blue) and membrane (red). Oriented streamlines illustrate fluid velocity in the $xy$-plane colored by magnitude in units of $10\,\mu$m/s. All snapshots have been zoomed in for visibility and therefore do not show all of $\Omega$.
    \textbf{Parameters}: Initially spherical membrane of radius of $5.4\mu$m and half-spherical cap droplet of radius $2.5\mu$m, centered on the membrane. Other parameters are $\varepsilon=0.02\mu$m, $K_B=8\cdot 10^{-19}\,\text{Nm}$, $K_A = 5\cdot 10^{-3}$\,\text{N/m}, $P=10^{-7}$m$^2$s/kg, $\eta = 1\,\text{Pa}\cdot\text{s}$, $\rho = 10^3\,\text{kg/m}^3$. Axisymmetric simulations.}
    \label{fig:singleDroplpetNew}
\end{figure}

\textbf{Case 1: Adhesion} (\autoref{fig:singleDroplpetNew} top row). 
Here $\sigma_0$ is chosen smaller than the other two tensions. As a result, the droplet indents the membrane slightly. During the indentation, all three fluids (droplet, ambient, intramembrane) are set in motion. This motion decays as the system reaches a (nearly) stationary state around $t=18$s.  
The observed contact angle of approximately $\alpha=135^{\circ}$ is lower than the theoretical prediction of $151^{\circ}$ from Eq.~\eqref{eq: angle alpha Neumann}. This is due to the fact that droplet and membrane radius are of the same order of magnitude, hence, Neumann's law does not hold. 

\textbf{Case 2: Lens shape} (\autoref{fig:singleDroplpetNew} middle row). 
In the lens shape case, $\sigma_1$ is smaller than the other two surface tensions. Hence, the droplet tends to cover more of the membrane's surface, leading to a smaller contact angle ($\theta\approx 85^{\circ}$) and the lens-like shape of the droplet in the stationary state.

\textbf{Case 3: Wrapping} (\autoref{fig:singleDroplpetNew} bottom row). 
For this test case, the values of the surface tensions are chosen such that the capillary forces of the droplet are dominant and there is a minimum influence of the membrane tension at the contact surface between droplet and membrane. The droplet sinks completely into the membrane and gets wrapped by it. Since the present ALE model is not capable to describe topological changes of the membrane, the simulations were stopped at $t=18\,$s before a stationary state is reached. 
As predicted by theory\cite{tiemei2022}, the droplet should be completely wrapped such that endocytosis happens. 
In this case, it is clearly observable, that the volume enclosed by the membrane decreases as the droplet sinks into it. This effect occurs due to the interplay between in-plane elasticity (high $K_A$) and membrane permeability. The increase of membrane area which is necessary to wrap the droplet is opposed by the dilational elasticity, leading to a high pressure inside the membrane, pushing out fluid such that its volume decreases.

\begin{figure}[!ht]
    \centering
    \begin{tabular}{ccccc}
        {\includegraphics[width=0.19\textwidth]{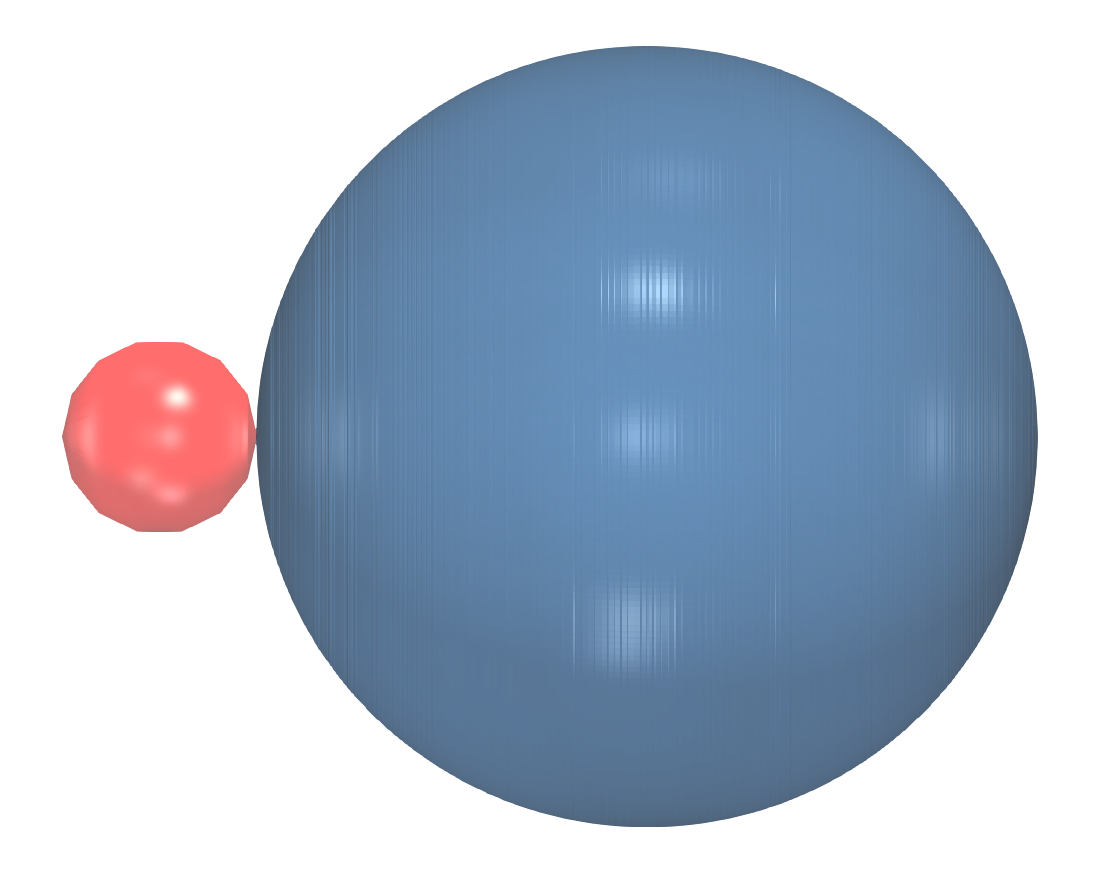}}  &
        {\includegraphics[width=0.19\textwidth]{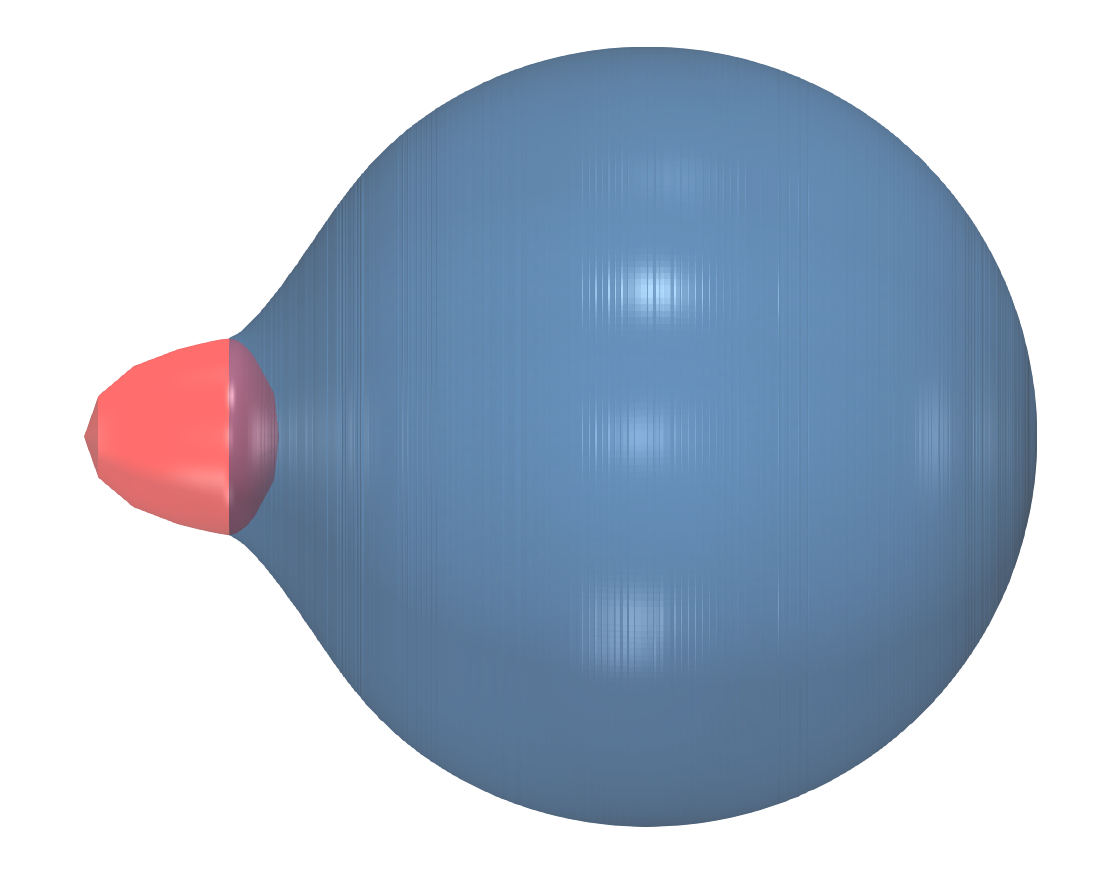}}&
        {\includegraphics[width=0.19\textwidth]{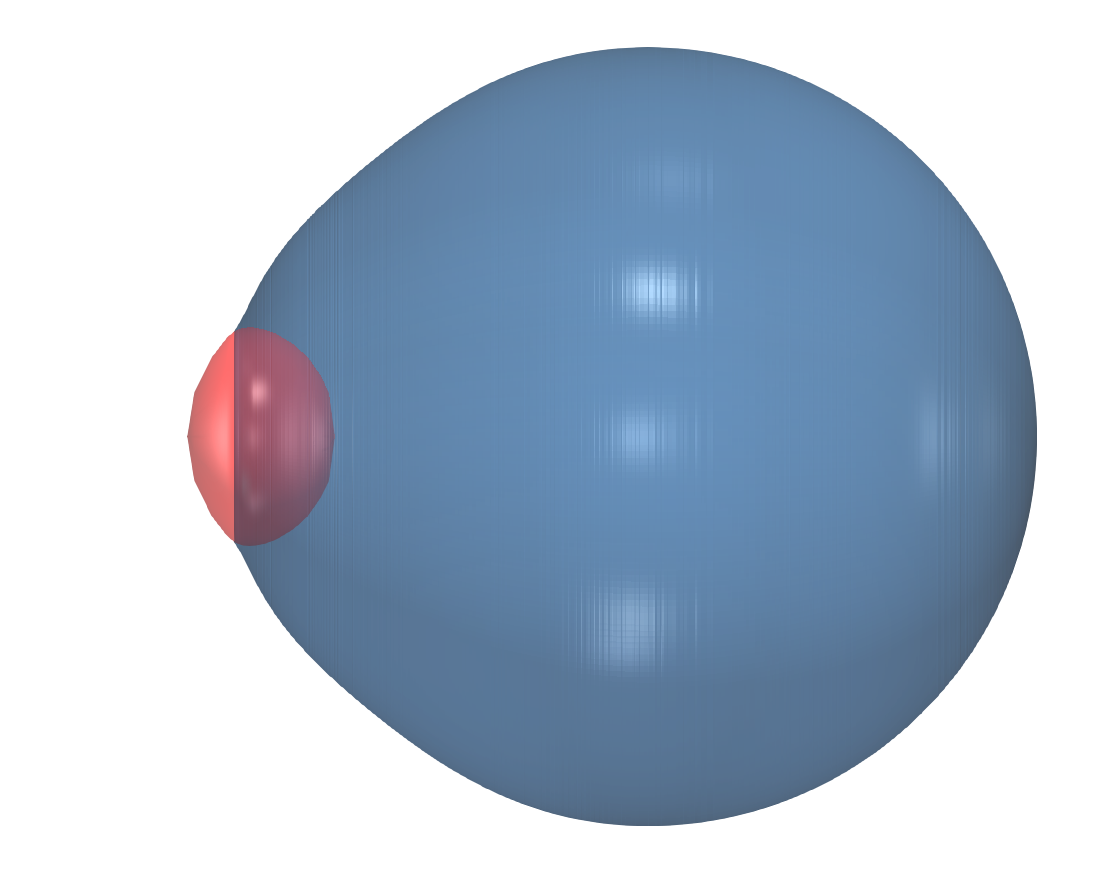}}  &
        {\includegraphics[width=0.19\textwidth]{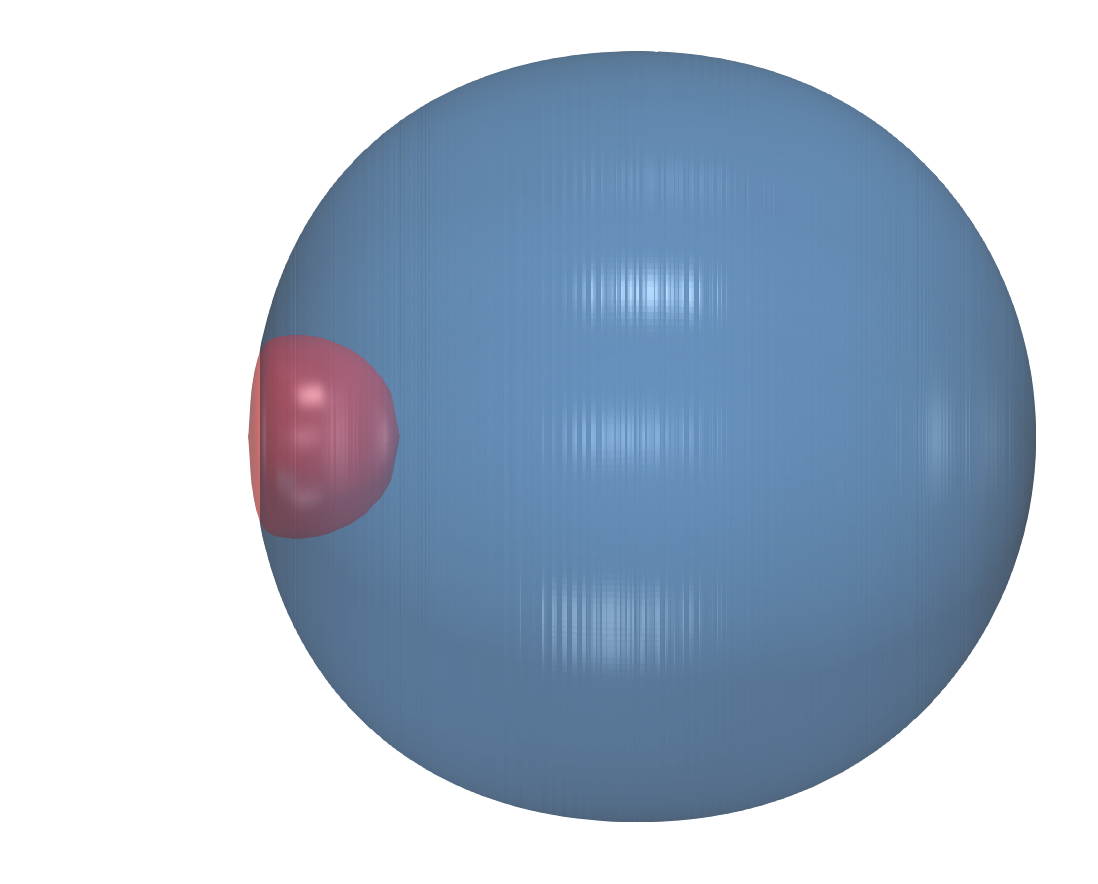}}  &
        {\includegraphics[width=0.19\textwidth]{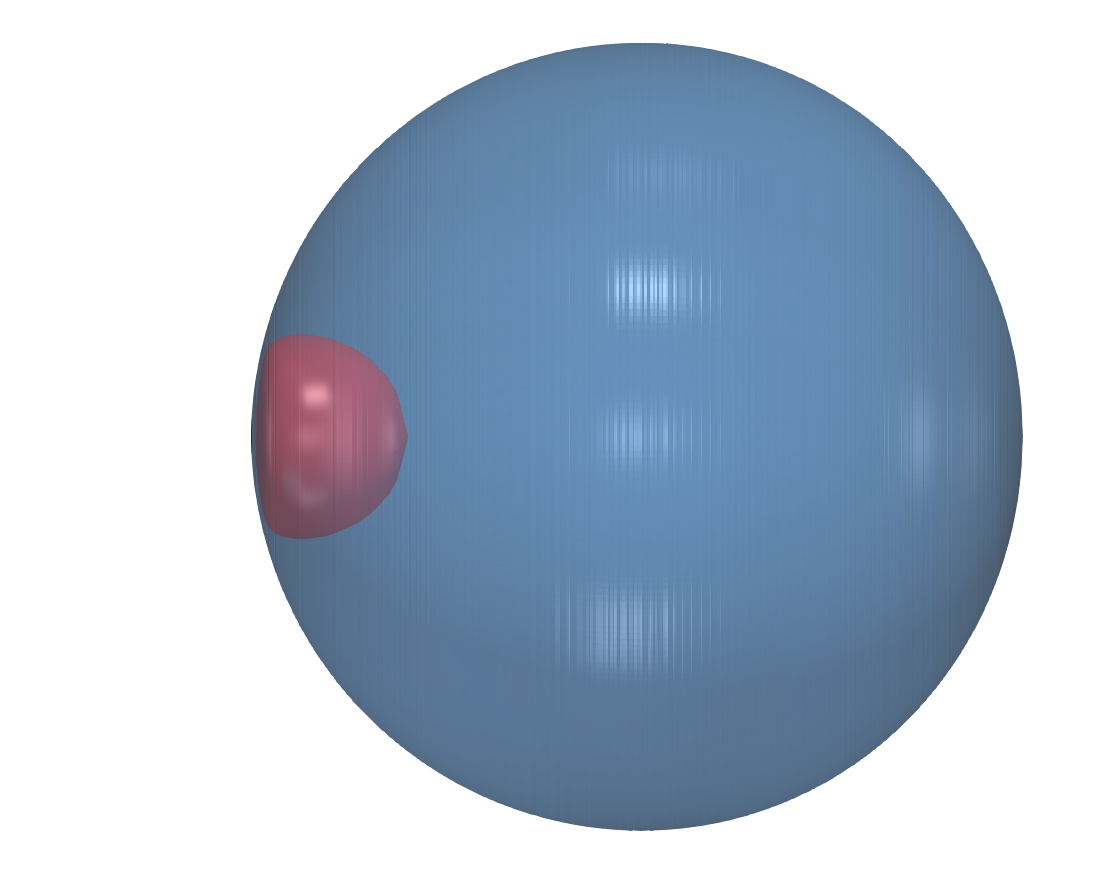}}  \\
        \footnotesize $t=0~$s & 
        \footnotesize $t=0.5~$s & 
        \footnotesize $t=1~$s & 
        \footnotesize $t=2~$s & 
        \footnotesize $t=5~$s 
    \end{tabular}    
    \caption{Inverted endocytosis: A small spherical vesicle (red) is absorbed by a larger drop (blue). Time evolution of axisymmetric simulation. Vesicle can change shape due to imposed permeability. \textbf{Parameters}: $\sigma_{\text{f}} = 30\,\mu\text{N/m}, ~\sigma_0 = 31\,\mu\text{N/m}, ~\sigma_1 = 1\,\mu\text{N/m}, K_B=8\cdot 10^{-17}\,\text{Nm}, K_A = 5\cdot 10^{-3}\,\text{N/m}, P=10^{-7}$m$^2$s/kg, initial radii $1.25\,\mu$m (vesicle), $5\,\mu$m (drop). Axisymmetric simulation.}
    \label{fig:reverseEndocytosis}
\end{figure}

\textbf{Inverted endocytosis}. 
An opposite case of inverted endocytosis is illustrated in \autoref{fig:reverseEndocytosis}. Here, a small vesicle is absorbed into a larger droplet, driven by surface tension forces. The surface tensions are chosen such that the extreme case of a contact angle of $\theta=0^{\circ}$ is approached. 
In this case, the full endocytosis can be simulated, since the topological transition happens now for the \textit{diffuse} fluid-fluid interface instead of the membrane.


\subsubsection{Multiple droplets} 
\begin{figure}[!ht]
    \centering
    \begin{tabularx}{\textwidth}{cc}
      \noindent\parbox[c]{0.29\textwidth}{\includegraphics[width=0.29\textwidth]{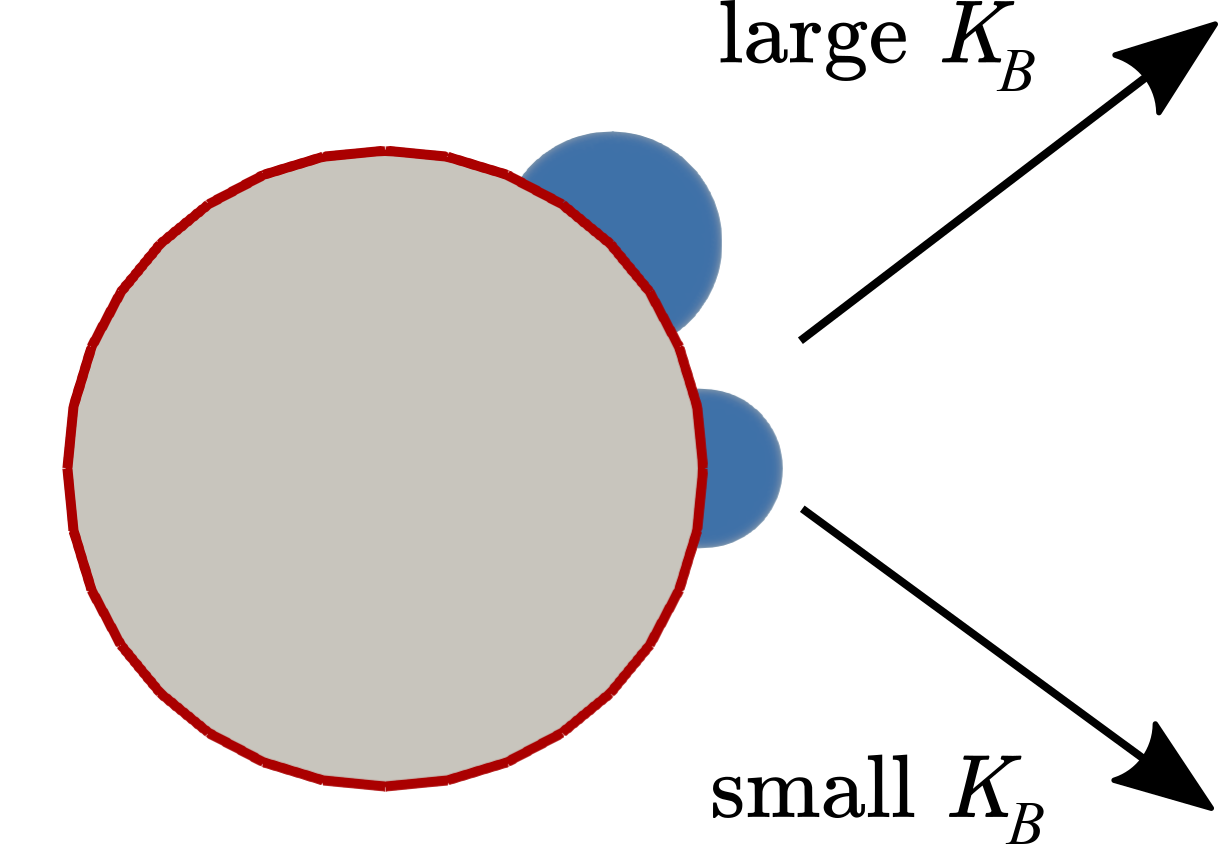}}   & 
      \begin{tabular}{ccc}
       \includegraphics[width=0.2\textwidth]{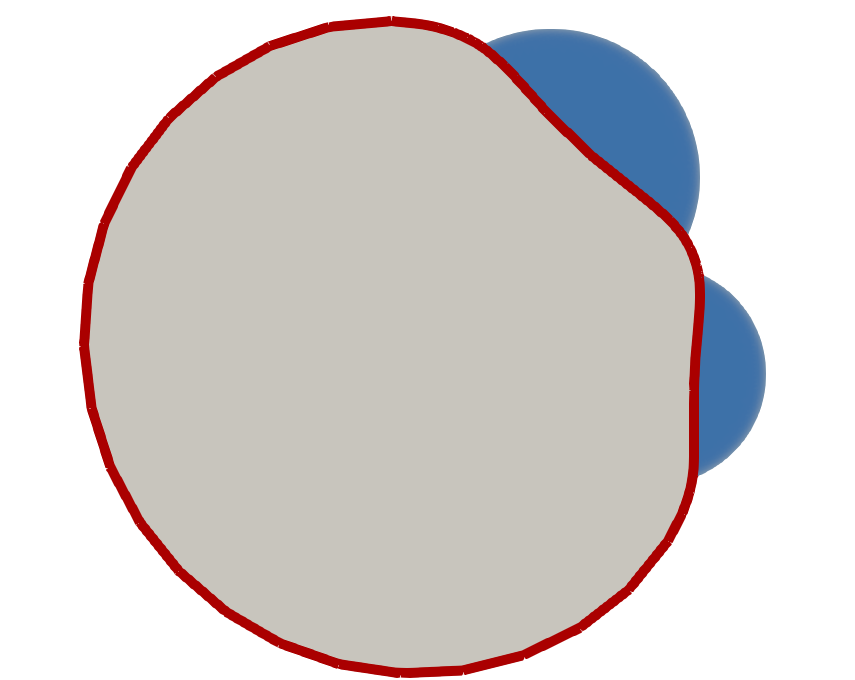}  & \includegraphics[width=0.2\textwidth]{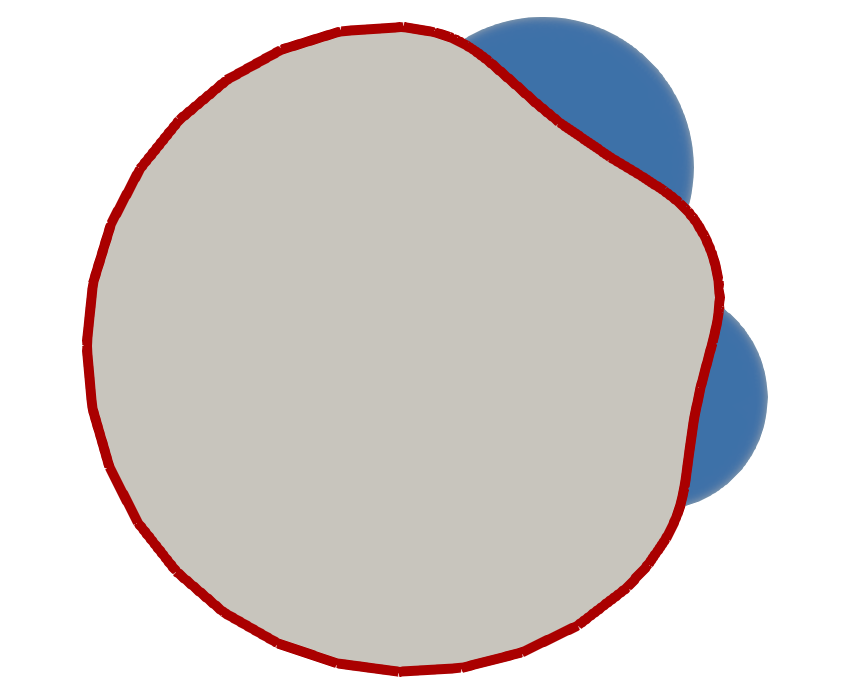} & 
       \includegraphics[width=0.2\textwidth]{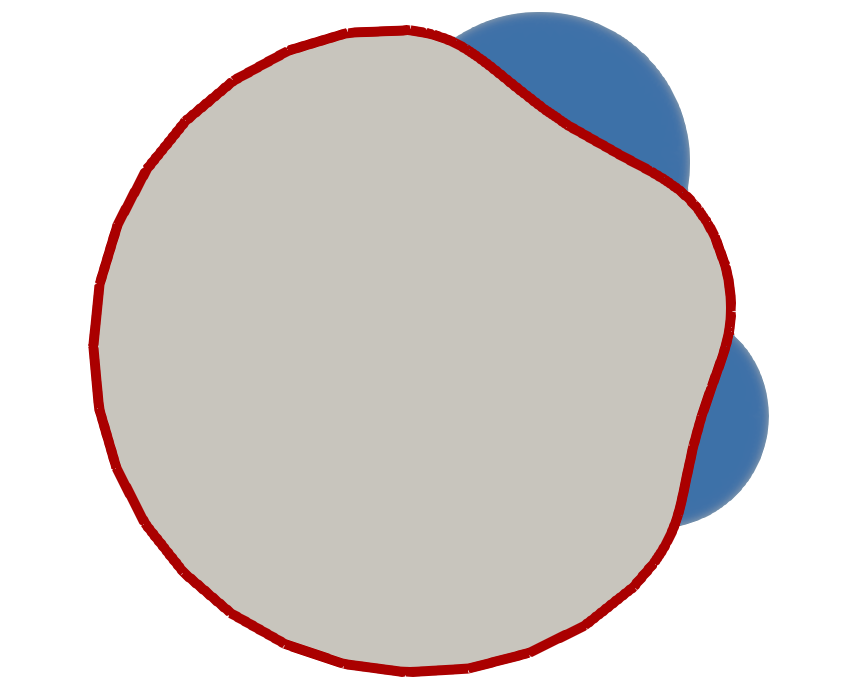} \\
        \footnotesize $t=10~$s & \footnotesize $t=50~$s & 
        \footnotesize $t=130~$s\\
         \includegraphics[width=0.2\textwidth]{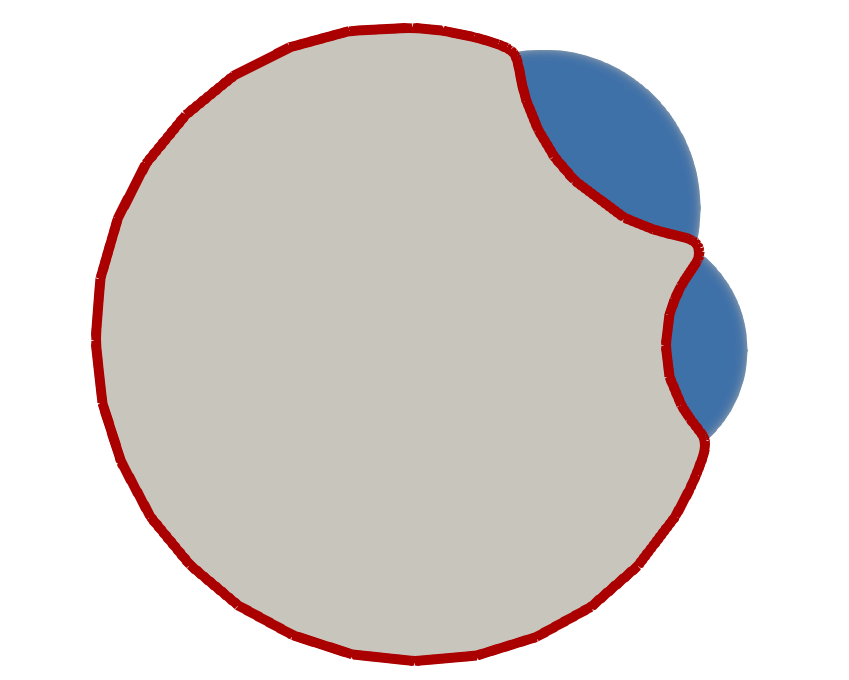}  & \includegraphics[width=0.2\textwidth]{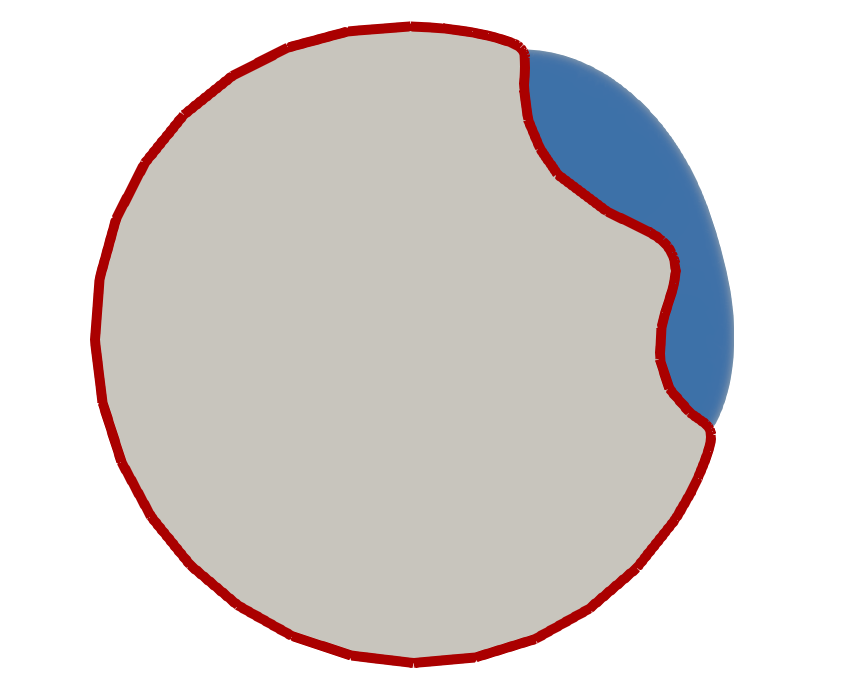} & 
         \includegraphics[width=0.2\textwidth]{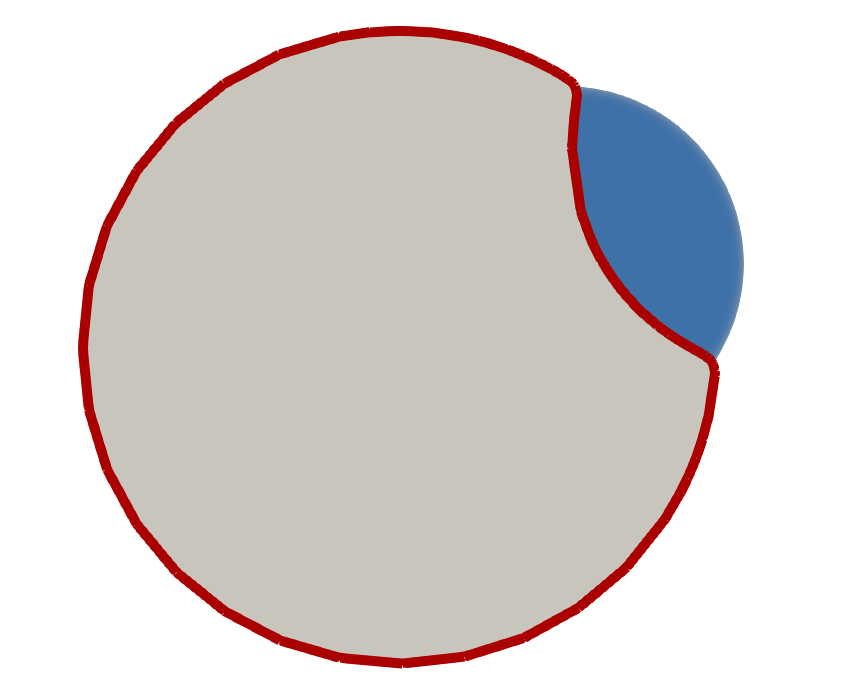} \\
         \footnotesize $t=10~$s & \footnotesize $t=13~$s & 
         \footnotesize $t=50~$s
      \end{tabular}
    \end{tabularx}
    \caption{Inverted Cheerios effect: Droplets  repel each other mediated by membrane bending (top, $K_B=8\cdot 10^{-17}\,\text{Nm}$). For low bending stiffness, the effect is not present and droplets coalesce (bottom, $K_B=8\cdot 10^{-19}\,\text{Nm}$). \textbf{Parameters}: Initial membrane radius $10\,\mu$m. Two droplets are placed as spherical caps with radius $r_1=2.5\,\mu$m and $r_2=3.5\,\mu$m at the membrane, where the centers of the two drops are $c_1=(10,0)^T$ and $c_1=(7.1,7.1)^T$. The remaining parameters have been chosen equally to the lens shape case of the single droplet simulations.}
    \label{fig:twoDroplets}
\end{figure}

\noindent
Another interesting test scenario is the inverted cheerios effect. 
This effect, which has been reported experimentally \cite{Karpitschka2016} and numerically \cite{mokbel2021}, describes the mutual attraction or repulsion of droplets, mediated by elasticity of their underlying substrate. 
Here, we perform simulations in 2D with an initially spherical membrane, where the two drops are represented as spherical caps, placed outside the membrane with a small distance in between. An illustration of the resulting dynamics is depicted in \autoref{fig:twoDroplets}, for two different values of bending stiffness. 
Other parameters are chosen equal to the previous lens-shape case. Accordingly, at early times we observe a similar evolution for the two droplets as before, \autoref{fig:twoDroplets}. The droplets indent the membrane until the forces on the membrane due to bending stiffness and surface tension are (nearly) in equilibrium around $t=10~$s. However, in the case of larger bending stiffness, the droplets are driven apart from each other due to the high membrane curvature between the droplets \autoref{fig:twoDroplets}(top row). This membrane-mediated droplet repulsion happens on a much slower time scale than the initial indentation of the membrane.
Contrary, in the case of smaller bending stiffness, \autoref{fig:twoDroplets}(bottom row), the distance between the two droplets decreases until both droplets merge and, as a result, the stationary shape is a single lens shaped droplet comparable to that in \autoref{fig:singleDroplpetNew}.

\subsubsection{Phase separation around a membrane}
\noindent
Finally, we investigate phase separation around a membrane. We study three different situations, namely phase separation on the inside, outside, and on both sides of a vesicle. For each of these cases, three different parameter configurations are studied: neutral ($\theta = 90^\circ$), high wettability ($\theta = 60^\circ$), and low wettability ($\theta = 120^\circ$).

\begin{figure}[!ht]
    \centering
   outside only \\
   \begin{tabular}{cccc}
   {\includegraphics[width=0.24\textwidth]{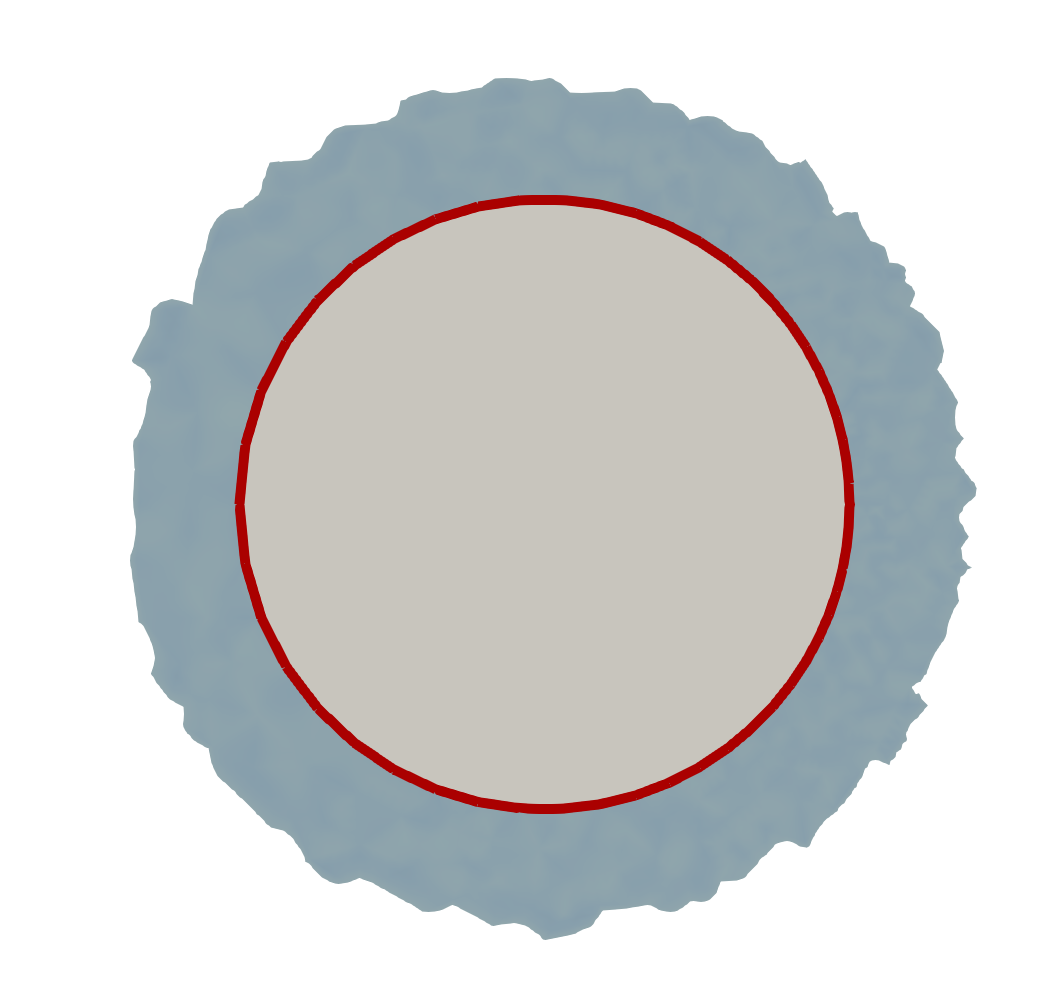}}
   &{\includegraphics[width=0.24\textwidth]{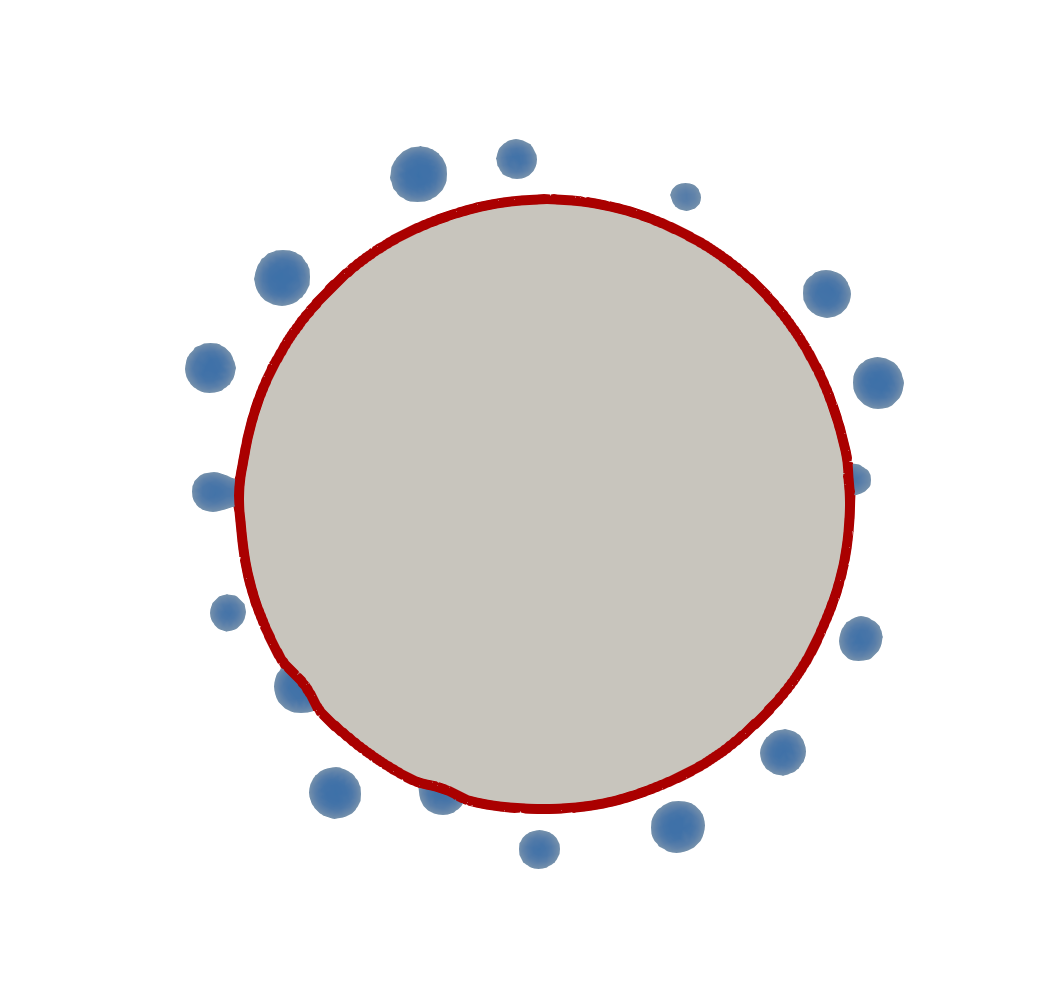}}
    &{\includegraphics[width=0.24\textwidth]{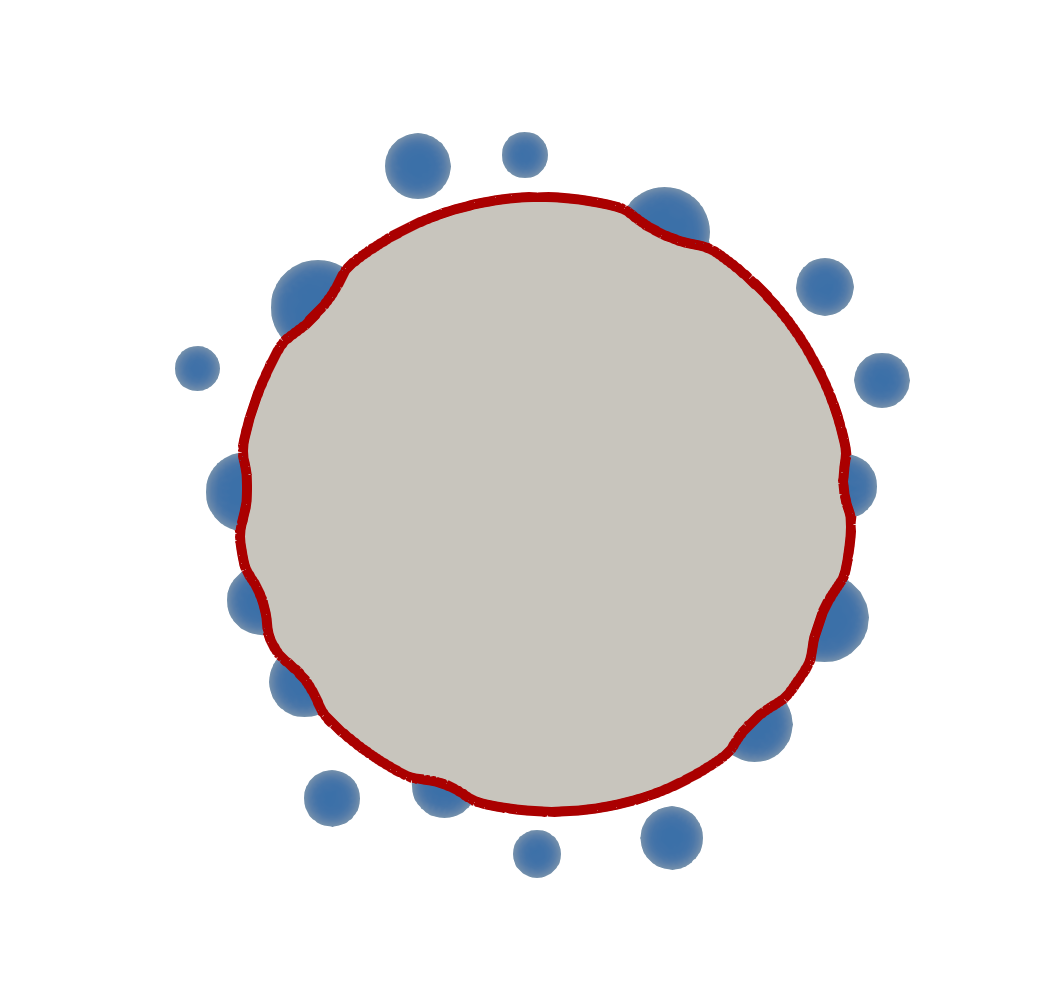}}
   &{\includegraphics[width=0.24\textwidth]{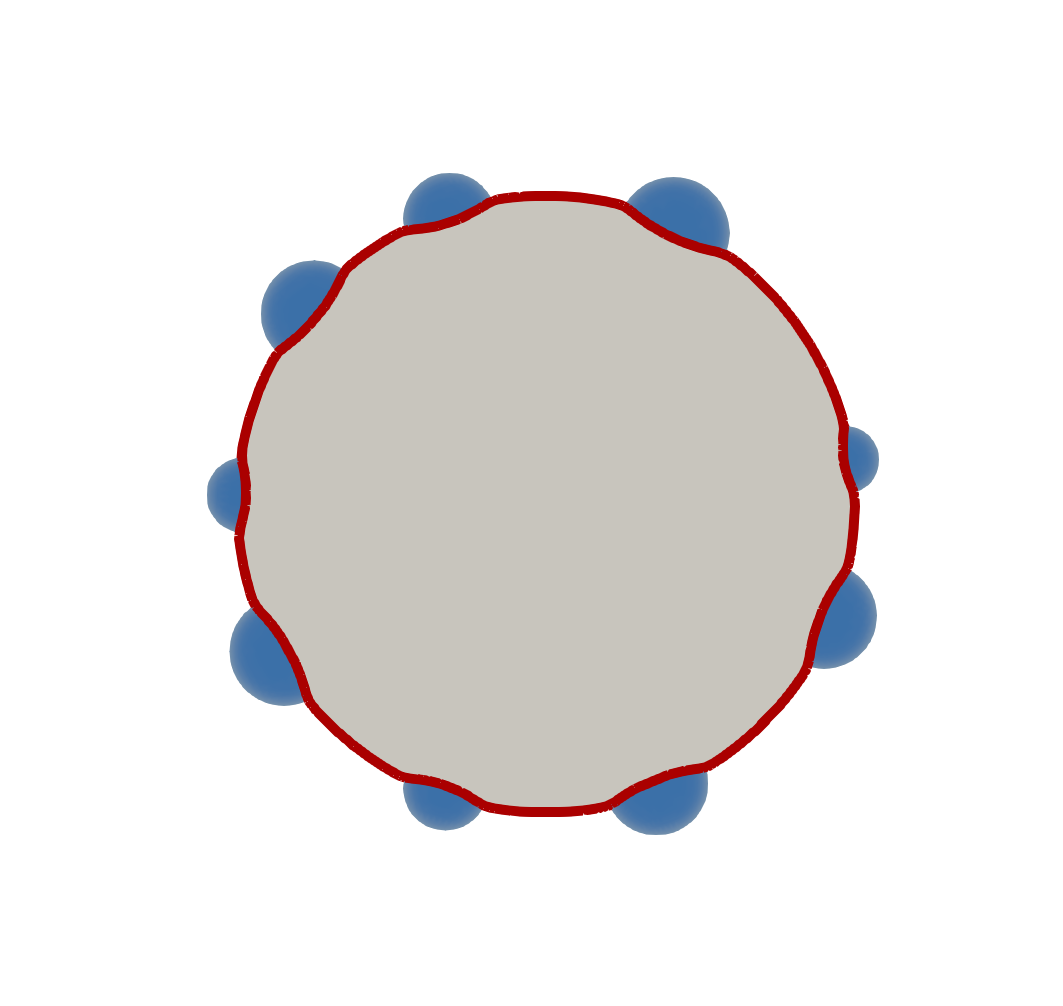}} 
   \end{tabular}
   inside only \\
   \begin{tabular}{cccc}
   {\includegraphics[width=0.24\textwidth]{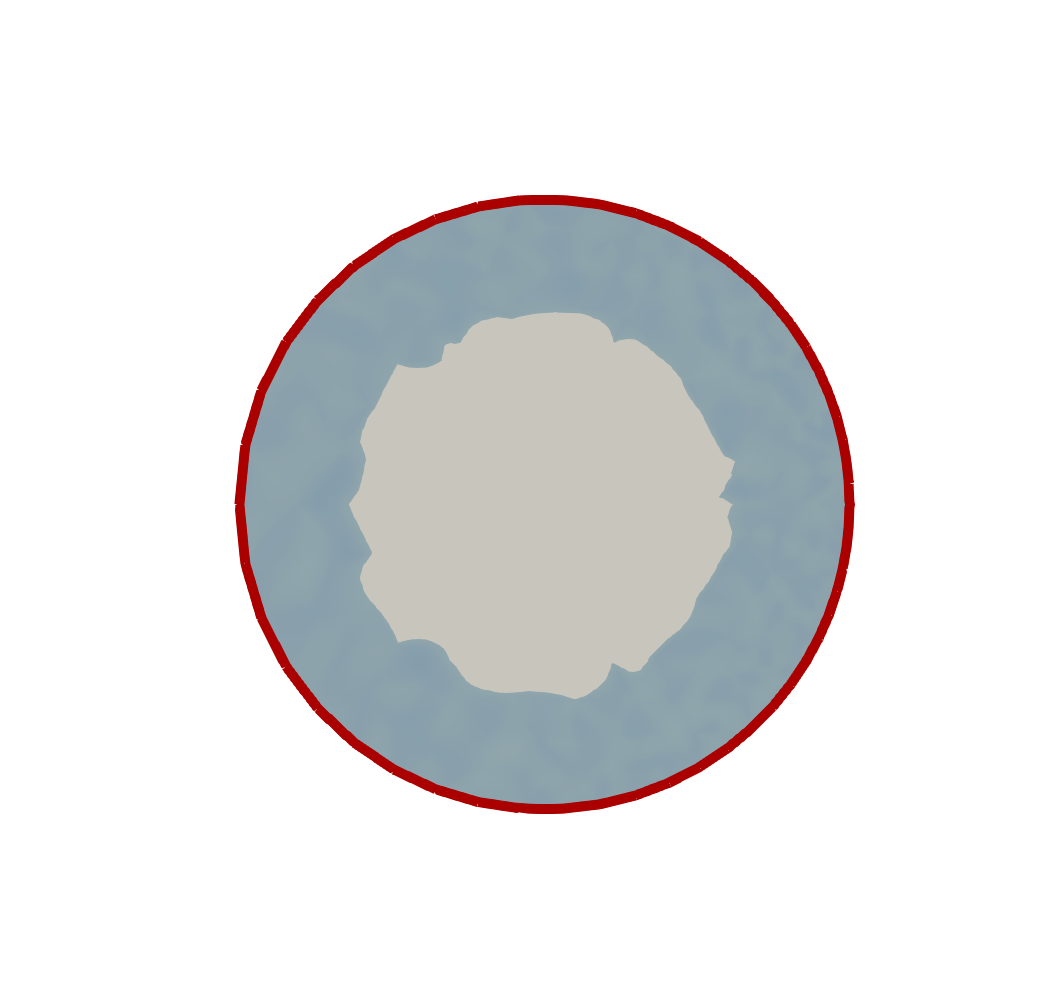}}
   & {\includegraphics[width=0.24\textwidth]{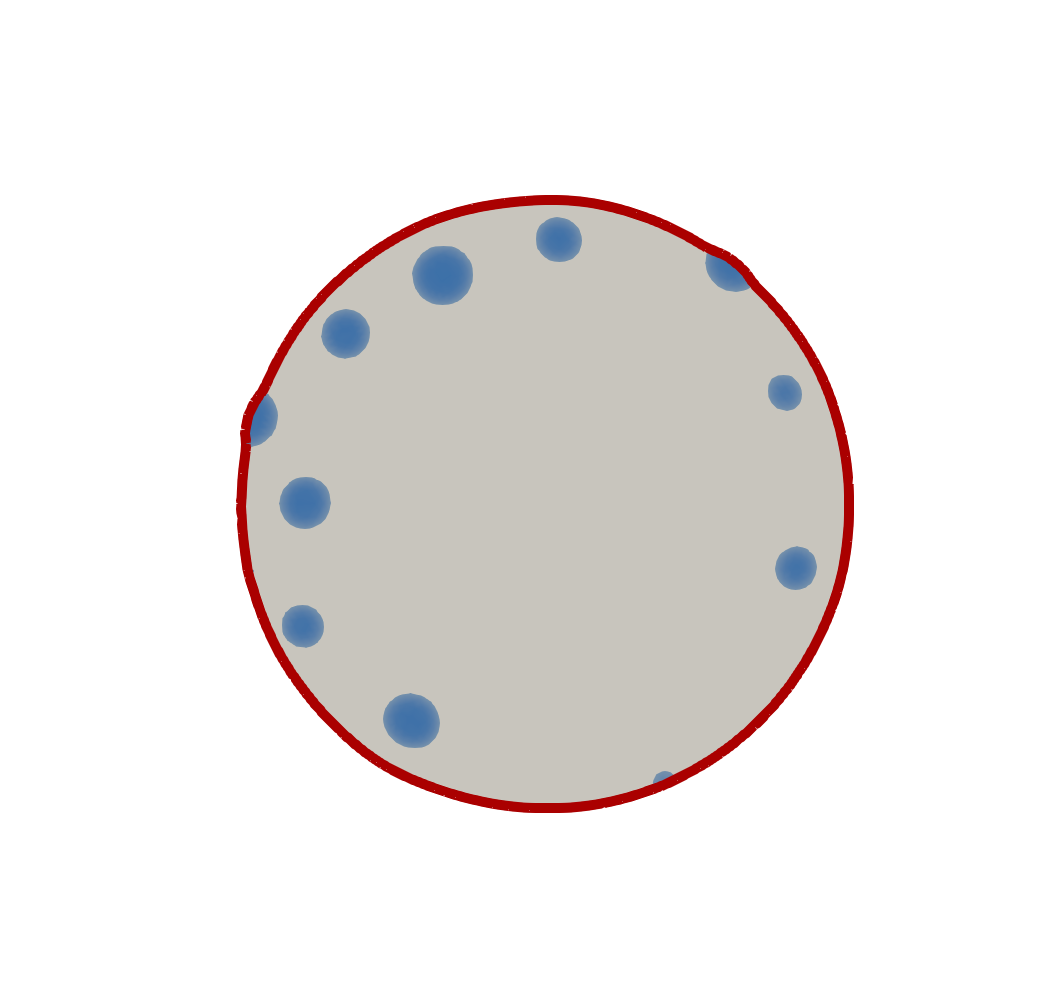}}
   & {\includegraphics[width=0.24\textwidth]{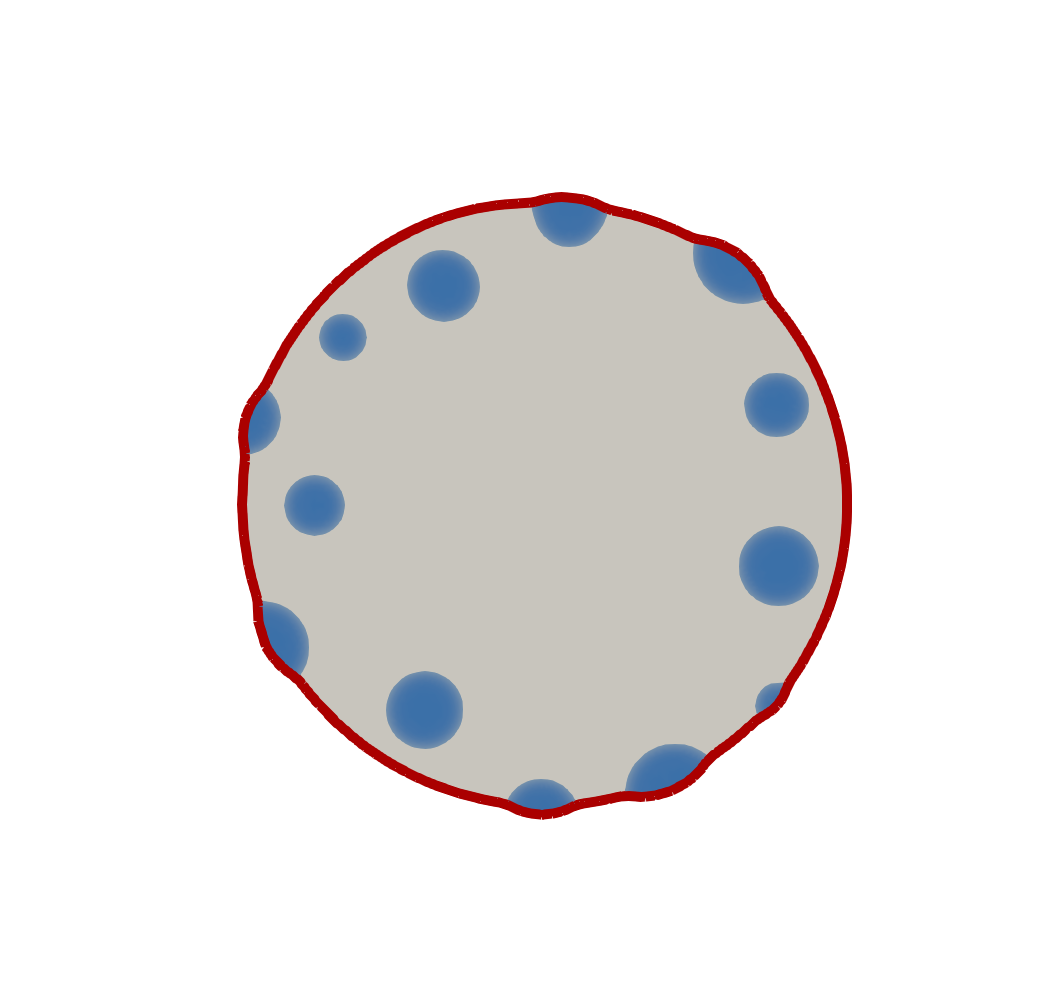}}
   & {\includegraphics[width=0.24\textwidth]{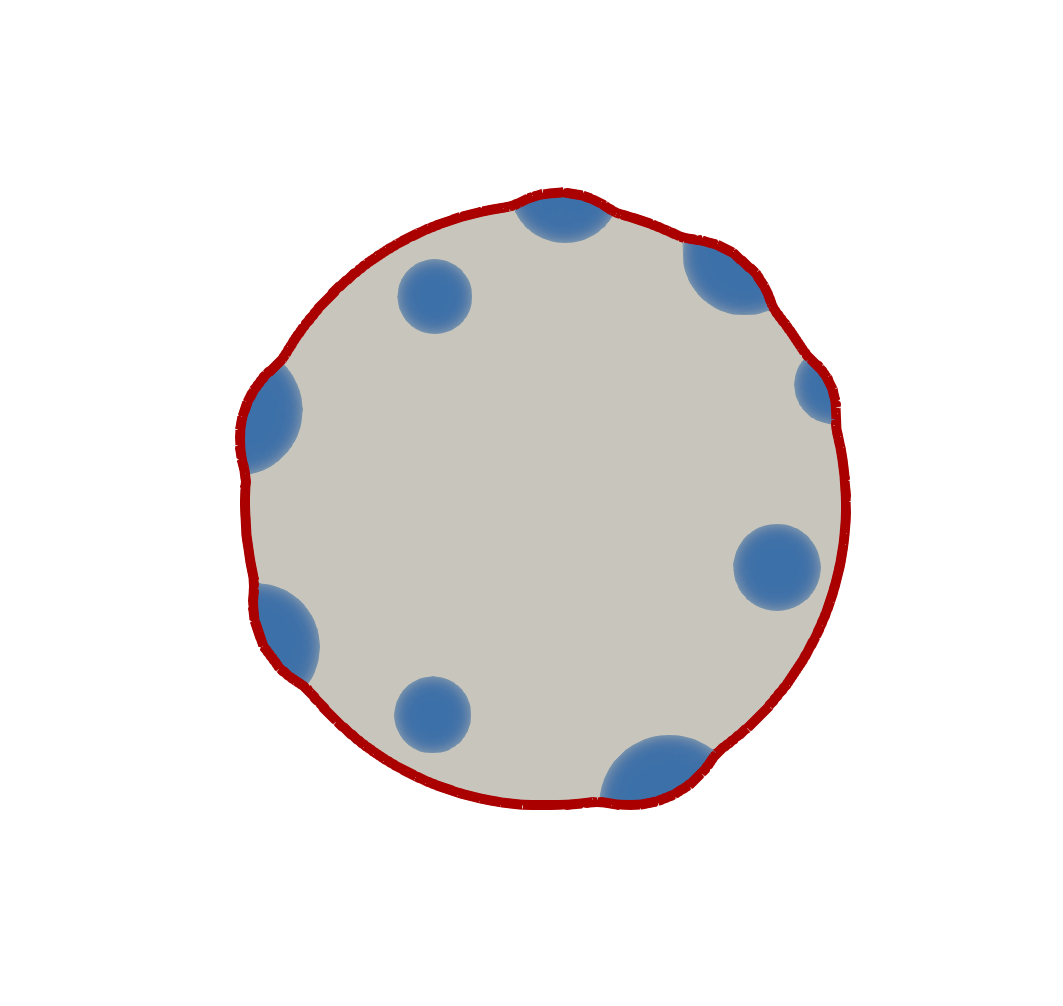}} 
   \end{tabular}
    inside and outside \\
    \begin{tabular}{cccc}
        {\includegraphics[width=0.24\textwidth]{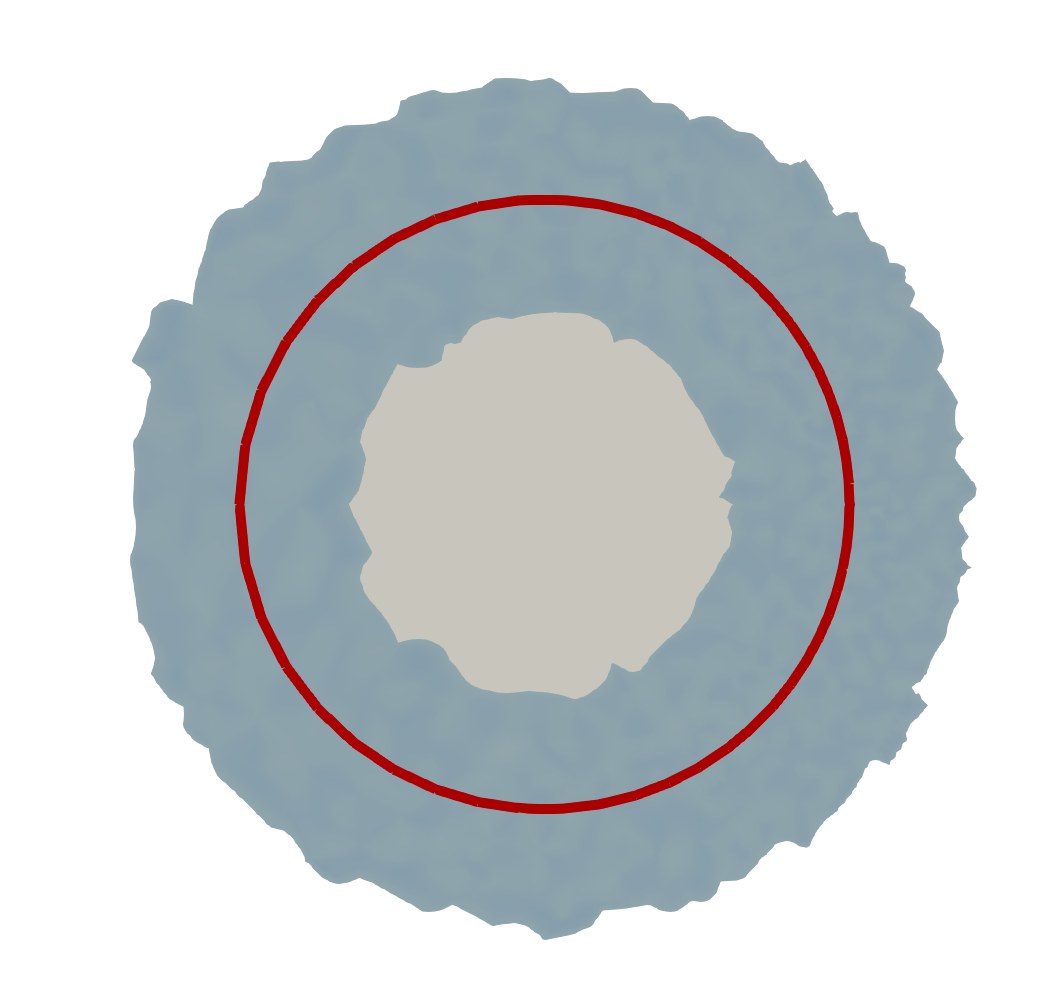}} &
   {\includegraphics[width=0.24\textwidth]{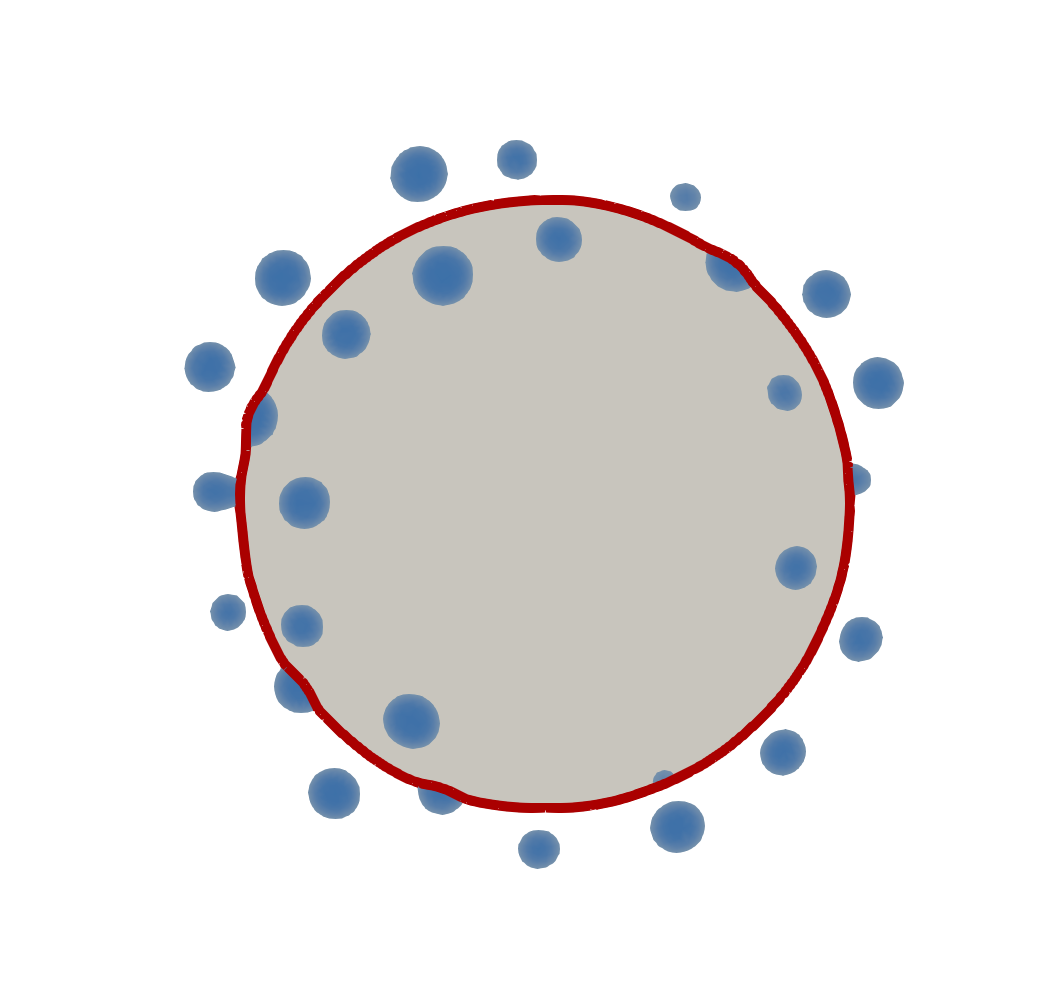}}
   & {\includegraphics[width=0.24\textwidth]{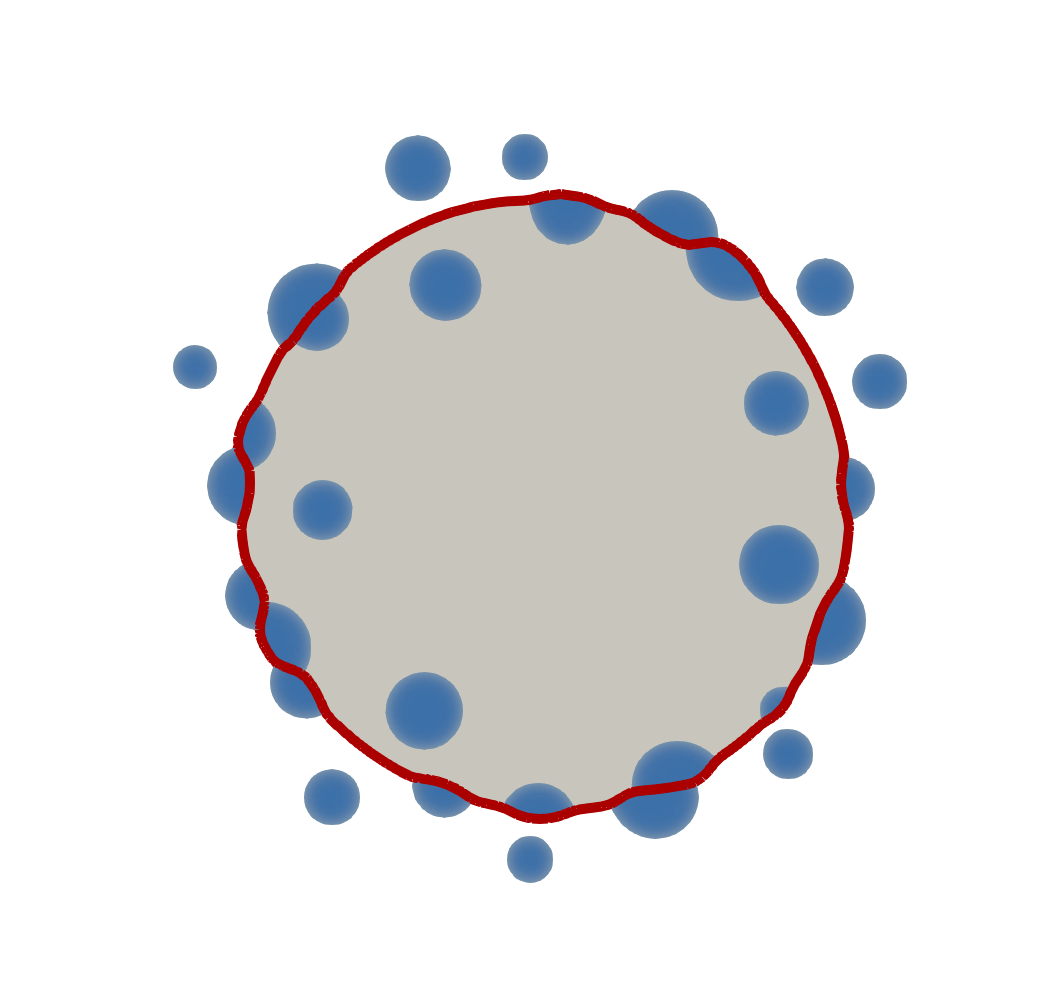}}
   & {\includegraphics[width=0.24\textwidth]{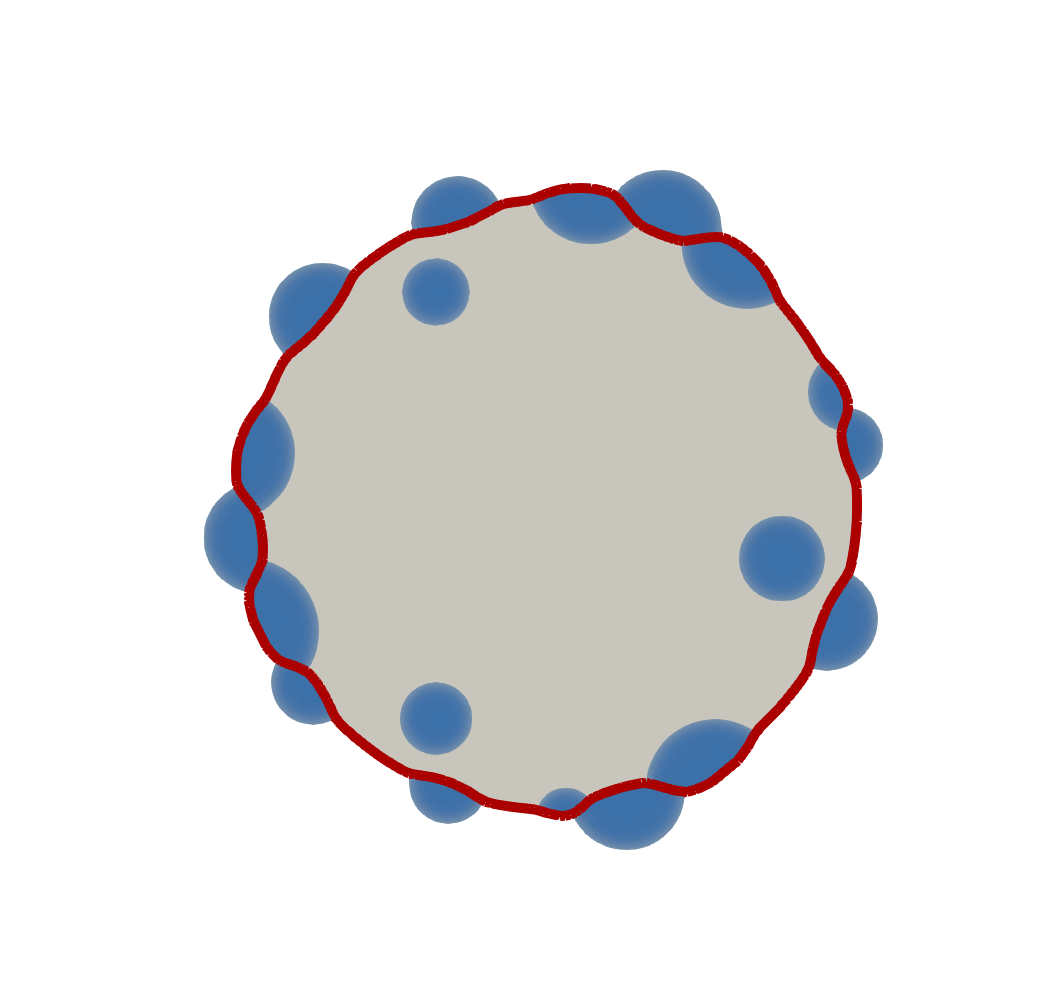}} \\
    \footnotesize $t=0~$s & \footnotesize $t=0.5~$s &  \footnotesize $t=2~$s & \footnotesize $t=10~$s
    \end{tabular}
    \caption{Droplet/ambient phase separation outside (top row), inside (middle row), and on both sides (bottom row) of a spherical membrane with contact angle $\theta=90^\circ$. In the light blue region at the initial state $t=0~$s, the phase-field value is prescribed as random number $\in[0.2, 0.3]$ in each grid point. 
    For the other time points, the droplets ($\phi\geq 0.5$) are illustrated in blue, the membrane in red and the ambient fluids in white (outside) and gray (inside). After phase separation, coarsening sets in, such that small droplets vanish (Ostwald ripening) and larger droplets grow. All droplets interact with the membrane and with each other. \textbf{Parameters}:  $\sigma_{\text{f}} = 15\,\mu\text{N/m}$, $\sigma_0 = 30\,\mu\text{N/m}$, $\sigma_1 = 30\,\mu\text{N/m}$, $K_B=8\cdot 10^{-20}\,\text{Nm}$, $K_A = 5\cdot 10^{-5}$\,\text{N/m}, $\varepsilon = 0.025\,\mu$m, and $P=0\,$m$^2$s/kg. Initial membrane radius is $2.5\,\mu$m. Random initial values for $\phi$ are set in a ring with internal radius $1.5\mu$m and/or external radius $3.5\,\mu$m.}
    \label{fig:phaseSeparation}
\end{figure}

\begin{figure}[!ht]
    \centering
   outside only \\
   \begin{tabular}{cccc}
   {\includegraphics[width=0.24\textwidth]{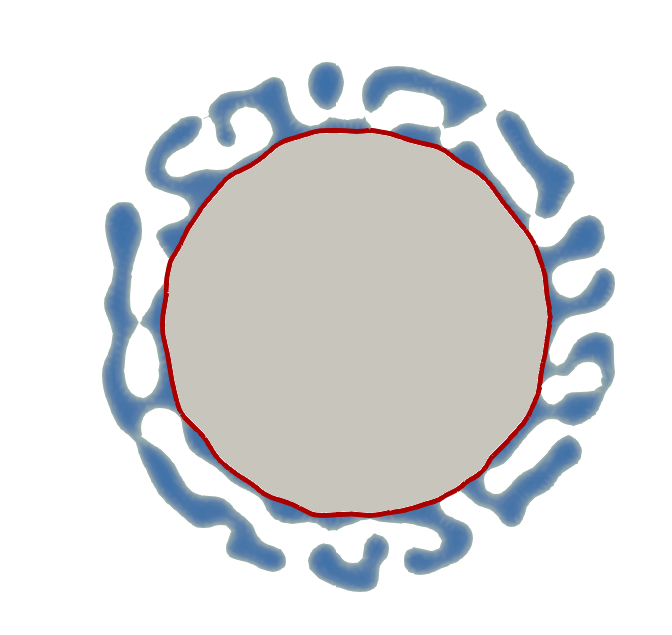}}
   &{\includegraphics[width=0.24\textwidth]{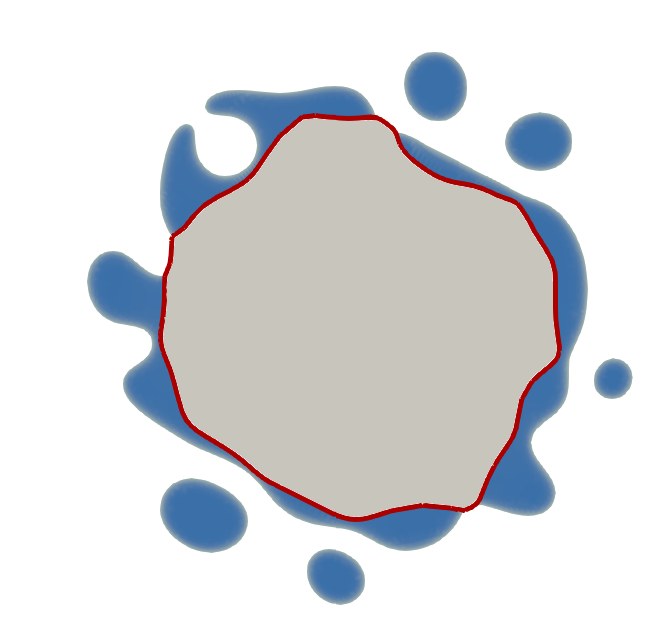}}
    &{\includegraphics[width=0.24\textwidth]{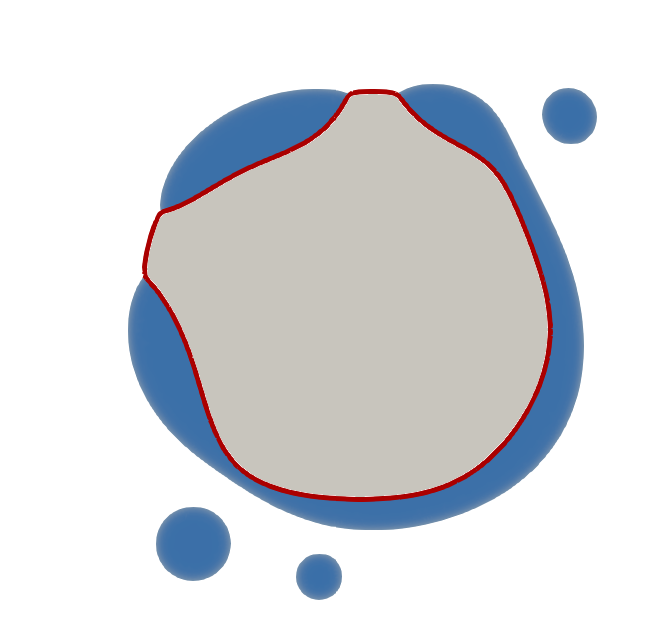}}
   &{\includegraphics[width=0.24\textwidth]{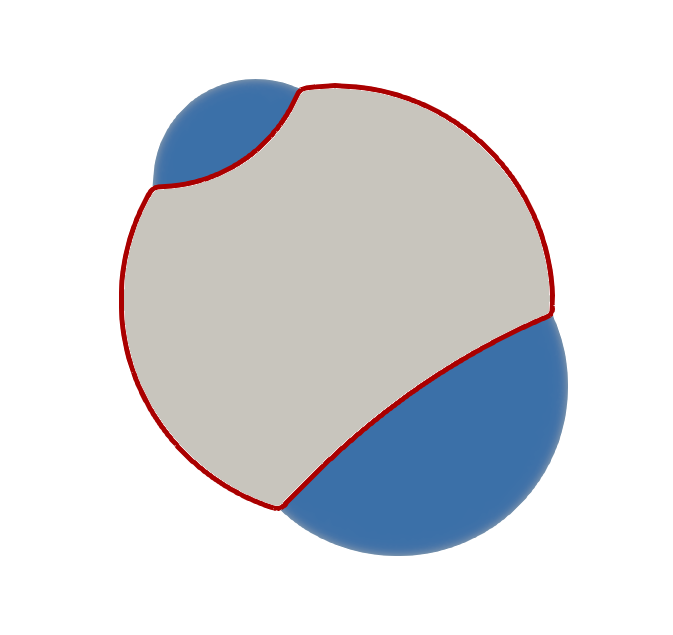}} 
   \end{tabular}
   inside only \\
   \begin{tabular}{cccc}
   {\includegraphics[width=0.24\textwidth]{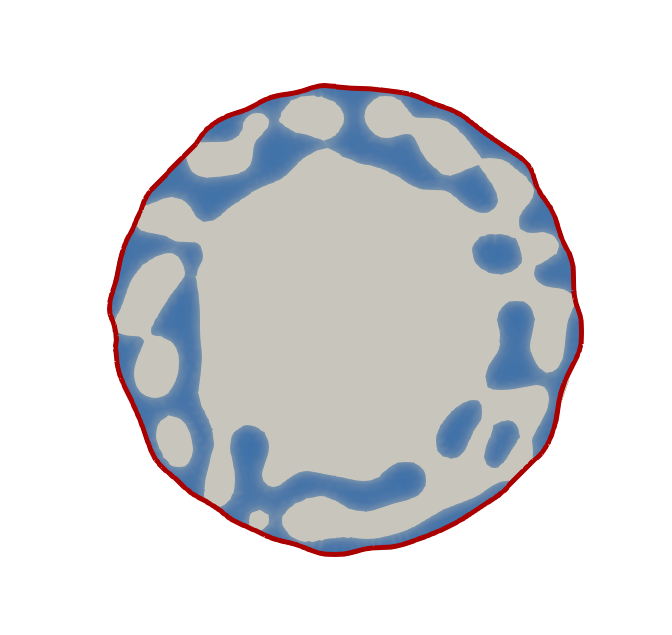}}
   & {\includegraphics[width=0.24\textwidth]{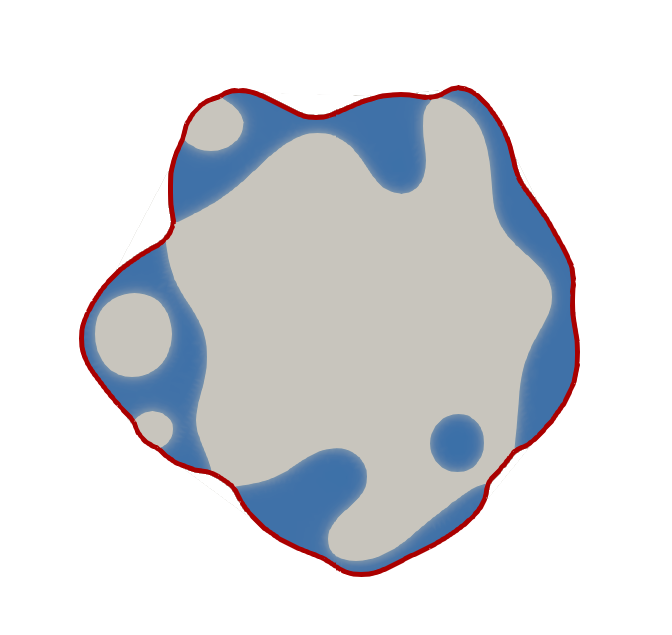}}
   & {\includegraphics[width=0.24\textwidth]{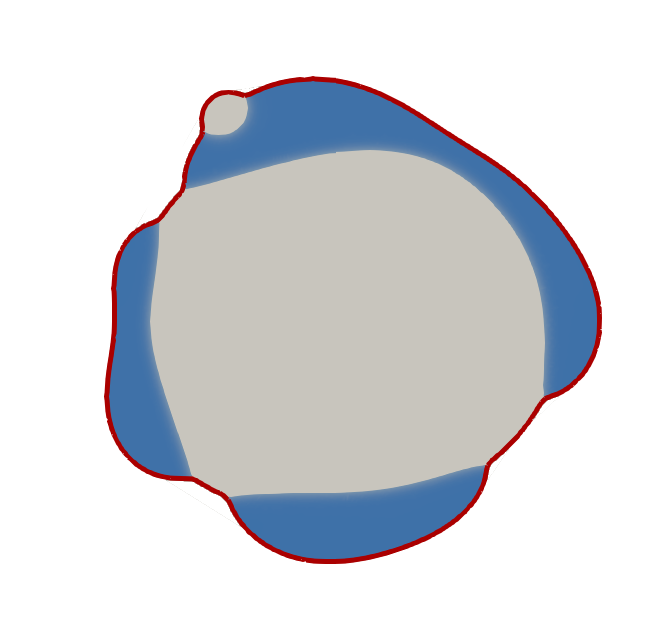}}
   & {\includegraphics[width=0.24\textwidth]{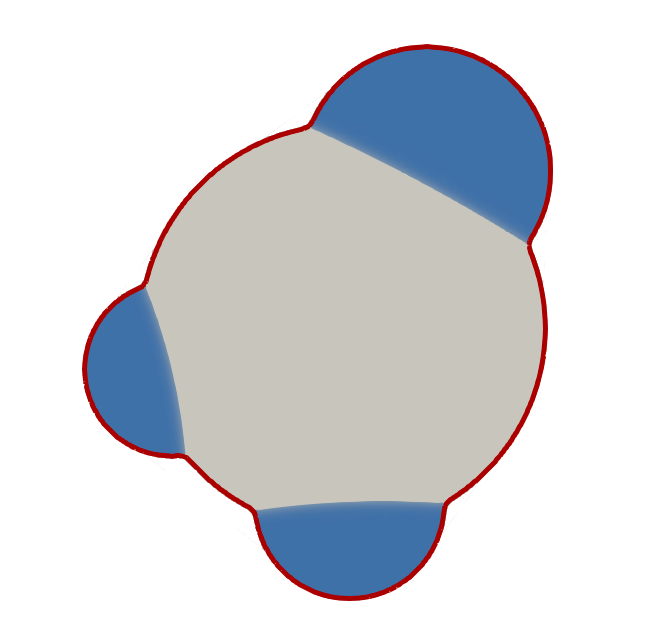}} 
   \end{tabular}
    inside and outside \\
    \begin{tabular}{cccc}
        {\includegraphics[width=0.24\textwidth]{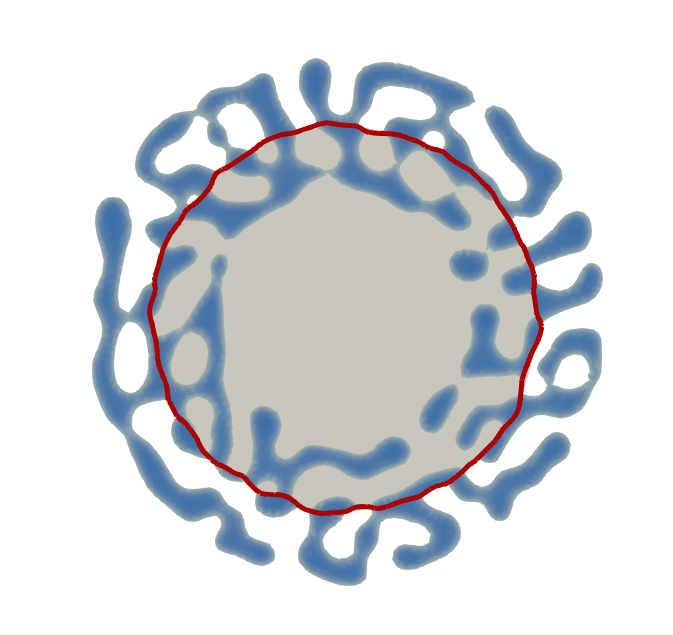}} &
   {\includegraphics[width=0.24\textwidth]{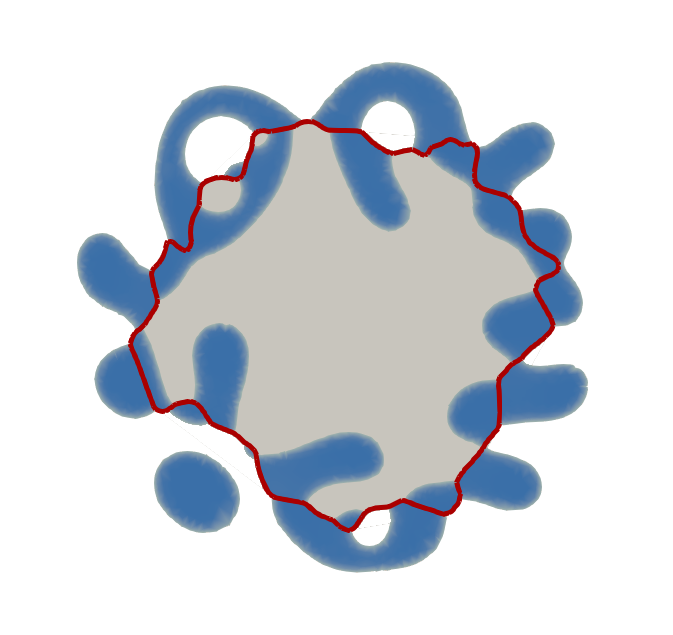}}
   & {\includegraphics[width=0.24\textwidth]{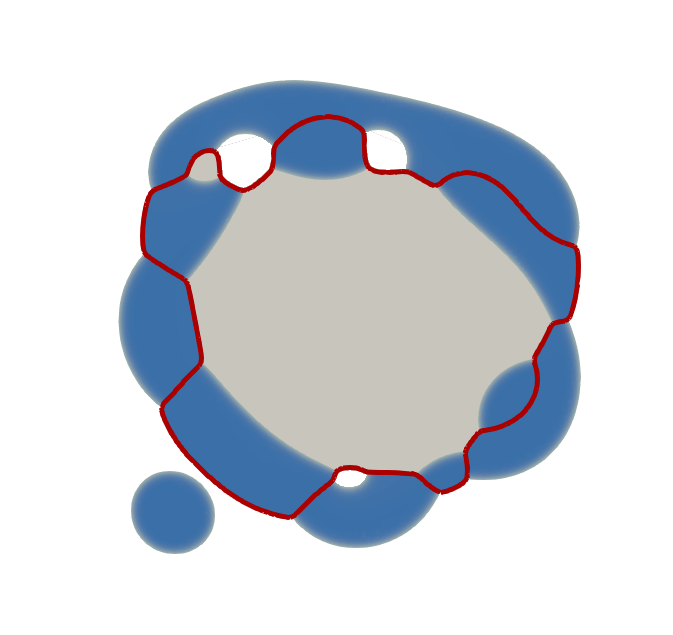}}
   & {\includegraphics[width=0.24\textwidth]{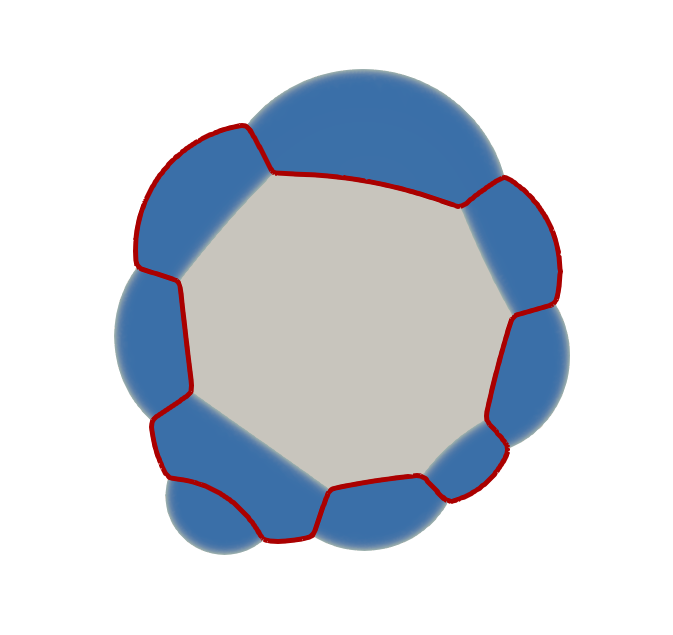}} \\
    \footnotesize $t=0.05~$s & \footnotesize $t=0.25~$s &  \footnotesize $t=1~$s & \footnotesize $t=8~$s
    \end{tabular}
    \caption{Droplet/ambient phase separation outside (top row), inside (middle row), and on both sides (bottom row) of a spherical membrane with contact angle $\theta=60^\circ$. 
    The initial phase-field value is chosen to be a random number $\in[0.2, 0.8]$ on a ring around the membrane. 
    The droplets ($\phi\geq 0.5$) are illustrated in blue, , the membrane in red and the ambient fluids in white (outside) and gray (inside). \textbf{Parameters}:  $\sigma_{\text{f}} = 30\,\mu\text{N/m}$, $\sigma_0 = 15\,\mu\text{N/m}$, $\sigma_1 = 30\,\mu\text{N/m}$, $K_B=8\cdot 10^{-20}\,\text{Nm}$, $K_A = 5\cdot 10^{-5}$\,\text{N/m}, $\varepsilon = 0.025\,\mu$m, and $P=0\,$m$^2$s/kg. Initial membrane radius is $2.5\,\mu$m. Random values for the phase field are set in a ring with internal radius $1.5\mu$m and/or external radius $3.5\,\mu$m.}
    \label{fig:phaseSeparationTheta60}
\end{figure}

\begin{figure}[!ht]
    \centering
   outside only \\
   \begin{tabular}{cccc}
   {\includegraphics[width=0.24\textwidth]{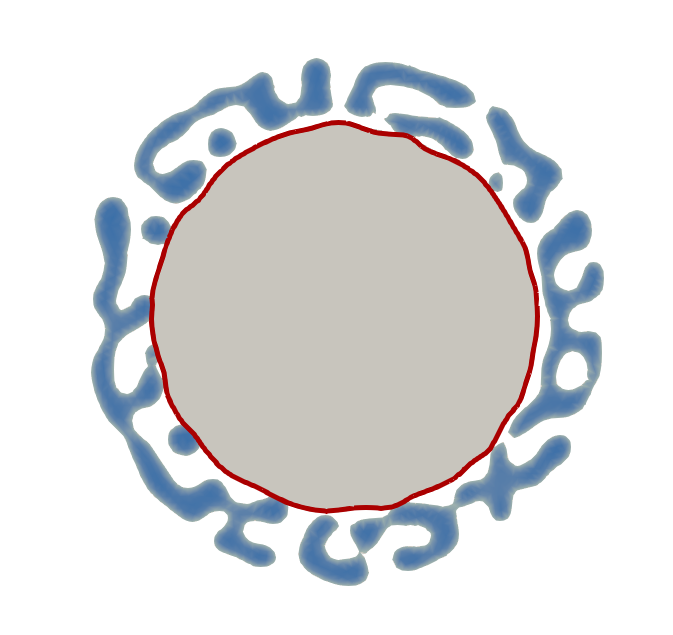}}
   &{\includegraphics[width=0.24\textwidth]{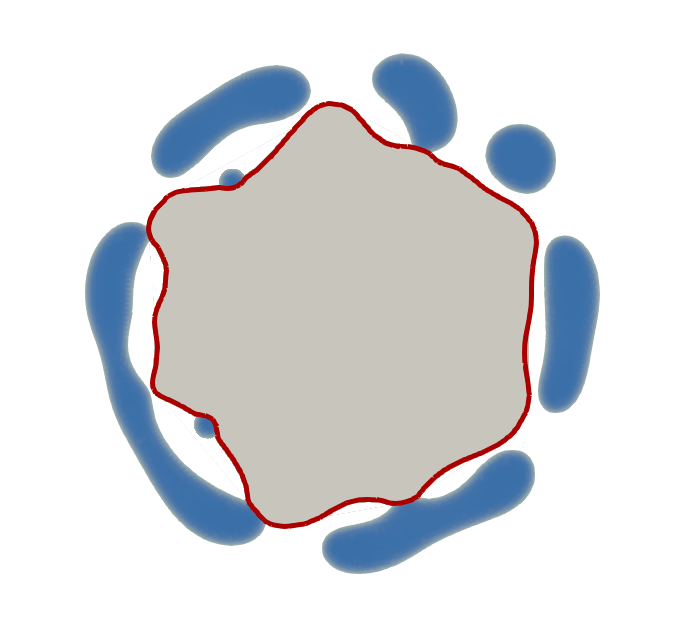}}
    &{\includegraphics[width=0.24\textwidth]{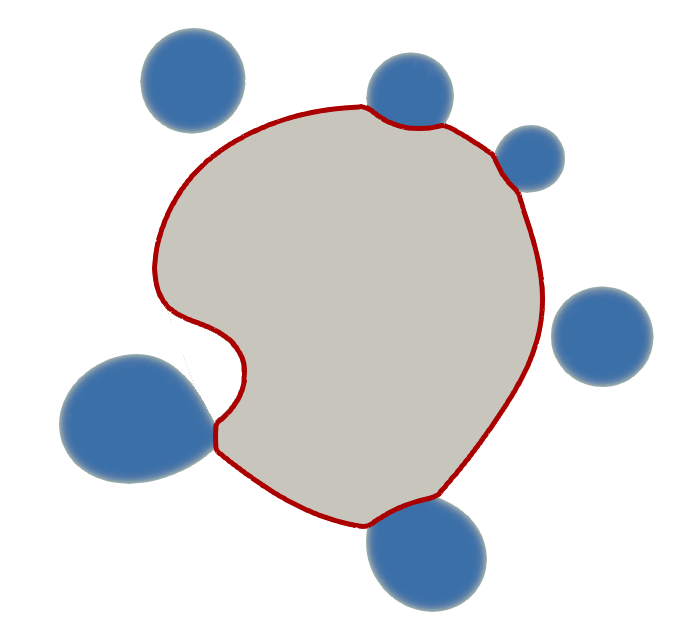}}
   &{\includegraphics[width=0.24\textwidth]{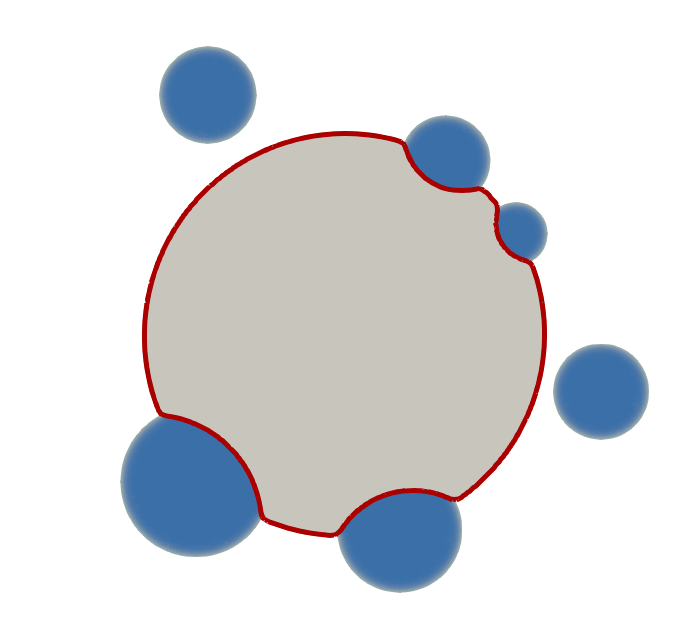}} 
   \end{tabular}
   inside only \\
   \begin{tabular}{cccc}
   {\includegraphics[width=0.24\textwidth]{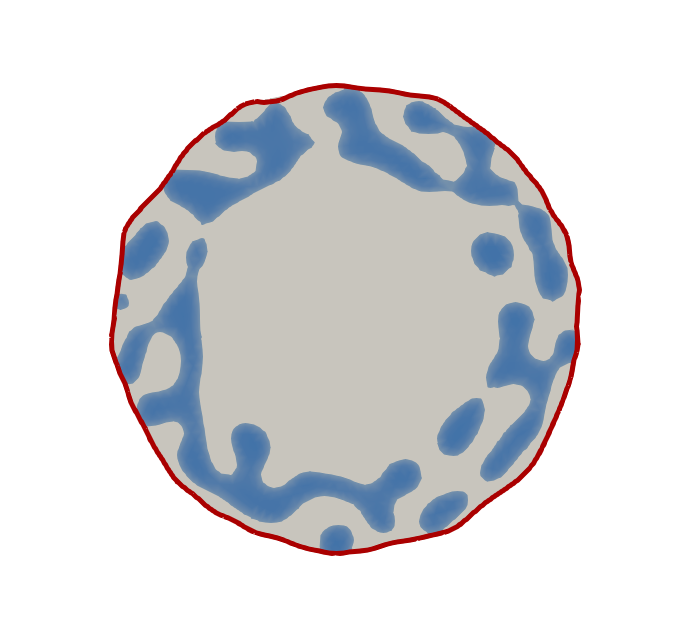}}
   & {\includegraphics[width=0.24\textwidth]{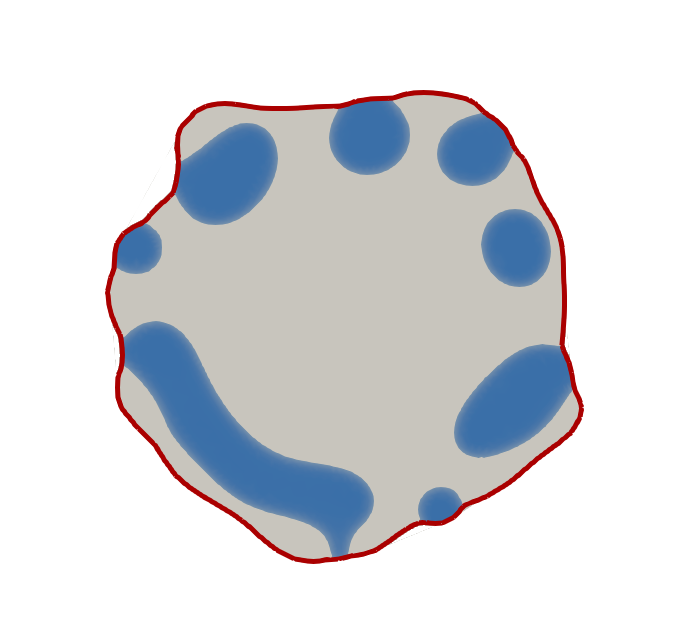}}
   & {\includegraphics[width=0.24\textwidth]{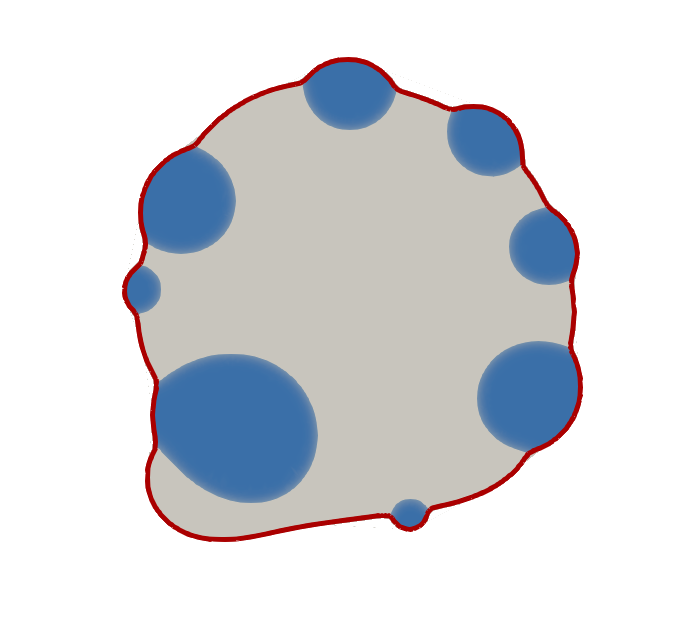}}
   & {\includegraphics[width=0.24\textwidth]{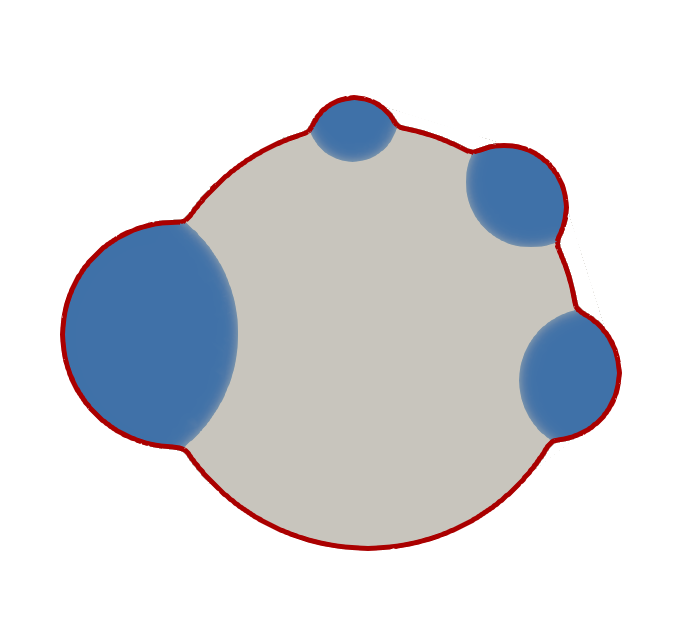}} 
   \end{tabular}
    inside and outside \\
    \begin{tabular}{cccc}
        {\includegraphics[width=0.24\textwidth]{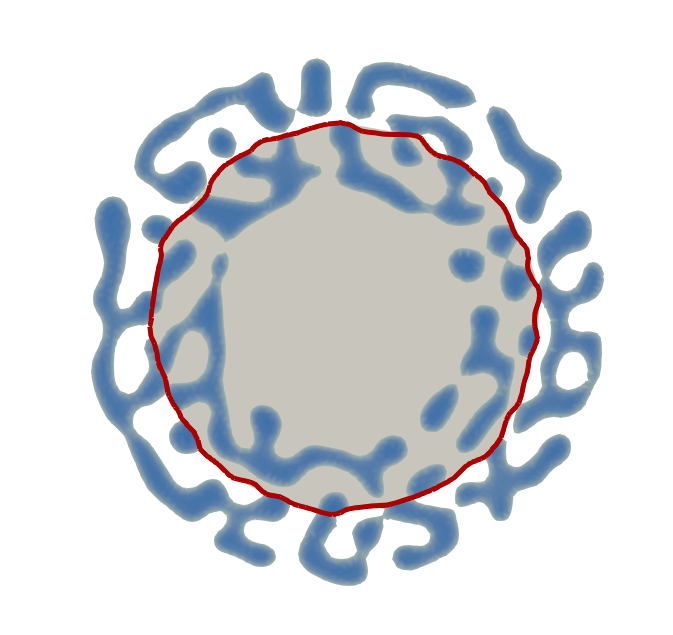}} &
   {\includegraphics[width=0.24\textwidth]{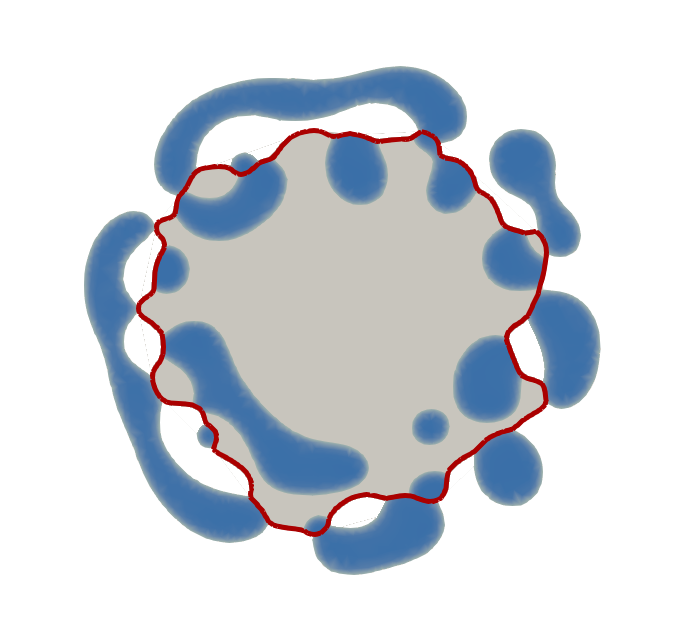}}
   & {\includegraphics[width=0.24\textwidth]{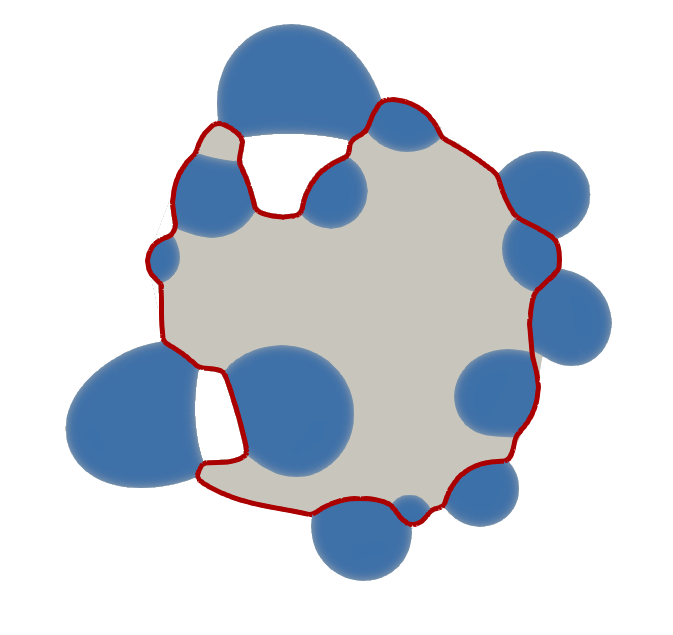}}
   & {\includegraphics[width=0.24\textwidth]{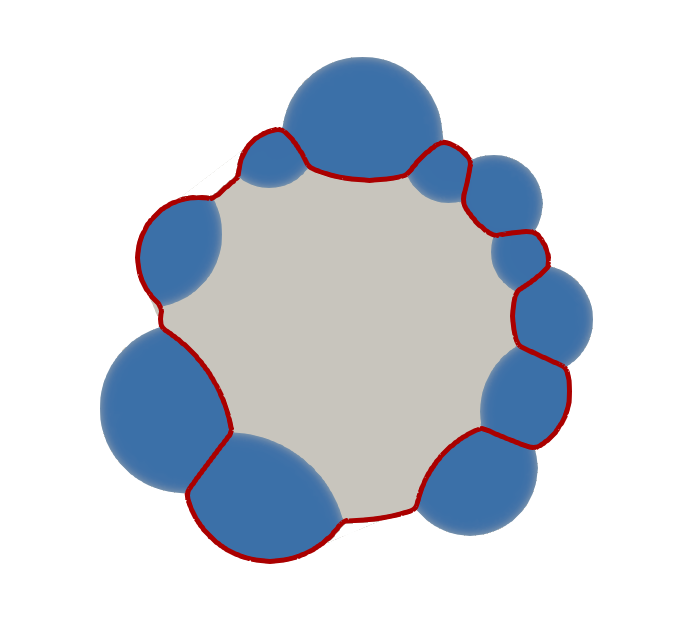}} \\
    \footnotesize $t=0.05~$s & \footnotesize $t=0.25~$s &  \footnotesize $t=1~$s & \footnotesize $t=8~$s
    \end{tabular}
    \caption{Droplet/ambient phase separation outside (top row), inside (middle row), and on both sides (bottom row) of a spherical membrane with contact angle $\theta=120^\circ$.  
    The initial phase-field value is chosen to be a random number $\in[0.2, 0.8]$ on a ring around the membrane. 
    The droplets ($\phi\geq 0.5$) are illustrated in blue, , the membrane in red and the ambient fluids in white (outside) and gray (inside). \textbf{Parameters}:  $\sigma_{\text{f}} = 30\,\mu\text{N/m}$, $\sigma_0 = 30\,\mu\text{N/m}$, $\sigma_1 = 15\,\mu\text{N/m}$, $K_B=8\cdot 10^{-20}\,\text{Nm}$, $K_A = 5\cdot 10^{-5}$\,\text{N/m}, $\varepsilon = 0.025\,\mu$m, and $P=0\,$m$^2$s/kg. Initial membrane radius is $2.5\,\mu$m. Random values for the phase field are set in a ring with internal radius $1.5\mu$m and/or external radius $3.5\,\mu$m.}
    \label{fig:phaseSeparationTheta120}
\end{figure}

In order to simulate phase separation, the phase field was initialized with random values around the membrane ($\phi\in [0.2,0.3]$ uniformly). An illustration of the resulting dynamics is shown in \autoref{fig:phaseSeparation} (neutral), \autoref{fig:phaseSeparationTheta60} (high wettability),  and \autoref{fig:phaseSeparationTheta120} (low wettability). 
In the neutral case ($\theta=90^\circ$), numerous small droplets initially appear in the proximity of the membrane at $t=0.5~$s. Subsequently, these droplets undergo growth by coalescence and Ostwald ripening. During this process, the membrane morphology increasingly adjusts, since larger droplets induce stronger deformations. 

In the cases of high and low wettability (Figs.~\ref{fig:phaseSeparationTheta60}, \ref{fig:phaseSeparationTheta120}), we initialize the order parameter with larger values ($\phi\in [0.2,0.8]$ uniformly), such that larger droplets emerge. 
Consequently, we observe enhanced membrane remodeling, accompanied by strong fluid motion.  
For high wettability ($\theta=60^\circ$). The rich dynamics also enable the emergence of transient droplets of the ambient fluid. 
Membrane remodeling generally seems to assist the coarsening and merging of droplets, such that only few droplets remain at final time.
However, when condensate is present on both sides of the membrane, the opposite behavior is observed. 
Droplets within and outside the membrane exhibit a tendency to attract each other forming zig-zag patterns along the membrane. In this case, the membrane slaloms between the droplets on either side, keeping droplets distant, thereby preventing further coarsening, \autoref{fig:phaseSeparationTheta60}, \ref{fig:phaseSeparationTheta120} (bottom row). 


Phenomenologically, these results fit well to the experimental observations shown in Reference \cite{halim2021}, supplementary movies S3 and S7.
However, more detailed studies are necessary to understand the rich phenomenology and dynamics. 

\section{Conclusion and Outlook} 
\noindent \label{sec:Conclusion}
In this work we presented a first numerical method to simulate the continuum dynamics of droplets interacting with deformable membranes via wetting. 
We derived a thermodynamically consistent model which couples bulk hydrodynamics with capillary forces as well as bending, tension, and stretching of a thin membrane. 
The model combines the advantages of the phase-field method for simulation of multi-phase flow with those of the arbitrary Lagrangian-Eulerian (ALE) method for an explicit description of the elastic surface. 
Diffusivity of the liquid-liquid interface stabilizes the motion of the three-phase contact line, prevents contact line singularity and locking, and enables simulation of topological changes, such as droplet coalescence and break-up. 
On the other hand, the fitted finite element approach used to represent the membrane exactly resolves the discontinuities of pressure, viscosity, and liquid concentration across the membrane.
Paired with adaptive mesh refinement, the method can accurately resolve the high curvature at the apparent membrane cusp observed at the three-phase contact line. 
In addition to grid movement by the ALE method, we introduced a remeshing algorithm which enables to track large deformations of membranes up to a topological change.

We validated the method by comparing simulations for single droplets to the theoretical results of shape equations. Very good agreement was found, even for the high membrane curvature at the three-phase contact point. 
In a series of numerical tests, we illustrated the capabilities of the proposed method in 2D and 3D axisymmetric scenarios. 
We provided a first simulation of the dynamics of mutual droplet/membrane remodeling. Moreover, an inverted endocytosis was simulated, where a smaller vesicle is absorbed into a larger drop by capillary forces. 
Additionally, we found an inverted Cheerios effect, evidenced by membrane-mediated repulsion of two droplets.
Finally, we provided simulations of liquid-liquid phase separation around a vesicle membrane. The rich dynamics of which illustrate that numerical simulations are indispensable to systematically study these highly non-linear phenomena.

Despite its potential for such toy problems, the presented method represents only a first step towards biologically relevant simulations of wetted membranes.
The biological system comprises a variety of complex features which are not yet accounted for, including spontaneous curvature, line tension, and membrane binding \cite{zhao2021thermodynamics}.
Recognizing that bending stiffness generally depends on droplet contact \cite{mangiarotti2023biomolecular} underlines the need for wetting-dependent bending stiffness. 
In this case also the Gaussian bending stiffness which is linked to $K_B$ \cite{hu2012determining}  is no longer negligible and poses conceptual challenges as additional coupling terms will arise in the phase-field evolution.
Moreover, to address complex non-axisymmetric geometries, will require full 3D simulations, the computational effort of which demands parallelization, new re-meshing strategies and highly stable time stepping.

A clear limitation of the present method is given by the grid-based approach which excludes topological changes of the membrane.  To handle membrane fusion and fission will require fundamentally different membrane representation and different numerical techniques. 
Finally, condensates are also known to interact with a multitude of filamentous structures inside of cells \cite{gouveia2022capillary}. One of the most exciting phenomena is found in the genome, where condensates can have tremendous effects on the spatial organization of DNA \cite{quail2021force}. The development of numerical methods for droplets interacting with slim filaments such as DNA will be an ambitious numerical challenge in the future.

\textbf{Acknowledgements}: SA acknowledges support from the German Research Foundation (grant AL1705/5). 
The authors gratefully acknowledge computing time on the high-performance computer at the NHR Center of TU Dresden. This center is jointly supported by the Federal Ministry of Education and Research and the state governments participating in the NHR (www.nhr-verein.de/unsere-partner).

  \bibliographystyle{siamplain} 
  \bibliography{bibliographie}




\end{document}


\maketitle

\begin{center}
    \textbf{SUPPLEMENTARY MATERIAL}
\end{center}

\section{Appendix I - Energy time derivative}\label{sec:AppendixEnergyTimeDerivative}

To derive a thermodynamically consistent system, we will compute the time variation of the total energy step by step. We denote the material time derivative by $\partial^{\bullet}_t=\partial_t+\textbf{v}\cdot\nabla$. From Reynold's transport theorem on a closed surface moving with velocity $\textbf{v}$, we obtain
\begin{align}
    d_t E_{\sigma_{\text{m}}}=\int_{\Gamma}\sigma_{\text{m}}'(\phi)\partial^{\bullet}_t\phi+\sigma_{\text{m}}(\phi)\nabla_{\Gamma}\cdot\textbf{v}\,\text{d}A
\end{align}
where $\nabla_{\Gamma}\cdot$ is the surface divergence. 
Using $\textbf{v}=\textbf{P}\textbf{v}+\left(\textbf{v}\cdot\textbf{n}\right)\textbf{n}$ with the surface projection operator $\textbf{P}=\textbf{I}-\textbf{n}\otimes\textbf{n}$, $\kappa=\nabla_{\Gamma}\cdot\textbf{n}$ 
and integration by parts gives
\begin{align}
    d_t E_{\sigma_{\text{m}}}=\int_{\Gamma}\sigma_{\text{m}} '(\phi)\partial^{\bullet}_t\phi+\textbf{v}\cdot\left(\kappa\textbf{n}\sigma_{\text{m}}(\phi)-\nabla_{\Gamma}\sigma_{\text{m}}(\phi)\right)\,\text{d}A.\label{EsigmamVariation}
\end{align}
For the droplet/ambient surface energy we can use Reynold's transport theorem for a scalar quantity $f(\textbf{x})$,
\begin{align}
    d_t\int_{\Omega_{\text{out}}} f\,\text{d}V
   & =\int_{\Omega_{\text{out}}}\partial_t f\,\text{d}V+\int_{\partial\Omega_{\text{out}}}\left(\textbf{v}\cdot\textbf{n}\right)f\,\text{d}A
    =\int_{\Omega_{\text{out}}}\partial^{\bullet}_tf+f\nabla\cdot\textbf{v} \,\text{d}V\nonumber\\
   & =\int_{\Omega_{\text{out}}}\partial^{\bullet}_tf\,\text{d}V,\label{eq:Reynolds}
\end{align}
where the fluid incompressibility has been used in the last step. 
This yields
\begin{align}
    d_t E_{\sigma_{\text{f}}}=\tilde{\sigma}_{\text{f}}\int_{\Omega_{\text{out}}}\varepsilon\nabla\phi\partial^{\bullet}_t\nabla\phi+\frac{1}{\varepsilon}W'(\phi)\partial^{\bullet}_t\phi\,\text{d}V.
\end{align}
Consequently, using $\partial^{\bullet}_t\nabla\phi=\nabla\partial^{\bullet}_t\phi-\Sigma_i\nabla\textbf{v}_i\cdot\nabla_i\phi$ and integration by parts twice we obtain
\begin{align}
    d_tE_{\sigma_{\text{f}}} &= \tilde{\sigma}_{\text{f}}\int_{\Omega_{\text{out}}}\varepsilon\nabla\phi\nabla\partial^{\bullet}_t\phi + \frac{1}{\varepsilon}W'(\phi)\partial^{\bullet}_t\phi-\varepsilon\nabla\phi\cdot\nabla\textbf{v}\cdot\nabla\phi\,\text{d}V \\
    &=\int_{\Omega_{\text{out}}}\partial^{\bullet}_t\phi\frac{\delta E_{\sigma_{\text{f}}}}{\delta\phi}-\textbf{v}\cdot\nabla\cdot\textbf{S}_{\sigma_{\text{f}}}\,\text{d}V + \int_{\Gamma}\textbf{v}\cdot\textbf{S}_{\sigma_{\text{f}}}\cdot\textbf{n}+\tilde{\sigma}_{\text{f}}\varepsilon\textbf{n}\cdot\nabla\phi\partial^{\bullet}_t\phi\,\text{d}A\,,\label{EsigmafVariation}
\end{align}
where we introduced the chemical potential $\frac{\delta E_{\sigma_{\text{f}}}}{\delta\phi}:=\tilde{\sigma}_{\text{f}}\left(\frac{1}{\varepsilon}W'(\phi)-\varepsilon\Delta\phi\right)$ and the capillary stress $\textbf{S}_{\sigma_{\text{f}}}:=-\tilde{\sigma}_{\text{f}}\varepsilon\nabla\phi\otimes\nabla\phi$. Anticipating the role of $\textbf{S}_{\sigma_{\text{f}}}$ as a stress we may write the momentum and mass balance as
\begin{align}
    \rho\partial^{\bullet}_t\textbf{v}&=\nabla\cdot\left(\textbf{S}+\chi_{\text{out}}\textbf{S}_{\sigma_{\text{f}}}\right) &\text{ in }\Omega_{\text{in}}\cup\Omega_{\text{out}} \label{momentumBalance} \\
    \nabla\cdot\textbf{v}&=0 &\text{ in }\Omega_{\text{in}}\cup\Omega_{\text{out}}
\end{align}
with the stress tensor composed of pressure and viscous contribution, $\textbf{S}:=-p\textbf{I}+\eta\left(\nabla\textbf{v}+\nabla\textbf{v}^T\right)$, and $\chi_{\text{out}}$ being the characterisic function of $\Omega_{\text{out}}$, i.e.\ $\chi_{\text{out}}=1$ in $\Omega_{\text{out}}$ and $\chi_{\text{out}}=0$ in $\Omega\setminus\Omega_{\text{out}}$. Let us note here that $p$ and $\eta$, unlike $\textbf{v}$, may be discontinuous across the membrane, e.g.\ $\eta=\eta_{\text{out}}(\phi)\chi_{\text{out}}+\eta_{\text{in}}\chi_{\text{in}}$. Moreover, we assume $\rho$ to be constant in the following. For the kinetic energy, we obtain
\begin{align}
    d_t E_{\text{kin}}&=\int_{\Omega_{\text{in}}\cup\Omega_{\text{out}}}\rho\textbf{v}\cdot\partial^{\bullet}_t\textbf{v}\,\text{d}V \\
    &{\underset{\scriptsize\ref{momentumBalance}}{=}}\int_{\Omega_{\text{in}}\cup\Omega_{\text{out}}}\textbf{v}\cdot\nabla\cdot\textbf{S}\,\text{d}V+\int_{\Omega_{\text{out}}}\textbf{v}\cdot\nabla\cdot\textbf{S}_{\sigma_{\text{f}}}\,\text{d}V \\
    &= \int_{\Omega_{\text{in}}\cup\Omega_{\text{out}}}-\frac{\eta}{2}\lVert\nabla\textbf{v}+\nabla\textbf{v}^T\rVert^2_{\text{F}}\,\text{d}V+\int_{\Omega_{\text{out}}}\textbf{v}\cdot\nabla\cdot\textbf{S}_{\sigma_{\text{f}}}\,\text{d}V+\int_{\Gamma}\left[\textbf{S}\cdot\textbf{n}\right]_{\text{in}}^{\text{out}} \cdot\textbf{v}\,\text{d}A\,,\label{EkinVariation}
\end{align}
where $\lVert\cdot\rVert_{\text{F}}$ denotes the Frobenius inner product and $\left[\cdot\right]_{\text{in}}^{\text{out}}$ the jump of the quantity in brackets across $\Gamma$. For \ref{EkinVariation} we used integration by parts twice, $\nabla\cdot\textbf{v}=0$ and neglected the occuring boundary integrals indicating that there is no energy inflow.

Summing up \ref{EsigmamVariation}, \ref{EsigmafVariation}, and \ref{EkinVariation}, we obtain for the total energy
\begin{align}
    d_t E=\int_{\Gamma}\partial^{\bullet}_t\phi\left(\tilde{\sigma}_{\text{f}}\varepsilon\textbf{n}\cdot\nabla\phi+\sigma_{\text{m}}'(\phi)\right)+\textbf{v}\cdot\left(\kappa\textbf{n}\sigma_{\text{m}}(\phi)-\nabla_{\Gamma}\sigma_{\text{m}}(\phi)\right) \notag \\
    + \left[\textbf{S}\cdot\textbf{n}+\chi_{\text{out}}\textbf{S}_{\sigma_{\text{f}}}\cdot\textbf{n}\right]^{\text{out}}_{\text{in}}\cdot\textbf{v}+\textbf{v}\cdot\left(\frac{\delta E_{\text{bend}}}{\delta\Gamma}+\frac{\delta E_{\text{stretch}}}{\delta\Gamma}\right)\,\text{d}A \notag \\
    +\int_{\Omega_{\text{out}}}\partial^{\bullet}_t\phi\frac{\delta E_{\sigma_{\text{f}}}}{\delta\phi}\,\text{d}V-\int_{\Omega_{\text{in}}\cup\Omega_{\text{out}}}\frac{\eta}{2}\lVert\nabla\textbf{v}+\nabla\textbf{v}^T\rVert^2_{\text{F}}\,\text{d}V. \label{eq:dtE}
\end{align}

It remains to specify $\frac{\delta E_{\text{bend}}}{\delta\Gamma}$ and $\frac{\delta E_{\text{stretch}}}{\delta\Gamma}$. 

The specific form of $ E_{\text{stretch}}$ may depend on the physical constitution of the membrane. For example for a lipid bilayer membrane the shear modulus would be zero, such that only surface dilation contributes to the energy. We compute the time derivative of $ E_{\text{stretch}}$  as
\begin{align}
    d_t E_{\text{stretch}}&=K_A\int_{\Gamma}\left(J-1\right)\underbrace{\partial^{\bullet}_tJ}_{=J\nabla_{\Gamma}\cdot\textbf{v}}+\frac{1}{2}\left(J-1\right)^2\nabla_{\Gamma}\cdot\textbf{v}\,\text{d}A \\
    &=K_A\int_{\Gamma}\nabla_{\Gamma}\cdot\textbf{v}\left(J^2-J+\frac{1}{2}J^2-J+\frac{1}{2}\right)\,\text{d}A \\
    &\approx K_A\int_{\Gamma}\textbf{v}\cdot\left(\kappa\textbf{n}\left(J-1\right)-\nabla_{\Gamma}\left(J-1\right)\right)\,\text{d}A\,.\label{eq:EstretchVariation}
\end{align}
Note, that the term in brackets was simplified by linearization around $J_0=1$ in the last step. This approximation is quite accurate, as the stretching of a membrane is typically very small ($J\approx 1$). 
Hence, we define
\begin{align}
    \frac{\delta E_{\text{stretch}}}{\delta\Gamma}=K_A\kappa\textbf{n}(J-1)-K_A\nabla_{\Gamma}(J-1).
\end{align}
Moreover, the bending energy is given by \cite{lowengrub}
\begin{align}
    \frac{\delta E_{\text{bend}}}{\delta\Gamma}=K_B\left(\Delta_{\Gamma}\kappa-2K_g\kappa+\frac{1}{2}\kappa^3 \right)\textbf{n}\,,\label{eq:EbendVariation}
\end{align}
with the Gaussian curvature $K_g$ and the surface Laplacian $\Delta_{\Gamma}$. 

\section{Appendix II - Derivation of the weak formulation of the bending force}
We show here the correctness of the given weak formulation of the bending force from Dziuk \cite{dziuk08}:
\begin{align}
    \int_\Gamma \frac{\delta E_{\text{bend}}}{\delta\Gamma}{\textbf{\textpsi}} = &K_B\int_\Gamma \frac{1}{2}|\textbf{\textkappa}|^2\nabla_\Gamma\cdot\textbf{\textpsi} +  \nabla_\Gamma{\textbf{\textkappa}}:\nabla_\Gamma\textbf{\textpsi} +  \nabla_\Gamma\cdot\textbf{\textkappa}\nabla_\Gamma\cdot\textbf{\textpsi}  \nonumber\\ &~~~~~~~~-  \left(\nabla_\Gamma\textbf{\textpsi} + \nabla_\Gamma\textbf{\textpsi}^T\right)\mathbf{P}:\nabla_\Gamma\textbf{\textkappa}~\mathrm{d}A.\label{eq:bendingForceWeakForm}
\end{align}
Therefore, take the alternative version of \autoref{eq:EbendVariation}
\begin{align}
    \frac{\delta E_{\text{bend}}}{\delta\Gamma}=K_B\left(\Delta_{\Gamma}\kappa-\frac{1}{2}\kappa^3  + \kappa\lVert\nabla_\Gamma\mathbf{n}\rVert_{\text{F}}^2\right)\textbf{n},
\end{align}
multiply it with test functions $\textbf{\textpsi}$ and integrate over $\Gamma$
\begin{align}
    \int_\Gamma\frac{\delta E_{\text{bend}}}{\delta \Gamma}\cdot\textbf{\textpsi} &= K_B\int_\Gamma\Delta_\Gamma\kappa\mathbf{n}\cdot\textbf{\textpsi} - \frac{1}{2}\kappa^3\mathbf{n}\cdot\textbf{\textpsi} + \kappa\lVert\nabla_\Gamma\mathbf{n}\rVert_{\text{F}}^2\mathbf{n}\cdot\textbf{\textpsi} 
\end{align}
It holds that
\begin{align}
    \lVert\nabla_\Gamma\mathbf{n}\rVert_{\text{F}}^2\mathbf{n} = \nabla_\Gamma\kappa - \Delta_\Gamma\mathbf{n},
\end{align}
which we replace in the above equation to get
\begin{align}
    \int_\Gamma\frac{\delta E_{\text{bend}}}{\delta \Gamma}\cdot\textbf{\textpsi} &= K_B\int_\Gamma\Delta_\Gamma\kappa\mathbf{n}\cdot\textbf{\textpsi} - \frac{1}{2}\kappa^3\mathbf{n}\cdot\textbf{\textpsi} + \kappa \nabla_\Gamma\kappa\cdot\textbf{\textpsi} - \kappa\Delta_\Gamma\mathbf{n}\cdot\textbf{\textpsi}. \label{eq:adaptedBendingWeak}
\end{align}
Now, we replace the second and third term by
\begin{align}
    \int_\Gamma -\frac{1}{2}\kappa^3\mathbf{n}\cdot\textbf{\textpsi} + \kappa \nabla_\Gamma\kappa\cdot\textbf{\textpsi} = -\frac{1}{2}\int_\Gamma \kappa^2\nabla_\Gamma\cdot\textbf{\textpsi},\label{eq:firstReplacement}
\end{align}
due to the following considerations. By the product rule, it is
\begin{align}
    \int_\Gamma \frac{1}{2} \nabla_\Gamma\cdot\left(\kappa^2\textbf{\textpsi}\right) = \int_\Gamma \frac{1}{2}\kappa^2\nabla_\Gamma\cdot\textbf{\textpsi} + \kappa\nabla_\Gamma\kappa\cdot\textbf{\textpsi},\label{eq:lemma}
\end{align}
and, by Gauss' theorem, it is
\begin{align}
    \int_\Gamma \nabla_\Gamma\cdot \bm{f} - \kappa \bm{f}\cdot \mathbf{n} = \int_{\partial\Gamma} \bm{f}\cdot\bm{\nu}\label{eq:Gauss}
\end{align}
for a vector field $\bm{f}$ and the conormal $\bm{\nu}$ to the boundary of $\Gamma$. In our case, the term on the right hand side vanishes, since the surface is closed. Using \autoref{eq:lemma} and \autoref{eq:Gauss} yields \autoref{eq:firstReplacement}. Additionally, by recalling that $\textbf{\textkappa}=-\kappa\mathbf{n}$ it is
\begin{align}
\nabla_\Gamma\cdot\textbf{\textkappa} = -\nabla_\Gamma\cdot\left(\kappa\mathbf{n}\right) = -\nabla_\Gamma\kappa\cdot\mathbf{n} - \kappa^2 = -\kappa^2  \label{eq:divKappaVEcEqualsMinusKappaSquared} 
\end{align}
Hence, we can obtain
\begin{align}
    -\int_\Gamma \frac{1}{2}\kappa^2\nabla_\Gamma\cdot\textbf{\textpsi} = \int_\Gamma \nabla_\Gamma\cdot\textbf{\textkappa}\nabla_\Gamma\cdot\textbf{\textpsi} +\int_\Gamma \frac{1}{2}\left|\textbf{\textkappa}\right|^2\nabla_\Gamma\cdot\textbf{\textpsi}  \label{eq:intermediateResult}
\end{align}
where the right hand side are exactly the first and third term in the weak form of the bending force \autoref{eq:bendingForceWeakForm}. Note that both, \autoref{eq:divKappaVEcEqualsMinusKappaSquared} as well as $\kappa^2 = \left|\textbf{\textkappa}\right|^2$ hold. The usage of both terms combined as in the right hand side of \autoref{eq:intermediateResult} is used for consistency with the literature \cite{dziuk08,barrett2020}.

Next, consider the first and last term of \autoref{eq:adaptedBendingWeak}. Here, knowing that $\textbf{P}\mathbf{n}=\mathbf{n}^T\textbf{P}=0$, the following identity holds:
\begin{align}
    \nabla_\Gamma\kappa\otimes\mathbf{n}:\nabla_\Gamma\textbf{\textpsi}\textbf{P} = \nabla_\Gamma\kappa^T\nabla_\Gamma\textbf{\textpsi}\textbf{P}\mathbf{n} = 0 \nonumber\\
    \nabla_\Gamma\kappa\otimes\mathbf{n}:(\nabla_\Gamma\textbf{\textpsi})^T\textbf{P} = \nabla_\Gamma\kappa^T(\nabla_\Gamma\textbf{\textpsi})^T\textbf{P}\mathbf{n} = 0
\end{align}
Using integration by parts for a surface without boundary, the properties of the Frobenius inner product, the fact that $\nabla_\Gamma\mathbf{n}$ is symmetrical with $\textbf{P}\nabla_\Gamma\mathbf{n}=\nabla_\Gamma\mathbf{n}\textbf{P} = \nabla_\Gamma\mathbf{n}$, and $\textbf{\textkappa}=-\kappa\mathbf{n}$, and the product rule we get
\begin{align}
    \int_\Gamma \Delta_\Gamma\kappa\mathbf{n}\cdot\textbf{\textpsi} - \kappa\Delta_\Gamma\mathbf{n}\cdot\textbf{\textpsi} &= 
    \int_\Gamma -\nabla_\Gamma\kappa\cdot\nabla_\Gamma\left(\mathbf{n}\cdot\textbf{\textpsi}\right) + \nabla_\Gamma\mathbf{n}:\nabla_\Gamma\left(\kappa\textbf{\textpsi}\right) \nonumber\\ 
    &= \int_\Gamma-\nabla_\Gamma\kappa\cdot\left(\nabla_\Gamma\mathbf{n}\textbf{\textpsi} + \nabla_\Gamma\textbf{\textpsi}\mathbf{n}\right) + \nabla_\Gamma\mathbf{n} : \left(\nabla_\Gamma\kappa\otimes\textbf{\textpsi} + \kappa\nabla_\Gamma\textbf{\textpsi}\right) \nonumber\\
     &=\int_\Gamma -\nabla_\Gamma\kappa^T\nabla_\Gamma\mathbf{n}\textbf{\textpsi} - \nabla_\Gamma\kappa^T\nabla_\Gamma\textbf{\textpsi}\mathbf{n} + \nabla_\Gamma\kappa^T\nabla_\Gamma\mathbf{n}\textbf{\textpsi} \nonumber\\
     &~~~~~~~~+ \kappa\nabla_\Gamma\mathbf{n}:\nabla_\Gamma\textbf{\textpsi}\nonumber\\
     &= \int_\Gamma -\nabla_\Gamma\kappa\otimes\mathbf{n}:\nabla_\Gamma\textbf{\textpsi} + \kappa\nabla_\Gamma\mathbf{n}:\nabla_\Gamma\textbf{\textpsi}\nonumber\\
     &= \int_\Gamma \nabla_\Gamma\textbf{\textkappa}:\nabla_\Gamma\textbf{\textpsi} + 2\kappa\nabla_\Gamma\mathbf{n}:\nabla_\Gamma\textbf{\textpsi}\nonumber\\ 
     & = \int_\Gamma \nabla_\Gamma\textbf{\textkappa}:\nabla_\Gamma\textbf{\textpsi} + \kappa\nabla_\Gamma\mathbf{n}:\left(\nabla_\Gamma\textbf{\textpsi} + \nabla_\Gamma\textbf{\textpsi}^T\right)\textbf{P} \nonumber\\
     & = \int_\Gamma \nabla_\Gamma\textbf{\textkappa}:\nabla_\Gamma\textbf{\textpsi} + \left(\kappa\nabla_\Gamma\mathbf{n} + \nabla_\Gamma\kappa\otimes\mathbf{n}\right):\left(\nabla_\Gamma\textbf{\textpsi} + \nabla_\Gamma\textbf{\textpsi}^T\right)\textbf{P} \nonumber\\     
     &= \int_\Gamma\nabla_\Gamma\textbf{\textkappa}:\nabla_\Gamma\textbf{\textpsi} - \int_\Gamma \nabla_\Gamma\textbf{\textkappa}:\left(\nabla_\Gamma\textbf{\textpsi} + (\nabla_\Gamma\textbf{\textpsi})^T\right)\textbf{P}
\end{align}
which is exactly the second and last term in \autoref{eq:bendingForceWeakForm}.

In 2D, however, the third and fourth term in weak form \autoref{eq:bendingForceWeakForm} can be simplified due to the following observations:
\begin{align}
    \nabla_\Gamma\mathbf{n}:\nabla_\Gamma\mathbf{n} &= \kappa^2 \\
    \nabla_\Gamma\mathbf{n} &= \kappa\textbf{P}\\
    \textbf{P}:\nabla_\Gamma\textbf{\textpsi} &= \nabla_\Gamma\cdot\textbf{\textpsi}
\end{align}
It is then
\begin{align}
    \int_\Gamma \nabla_\Gamma\cdot\textbf{\textkappa}\nabla_\Gamma\cdot\textbf{\textpsi} - \nabla_\Gamma\textbf{\textkappa}:\left(\nabla_\Gamma\textbf{\textpsi} + \nabla_\Gamma\textbf{\textpsi}^T\right)\textbf{P} &=  \int_\Gamma -\kappa^2\nabla_\Gamma\cdot\textbf{\textpsi} + 2\kappa\nabla_\Gamma\mathbf{n}:\nabla_\Gamma\textbf{\textpsi}\nonumber\\
    &= \int_\Gamma -\kappa^2\nabla_\Gamma\cdot\textbf{\textpsi} + 2\kappa^2\textbf{P}:\nabla_\Gamma\textbf{\textpsi} \nonumber\\
    &= \int_\Gamma -\kappa^2\nabla_\Gamma\cdot\textbf{\textpsi} + 2\kappa^2\nabla_\Gamma\cdot\textbf{\textpsi}\nonumber\\
    &= \int_\Gamma \kappa^2\nabla_\Gamma\cdot\textbf{\textpsi}\nonumber\\
    &= \int_\Gamma -\nabla_\Gamma\cdot\textbf{\textkappa}\nabla_\Gamma\cdot\textbf{\textpsi}
\end{align}

\section{Appendix III - Derivation of the shape equations}\label{sec:shapeEquations}
In the following we derive the shape equations that describe a single liquid droplet binding to a membrane vesicle in the stationary state. These equations are used to validate the numerical method in Section 4.1 of the main text.
Kusumaatmaja et al. have studied a similar system in three dimensions, where they found that for negligible line tension the local contact angle between droplet and membrane is given by the Young-Dupré law independent of the vesicle size \cite{kusumaatmaja2009,young1805,dupre1869}. In contrast, the effective contact angles that are determined for instance by optical microscopy, when the vesicle appears to exhibit a kink, can differ significantly from the local contact angle.\\
We here consider a two-dimensional, mirror-symmetric system. The contour of the vesicle is parametrized by the arclength $S$. The radial distance $R(S)$, the height $Z(S)$ and the tilt angle $\psi(S)$ are related via $dR/dS=\cos(\psi)$ and $dZ/dS=-\sin(\psi)$. We set $S=0$ at the center of the contact region between the vesicle and the droplet. $S_1$ denotes the contact point, where membrane, dense phase and dilute phase meet. $S_2$ is the lower center point of the vesicle. The upper surface of the droplet is described by a circle segment.

\begin{figure}[!ht]
    \centering
\includegraphics[width=0.24\textwidth]{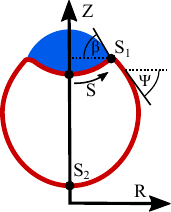}
    \caption{Schematic representation of a single droplet binding to a vesicle in arc-length parameterization}
    \label{fig:analytic_model_schematic_figure}
\end{figure}
The total energy of the system contains contributions from the bending energy, the surface tension acting at the various interfaces and the pressure difference between the inside of the vesicle and the dense and the dilute phase.\\
We model the bending energy  $E_{\rm bend}$ by a Helfrich energy term
\begin{equation}
	E_{\rm bend}  = 2\int_0^{S_2}dS \frac{K_B}{2} \kappa^2,
\end{equation}
where the total curvature reads $\kappa=\frac{d\psi}{dS}$ in the two-dimensional arc-length parametrization. We note that due to the system's symmetry integration is performed only along one half of the vesicle circumference, \textit{i.e.} between $S=0$ and $S_2$.  We neglect the contribution from the Gaussian curvature, since we consider a homogenous membrane for a system that does not under go topological changes. According to the Gauss–Bonnet theorem an integral over the Gaussian curvature merely adds a constant to the energy.\\
Furthermore, the tensions at the fuild-membrane interfaces $\sigma_0$, $\sigma_1$ and at the dense-dilute interface $\sigma_{\rm f}$ contribute to the total energy. 
To parametrize the length of the dense-dilute interface, we introduce the angle $\beta$ formed by the droplet surface with the horizontal at the contact point $S_1$. The energy contributions of the surface tensions $E_{\rm tens}$ thus read
\begin{equation}
	E_{\rm tens}  = 2\int_0^{S_1}dS \sigma_1 + 2\int_{S_1}^{S_2}dS \sigma_0 + \sigma_{\rm f} R_1\frac{2\beta}{\sin\beta},
\end{equation}
with $R_1=R(S_1)$. Finally, the contribution of pressure shall be discussed. The corresponding energy term  $E_{\rm pres}$ reads
\begin{equation}
	E_{\rm pres}  = \left(P_{1}-P_{0}\right)A_{\rm d} + \left(P_{\rm v}-P_{0}\right)A_{\rm v},
\end{equation}
with $P_{0}$, $P_{1}$, $P_{\rm v}$ the pressure in the dilute phase, the dense phase and the vesicle, while $A_{\rm d}$ and $A_{\rm V}$ denote the area, \textit{i.e.} the two-dimensional volume, of the droplet and vesicle, respectively. The vesicle area is obtained as
\begin{equation}
	A_{\rm v}  = 2\int_0^{S_2}dS\,R\sin\psi.
\end{equation}
The droplet area is divided into $A_{\rm d}^{\rm top}$ and $A_{\rm d}^{\rm bot}$, the area above and below the horizontal at the contact point $S_1$, with
\begin{equation}
	A_{\rm d}^{\rm top}  = R_1^2\frac{2\beta-\sin(2\beta)}{2\sin^2(\beta)}
\end{equation}
and 
\begin{equation}
	A_{\rm d}^{\rm bot}  = -2\int_0^{S_1}dS\,R\sin\psi.
\end{equation}
$E_{\rm pres}$ thus reads
\begin{equation}
	E_{\rm pres} = (P_{1}-P_{0})R_1^2\frac{2\beta-\sin(2\beta)}{2\sin^2(\beta)} + 2(P_{\rm v}-P_{1})\int_0^{S_1}dS\, R\sin\psi + 2(P_{\rm v}-P_{0})\int_{S_1}^{S_2}dS\, R\sin\psi.
\end{equation}
The pressure difference $P_{1}-P_{0}$ is related to the surface tension $\sigma_{\rm f}$ via the Laplace pressure, which leads to the expression
\begin{equation}
	P_{1}-P_{0}=-\sigma_{\rm f}\frac{\sin\beta}{R_1}.
	\label{eq:si_laplace_equation}
\end{equation}
The total energy $E$ of the system now reads
\begin{equation}
	E = E_{\rm bend} + E_{\rm tens} +E_{\rm pres}.
\end{equation}
To determine the energy minimizing shapes we use functional variation as described in detail in Reference \cite{julicher1996}. We introduce the generalized parameter $x$ and express $S(x)$ and subsequently $R(x)$, $Z(x)$ and $\psi(x)$ as function of $x$, with $S(x=0)=0$, $S(x_1)=S_1$ and $S(x_2)=S_2$. The functional  $\hat{E}$ relevant for our system is
\begin{subequations}
\begin{equation}
	\hat{E}=L + \int_0^{x_1}dx\mathcal{L}_{1} + \int_{x_1}^{x_2}dx\mathcal{L}_{0},\text{ with} 
\end{equation}    
\begin{equation}
	L = \sigma_{\rm f} R_1\frac{2\beta}{\sin\beta} + (P_{1}-P_{0})R_1^2\frac{2\beta-\sin(2\beta)}{2\sin^2(\beta)} +\lambda\left(R_1 + \frac{\sigma_{\rm f}}{P_{1}-P_{0}}\sin\beta\right),
\end{equation}
\begin{equation}
	\mathcal{L}_{1} = S^\prime K_B\left(\frac{\psi^\prime}{S^\prime}\right)^2 + 2S^\prime\sigma_1 + 2S^\prime\left(P_{\rm v}-P_{1}\right)R\sin\psi+\gamma\left(R^\prime - S^\prime\cos\psi\right),
\end{equation}
\begin{equation}
	\mathcal{L}_{0} = S^\prime K_B\left(\frac{\psi^\prime}{S^\prime}\right)^2 + 2S^\prime\sigma_0 + 2S^\prime\left(P_{\rm v}-P_{0}\right)R\sin\psi+\gamma\left(R^\prime - S^\prime\cos\psi\right),
\end{equation}
\end{subequations}
with $\lambda$ a Lagrange multiplier that enforces the Laplace pressure relation and $\gamma$ a Lagrange multiplier function that constrains the relation between $R$ and $\psi$. The derivative with respect to $x$ is denoted by a prime. The equilibrium shape is obtained, when the variation $\delta \hat{E}$ vanishes, where $\delta \hat{E}$ reads
\begin{align}
	\delta \hat{E}= &
	\int_0^{x_1}dx \left\{ \left[ \frac{\partial \mathcal{L}_{1}}{\partial \psi} - \frac{d}{dx}\frac{\partial \mathcal{L}_{1}}{\partial \psi^\prime} \right]\delta\psi +
	\left[ \frac{\partial \mathcal{L}_{1}}{\partial R} - \frac{d}{dx}\frac{\partial \mathcal{L}_{1}}{\partial R^\prime} \right]\delta R +
	\frac{\partial \mathcal{L}_{1}}{\partial \gamma}\delta\gamma -
	\frac{d}{dx}\frac{\partial \mathcal{L}_{1}}{\partial S^\prime}\delta S
	\right\}\\
	& + \frac{\partial \mathcal{L}_{1}}{\partial \psi^\prime}\delta\psi\bigg|_0^{x_1}
	+ \frac{\partial \mathcal{L}_{1}}{\partial R^\prime}\delta R\bigg|_0^{x_1}
	+ \frac{\partial \mathcal{L}_{1}}{\partial S^\prime}\delta S\bigg|_0^{x_1}\nonumber\\
	&+\int_{x_1}^{x_2}dx \left\{ \left[ \frac{\partial \mathcal{L}_{0}}{\partial \psi} - \frac{d}{dx}\frac{\partial \mathcal{L}_{0}}{\partial \psi^\prime} \right]\delta\psi +
	\left[ \frac{\partial \mathcal{L}_{0}}{\partial R} - \frac{d}{dx}\frac{\partial \mathcal{L}_{0}}{\partial R^\prime} \right]\delta R +
	\frac{\partial \mathcal{L}_{0}}{\partial \gamma}\delta\gamma -
	\frac{d}{dx}\frac{\partial \mathcal{L}_{0}}{\partial S^\prime}\delta S
	\right\}\nonumber\\
	& + \frac{\partial \mathcal{L}_{0}}{\partial \psi^\prime}\delta\psi\bigg|_{x_1}^{x_2}
	+ \frac{\partial \mathcal{L}_{0}}{\partial R^\prime}\delta R\bigg|_{x_1}^{x_2}
	+ \frac{\partial \mathcal{L}_{0}}{\partial S^\prime}\delta S\bigg|_{x_1}^{x_2}\nonumber\\
	&+\frac{\partial L}{\partial R_1} \delta R\bigg|_{x_1} + \frac{\partial L}{\partial \beta}\delta\beta+ \frac{\partial L}{\partial \lambda}\delta\lambda.\nonumber
\end{align}	
From $\delta \hat{E}=0$ we obtain the Euler-Lagrange equations
\begin{subequations}
\begin{equation}
	\frac{\partial \mathcal{L}_{1}}{\partial \psi} - \frac{d}{dx}\frac{\partial \mathcal{L}_{1}}{\partial \psi^\prime} =0,
\end{equation}    
\begin{equation}
	\frac{\partial \mathcal{L}_{1}}{\partial R} - \frac{d}{dx}\frac{\partial \mathcal{L}_{1}}{\partial R^\prime} =0,
\end{equation}
\begin{equation}
	\frac{\partial \mathcal{L}_{1}}{\partial \gamma}=0,
\end{equation}
\begin{equation}
	\frac{d}{dx}\frac{\partial \mathcal{L}_{1}}{\partial S^\prime}=0, \text{ with } i=0,1
\end{equation}
\end{subequations}
and boundary conditions that are discussed further below. From the Euler-Lagrange equations the following equations follow directly.
\begin{subequations}
\begin{equation}
	\frac{d^2\psi}{dS^2}=\frac{P_{\rm v}-P_i}{K_B}R\cos\psi +\frac{\gamma}{2K_B}\sin\psi
\end{equation}    
\begin{equation}
	\frac{d}{dS}\frac{\gamma}{2K_B}=\frac{P_{\rm v}-P_i}{K_B}\sin\psi
	 \label{eq:si_euler_lagrange_gamma}
\end{equation}
\begin{equation}
	\frac{dR}{dS}=\cos\psi
\end{equation}
\begin{equation}
	 \frac{d\mathcal{H}_i}{dS}=0, \text{with } \mathcal{H}_i=-\frac{1}{2}\left(\frac{d\psi}{dS}\right)^2+\frac{\sigma_1}{K_B} +  \frac{P_{\rm v}-P_i}{K_B}R\sin\psi -\frac{\gamma}{2K_B}\cos\psi,  \text{ for } i=0,1.
	 \label{eq:si_euler_lagrange_hamiltonian}
\end{equation}
\end{subequations}
Next, we derive the boundary conditions at the contact point $S_1$. Both the radius $R$ and the tilt angle $\psi$ have to be continuous at the contact point. From $\frac{\partial L}{\partial \lambda}=0$ we obtain the Laplace pressure relation Eq. \ref{eq:si_laplace_equation}. Inserting Eq. \ref{eq:si_laplace_equation} into $\frac{\partial L}{\partial \beta}=0$ leads to 
\begin{equation}
	\lambda \frac{\sigma_{\rm f}}{P_{1}-P_{0}}\cos\beta =0,
	\label{eq:si_lambda_is_zero}
\end{equation}
which implies $\lambda=0$. Using Eqs. \ref{eq:si_laplace_equation}, \ref{eq:si_lambda_is_zero} and $\delta R(x_1)\neq0$ we find the following boundary condition:
\begin{equation}
	\gamma(S_1)\bigg|_{1} - \gamma(S_1)\bigg|_{0} +2\sigma_{\rm f}\cos\beta=0.
	\label{eq:si_boundary_condition_gamma}
\end{equation}
From $\delta \psi(x_1)\neq0$ we find
\begin{equation}
	\frac{d\psi}{dS}(S_1)\bigg|_{1} -\frac{d\psi}{dS}(S_1)\bigg|_{0} =0.
	\label{eq:si_boundary_condition_mean_curvature}
\end{equation}
And from $\delta S(x_1)\neq0$  we find
\begin{equation}
	\mathcal{H}_{1}(S_1)-\mathcal{H}_{0}(S_1)=0.
	\label{eq:si_boundary_condition_hamiltonian}
\end{equation}
Eq. \ref{eq:si_boundary_condition_hamiltonian} together with Eqs. \ref{eq:si_laplace_equation}, \ref{eq:si_boundary_condition_gamma} and \ref{eq:si_boundary_condition_mean_curvature} leads to
\begin{equation}
	\cos\left(\beta-\psi(S_1)\right)=\frac{\sigma_0-\sigma_1}{\sigma_{\rm f}}.
\end{equation}
Finally from $\delta S(x_2)\neq0$ follows
\begin{equation}
	\mathcal{H}_{0}(S_2)=0,
\end{equation}
which implies
\begin{equation}
	\mathcal{H}_{1}=\mathcal{H}_{0}=0,
	\label{eq:si_hamiltonian_is_zero}
\end{equation}
where we use Eqs. \ref{eq:si_euler_lagrange_hamiltonian} and \ref{eq:si_boundary_condition_hamiltonian}. To describe a closed shape the radius and the tilt angle have to fulfill $R(S=0)=0$, $\psi(S=0)=0$ and $R(S_2)=0$, $\psi(S_2)=\pi$. In summary, the system is described by the following shape equations
\begin{subequations}
\begin{equation}
	\frac{dR}{dS}=\cos\psi	
\end{equation}
\begin{equation}
	\frac{dZ}{dS}=-\sin\psi		
\end{equation}
\begin{equation}
	\frac{d^2\psi}{dS^2}=	\begin{cases}
    \frac{P_{\rm v}-P_{1}}{K_B}R\cos\psi +\frac{\gamma}{2K_B}\sin\psi, & S<S_1\\
    \frac{P_{\rm v}-P_{0}}{K_B}R\cos\psi +\frac{\gamma}{2K_B}\sin\psi, & S>S_1
  \end{cases}
\end{equation}
\begin{equation}
	\frac{d\gamma}{dS}=	\begin{cases}
   2(P_{\rm v}-P_{1})\sin\psi, & S<S_1\\
   2(P_{\rm v}-P_{0})\sin\psi, & S>S_1,
  \end{cases}
\end{equation}
\end{subequations}
which have to fulfill the boundary conditions
\begin{subequations}
\begin{equation}
	R(0)=0;\quad Z(0)=0;\quad\psi(0)=0
\end{equation}
\begin{equation}
	\gamma(S_1)\bigg|_{0}=\gamma(S_1)\bigg|_{1} +2\sigma_{\rm f}\cos\beta;\quad \frac{d\psi}{dS}(S_1)\bigg|_{1} =\frac{d\psi}{dS}(S_1)\bigg|_{0};
\end{equation}
\begin{equation}
	 \cos\left(\beta-\psi(S_1)\right)=\frac{\sigma_0-\sigma_1}{\sigma_{\rm f}};\quad P_{1}-P_{0}=-\sigma_{\rm f}\frac{\sin\beta}{R(S_1)}
\end{equation}
\begin{equation}
	R(S_2)=0;\quad\psi(S_2)=\pi.
\end{equation}
\end{subequations}

\subsection{Large vesicle limit}
Suppose the size of the vesicle is significantly larger than the size of the droplet. In that case, the vesicle shape away from the contact region between the droplet and membrane appears as a planar surface. We then describe the vesicle as an infinite surface approaching a planar shape at infinity. For $S\to\infty$ the shape thus has to fulfill $\psi(S\to\infty)=0$, $R(S\to\infty)\to\infty$ and $d\psi/dS\big|_{S\to\infty}=0$. From $\mathcal{H}_{0}=0$ (Eq. \ref{eq:si_hamiltonian_is_zero}) it thus follows that $P_{\rm v}=P_{0}$ and $\gamma(S\to\infty)=2\sigma_0$. From $P_{\rm v}=P_{0}$ and Eq. \ref{eq:si_euler_lagrange_gamma} we furthermore find that $\gamma$ is constant for $S>S_1$. For the curvature beyond the contact point, this ultimately results in 
\begin{equation}
	\frac{d\psi}{dS}=\sqrt{\frac{2\sigma_0}{K_B}(1-\cos\psi)},\quad S>S_1
\end{equation}
The shape equations in the contact region of membrane and droplet ($S<S_1$) read 
\begin{subequations}
\begin{equation}
	\frac{dR}{dS}=\cos\psi	
\end{equation}
\begin{equation}
	\frac{dZ}{dS}=-\sin\psi		
\end{equation}
\begin{equation}
	\frac{d^2\psi}{dS^2}= -\frac{P_{1}-P_{0}}{K_B}R\cos\psi +\frac{\gamma}{2K_B}\sin\psi
\end{equation}
\begin{equation}
	\frac{d\gamma}{dS}=	-2(P_{1}-P_{0})\sin\psi
\end{equation}
\end{subequations}
with the boundary conditions
\begin{subequations}
\begin{equation}
	R(0)=0;\quad Z(0)=0;\quad\psi(0)=0
\end{equation}
\begin{equation}
	\gamma(S_1)=2\sigma_0-2\sigma_{\rm f}\cos\beta;\quad \frac{d\psi}{dS}\bigg|_{S_1}=\sqrt{\frac{2\sigma_0}{K_B}(1-\cos\psi(S_1))};
\end{equation}
\begin{equation}
	 \cos\left(\beta-\psi(S_1)\right)=\frac{\sigma_0-\sigma_1}{\sigma_{\rm f}};\quad P_{1}-P_{0}=-\sigma_{\rm f}\frac{\sin\beta}{R(S_1)}
\end{equation}
\end{subequations}
We use the elastocapillary length $\ell=\sqrt{K_B/\sigma_{\rm f}}$ to introduce the dimensionless variables $s=S/\ell$, $r=R/\ell$, $z=(Z-Z(S_1))/\ell$, $\bar{\gamma}=\gamma\ell^2/K_B$ and $\bar{\sigma}_0=\sigma_0/\sigma_{\rm f}$. Furthermore, we introduce the intrinsic contact angle $\theta_0$, which is defined as
\begin{equation}
	\theta_0 = \arccos\left(\frac{\sigma_0-\sigma_1}{\sigma_{\rm f}}\right).
\end{equation}
The shape equations in the contact region in dimensionless variables reads
\begin{subequations}
\begin{equation}
	\frac{dr}{ds}=\cos\psi	
\end{equation}
\begin{equation}
	\frac{dz}{ds}=-\sin\psi		
\end{equation}
\begin{equation}
	\frac{d^2\psi}{ds^2}= \frac{\sin(\theta_0+\psi_1)}{r_1}r\cos\psi +\frac{\bar{\gamma}}{2}\sin\psi
\end{equation}
\begin{equation}
	\frac{d\bar{\gamma}}{ds}=	2\frac{\sin(\theta_0+\psi_1)}{r_1}\sin\psi,
\end{equation}
\end{subequations}
with $\psi_1=\psi(s_1)$ and $r_1=r(s_1)$. The boundary conditions are
\begin{subequations}
\begin{equation}
	r(0)=0;\quad\psi(0)=0.
\end{equation}
\begin{equation}
	z(s_1)=0;\quad\bar{\gamma}(s_1)=2\bar{\sigma}_0-2\cos(\theta_0+\psi_1);\quad \frac{d\psi}{ds}\bigg|_{s_1}=\sqrt{2\bar{\sigma}_0(1-\cos\psi_1)};
\end{equation}
\end{subequations}

  \bibliographystyle{siamplain} 
  \bibliography{bibliographie}





%% file: ex_shared.tex
\makeatletter
\def\ps@pprintTitle{%
  \let\@oddhead\@empty
  \let\@evenhead\@empty
  \def\@oddfoot{\reset@font\hfil\thepage\hfil}
  \let\@evenfoot\@oddfoot
}
\makeatother

\usepackage{amssymb}
\usepackage[euler]{textgreek}
\usepackage{amsmath}
\usepackage{mathtools}
\usepackage{verbatim}
\usepackage{bm}
\usepackage[inkscapeformat=png]{svg}
\usepackage[greek,english]{babel}
\makeatletter
\let\oldtheequation\theequation
\renewcommand\tagform@[1]{\maketag@@@{\ignorespaces#1\unskip\@@italiccorr}}
\renewcommand\theequation{(\oldtheequation)}
\makeatother
\usepackage{multirow} 
\usepackage{multicol} 
\usepackage{tabularx} 
\usepackage{longtable} 
\usepackage{array}
\usepackage{todonotes}
\usepackage{float}
\usepackage{graphicx} 
\usepackage{color} 
\graphicspath{{images/}}
\DeclareGraphicsExtensions{.pdf,.png,.jpg} 
\usepackage{subfigure} 
\usepackage[all]{hypcap} 
\usepackage{soul}
 \usepackage{tikz}   

  \usetikzlibrary{arrows,calc,snakes}%

\usepackage{paralist}
\usepackage{bibgerm} 
\usepackage{cleveref}
\crefname{figure}{Fig.}{Figs.}
\crefname{subfigure}{Fig.}{Figs.}
\usepackage{breqn}

\addto\extrasenglish{%
}
\usepackage{lipsum}
\usepackage{amsfonts}
\usepackage{graphicx}
\usepackage{epstopdf}
\usepackage{algorithmic}
\ifpdf
  \DeclareGraphicsExtensions{.eps,.pdf,.png,.jpg}
\else
  \DeclareGraphicsExtensions{.eps}
\fi

\newsiamremark{remark}{Remark}
\newsiamremark{hypothesis}{Hypothesis}
\crefname{hypothesis}{Hypothesis}{Hypotheses}
\newsiamthm{claim}{Claim}

\headers{Wetting dynamics of liquid droplets
on deformable membranes}{M. Mokbel, D. Mokbel, S. Liese, C. A. Weber, and S. Aland}

\title{A simulation method for the wetting dynamics of liquid droplets
on deformable membranes
}

\author{Marcel Mokbel\thanks{Faculty of Mathematics and Informatics, TU Freiberg, Akademiestr. 6, 09599 Freiberg, Germany.}
\and Dominic Mokbel\footnotemark[2]
\and Susanne Liese\thanks{Faculty of Mathematics, Natural Sciences, and Materials Engineering: Institute of Physics, University of Augsburg,
Universitätsstr. 1, 86159 Augsburg, Germany.}
\and Christoph Weber\footnotemark[3]
\and Sebastian Aland\footnotemark[2] \thanks{Faculty of Informatics/Mathematics, HTW Dresden, Friedrich-List-Platz 1,  01069 Dresden, Germany. \\Center for Systems Biology Dresden, Pfotenhauerstr. 108, 01307 Dresden, Germany. \\sebastian.aland@math.tu-freiberg.de}}

\usepackage{amsopn}
